\documentclass[preprint2,trackchanges,appendixfloats]{aastex61_mod}

\usepackage[T1]{fontenc}
\usepackage[utf8]{inputenc}
\usepackage{multirow}
\usepackage{booktabs}
\usepackage{float}
\usepackage{xspace}

\newcommand{\nh}[3]{NH$_#1$~(#2,#3)}
\newcommand{\temp}[2]{T$_{#1#2}$}
\newcommand{\kms}{km\,s$^{-1}$\xspace}
\newcommand{\KMyr}{K\,Myr$^{-1}$\xspace}
\newcommand{\sqcm}{cm$^{-2}$\xspace}
\newcommand{\jybeam}{Jy\,beam$^{-1}$\xspace}

\defcitealias{Kruijssen2015}{K15}

\begin{document}

\hyphenation{kruijs-sen}

\received{07 May 2017}
\revised{17 October 2017}
\accepted{18 October 2017}
\published{-}
\submitjournal{\apjs}

\title{The Survey of Water and Ammonia in the Galactic Center (SWAG):\\ Molecular Cloud Evolution in the Central Molecular Zone}

\author[0000-0003-1104-2014]{Nico Krieger}
	\altaffiliation{email: krieger@mpia.de}
	\affiliation{Max-Planck-Institut f\"ur Astronomie, K\"onigstuhl 17, 69120 Heidelberg, Germany}

\author{J\"urgen Ott}
	\affiliation{National Radio Astronomy Observatory, P.O. Box O, 1003 Lopezville Road, Socorro, NM 87801, USA}

\author{Henrik Beuther}
	\affiliation{Max-Planck-Institut f\"ur Astronomie, K\"onigstuhl 17, 69120 Heidelberg, Germany}

\author{Fabian Walter}
	\affiliation{Max-Planck-Institut f\"ur Astronomie, K\"onigstuhl 17, 69120 Heidelberg, Germany}
	\affiliation{National Radio Astronomy Observatory, P.O. Box O, 1003 Lopezville Road, Socorro, NM 87801, USA}

\author{J.~M.~Diederik Kruijssen}
	\affiliation{Astronomisches Rechen-Institut, Zentrum f\"ur Astronomie der Universit\"at Heidelberg, M\"onchhofstraße 12-14, D-69120 Heidelberg, Germany}
	\affiliation{Max-Planck-Institut f\"ur Astronomie, K\"onigstuhl 17, 69120 Heidelberg, Germany}

\author{David S. Meier}
	\affiliation{New Mexico Institute of Mining and Technology, 801 Leroy Place, Socorro, NM 87801, USA}

\author{Elisabeth~A.~C. Mills}
	\affiliation{San Jose State University, 1 Washington Square, San Jose, CA 95192, USA}

	% in alphabetical order
\author{Yanett Contreras}
	\affiliation{Leiden Observatory, Leiden University, PO Box 9513, NL-2300 RA Leiden, the Netherlands}

\author{Phil Edwards}
	\affiliation{CSIRO Astronomy and Space Science, PO Box 76, Epping, NSW 1710, Australia}

\author{Adam Ginsburg}
	\affiliation{National Radio Astronomy Observatory, P.O. Box O, 1003 Lopezville Road, Socorro, NM 87801, USA}
	\affiliation{European Southern Observatory, Karl-Schwarzschild-Straße 2, 85748 Garching bei München, Germany}

\author{Christian Henkel}
	\affiliation{Max-Planck-Institut f\"r Radioastronomie, Auf dem H\"ugel 69, 53121 Bonn, Germany}
	\affiliation{Astronomy Department, King Abdulaziz University, P.O. Box 80203, Jeddah 21589, Saudi Arabia}

\author{Jonathan Henshaw}
	\affiliation{Max-Planck-Institut f\"ur Astronomie, K\"onigstuhl 17, 69120 Heidelberg, Germany}

\author{James Jackson}
	\affiliation{School of Mathematical and Physical Sciences, University of Newcastle, University Drive, Callaghan NSW 2308, Australia}

\author{Jens Kauffmann}
	\affiliation{Max-Planck-Institut f\"r Radioastronomie, Auf dem H\"ugel 69, 53121 Bonn, Germany}

\author{Steven Longmore}
	\affiliation{Astrophysics Research Institute, Liverpool John Moores University, 146 Brownlow Hill, Liverpool L3 5RF, UK}

\author{Sergio Mart\'{i}n}
	\affiliation{European Southern Observatory, Alonso de Córdova 3107, Vitacura Casilla 763 0355, Santiago, Chile}
	\affiliation{Joint ALMA Observatory, Alonso de Córdova 3107, Vitacura Casilla 763 0355, Santiago, Chile}

\author{Mark R. Morris}
	\affiliation{Department of Physics and Astronomy, UCLA, 430 Portola Plaza, Los Angeles, CA 90095-1547, USA}

\author{Thushara Pillai}
	\affiliation{Max-Planck-Institut f\"r Radioastronomie, Auf dem H\"ugel 69, 53121 Bonn, Germany}

\author{Matthew Rickert}
	\affiliation{Department of Physics and Astronomy and CIERA, Northwestern University, Evanston, IL 60208, USA}

\author{Erik Rosolowsky}
	\affiliation{Dept. of Physics, CCIS 4-181, University of Alberta, Edmonton, Alberta, T6G 2E1, Canada}

\author{Hiroko Shinnaga}
	\affiliation{Department of Physics, Kagoshima University, 1-21-35, Korimoto, Kagoshima, 890-0065, Japan}

\author{Andrew Walsh}
	\affiliation{International Centre for Radio Astronomy Research, Curtin University, Bentley, 6102, Australia}

\author{Farhad Yusef-Zadeh}
	\affiliation{Department of Physics and Astronomy and CIERA, Northwestern University, Evanston, IL 60208, USA}

\author{Qizhou Zhang}
	\affiliation{Harvard-Smithsonian Center for Astrophysics, 60 Garden Street, Cambridge, MA 02138}

\shorttitle{SWAG: Molecular Cloud Evolution in the CMZ}
\shortauthors{Krieger et al.}

\begin{abstract}
The Survey of Water and Ammonia in the Galactic Center (SWAG) covers the Central Molecular Zone (CMZ) of the Milky Way at frequencies between 21.2 and 25.4\,GHz obtained at the Australia Telescope Compact Array at $\sim 0.9$\,pc spatial and  $\sim 2.0$\,\kms spectral resolution.
In this paper, we present data on the inner $\sim 250$\,pc ($1.4^\circ$) between Sgr~C and Sgr~B2. We focus on the hyperfine structure of the metastable ammonia inversion lines (J,K) = (1,1) - (6,6) to derive column density, kinematics, opacity and kinetic gas temperature. In the CMZ molecular clouds, we find typical line widths of $8-16$\,\kms and extended regions of optically thick ($\tau > 1$) emission. Two components in kinetic temperature are detected at $25-50$\,K and $60-100$\,K, both being significantly hotter than dust temperatures throughout the CMZ.
We discuss the physical state of the CMZ gas as traced by ammonia in the context of the orbital model by \citet{Kruijssen2015} that interprets the observed distribution as a stream of molecular clouds following an open eccentric orbit. This allows us to statistically investigate the time dependencies of gas temperature, column density and line width. We find heating rates between $\sim 50$ and $\sim 100$\,K\,Myr$^{-1}$ along the stream orbit. No strong signs of time dependence are found for column density or line width. These quantities are likely dominated by cloud-to-cloud variations. Our results qualitatively match the predictions of the current model of tidal triggering of cloud collapse, orbital kinematics and the observation of an evolutionary sequence of increasing star formation activity with orbital phase.
\end{abstract}


\section{Introduction}\label{section: introduction}

\begin{figure*}
	\centering
	\includegraphics[width=\linewidth]{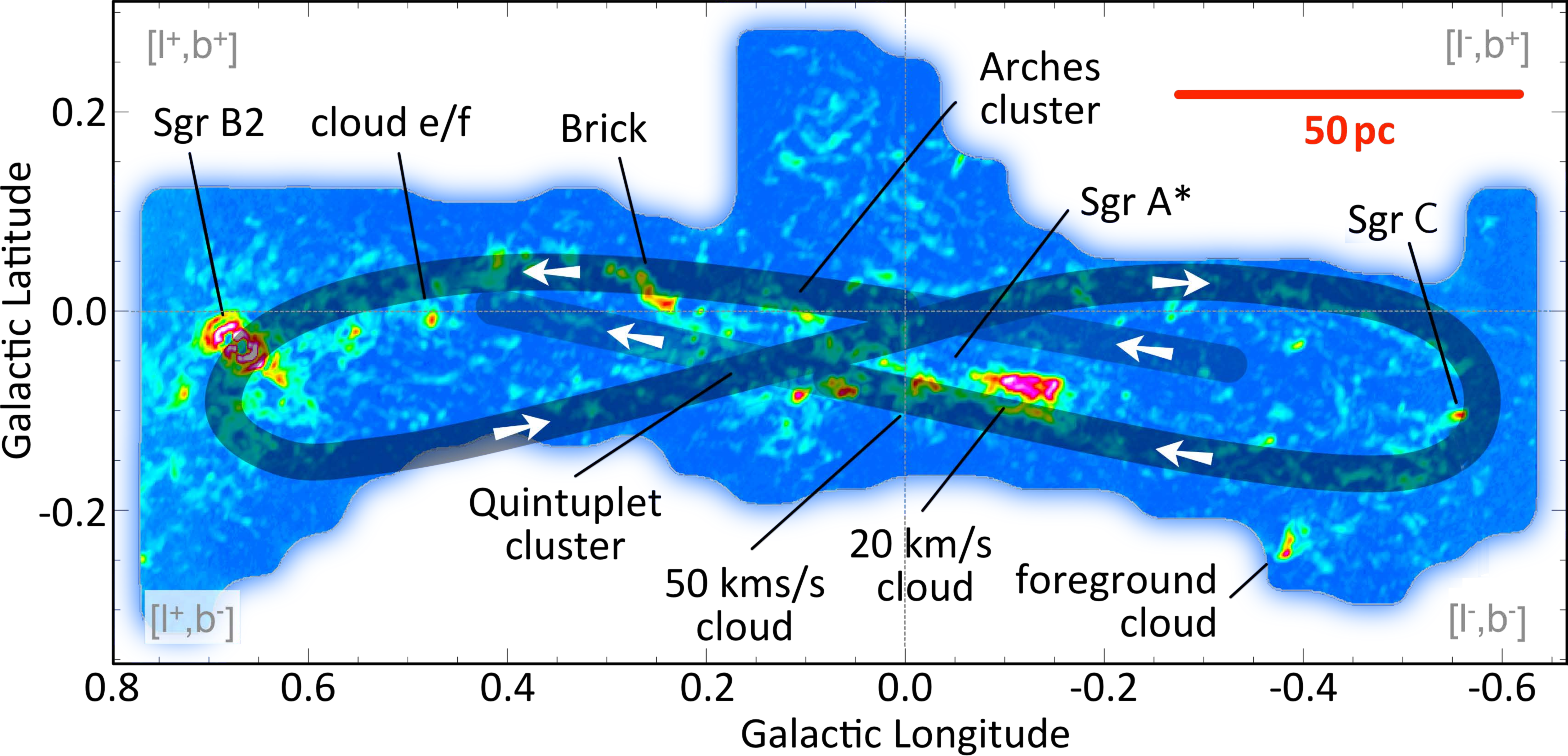}
	\caption{Overview of the GC. The gas (background, SWAG \nh333 peak intensity) resembles the shape of an $\infty$ symbol and is parametrized as a sequence of stream segments by \citetalias{Kruijssen2015}. The direction of motion is along the dust ridge ([$l^+, b^+$] denoting positive longitude and latitude) to Sgr B2, [$l^+, b^-$], [$l^-, b^+$] to Sgr C and [$l^-, b^-$]. Gas in the dust ridge [$l^+, b^+$] and [$l^-, b^-$] is mostly on the near side (in front of the GC) whereas [$l^+, b^-$] and [$l^-, b^+$] gas is mostly on the far side \citep{Bally2010,Molinari2011}. The most important sources are labeled.}
	\label{figure: GC overview}
\end{figure*}

The Galactic Center (GC) and the Central Molecular Zone (CMZ) in particular represent an environment with conditions that are not to be found anywhere else on large scales in the Milky Way.
The CMZ received its name due to the presence of a large reservoir of dense ($\gtrsim 10^4$\,\sqcm) and molecular gas of a few times $10^7$ \,M$_{\odot}$ \citep{Oka1998,Morris&Serabyn1996,Ferriere2007} and covers the central $\sim 500$\,pc of the GC region.
The large amount of molecular gas is found to be accompanied by a relatively high star formation rate (SFR) of $\sim 0.1$\,M$_\odot$\,yr$^{-1}$ \citep{Longmore2013a,Barnes2017}.
Star formation (SF) laws that build on the assumption of a constant gas depletion time of $1$\,Gyr can fit the observed SFR \citep{Bigiel2010,Leroy2008,Leroy2015}.
However, these relations are derived from gas at much lower densities than observed in the CMZ.
Density dependent SF laws, on the other hand, strongly overpredict the SFR \citep{Longmore2013a} at $\sim 0.4$\,M$_\odot$\,yr$^{-1}$ \citep{Kennicutt1998,Krumholz&McKee2005,Krumholz2012} to $\sim 0.8$\,M$_\odot$\,yr$^{-1}$ \cite{Lada2010,Lada2012}.
Thus, the star formation efficiency (SFE) in the CMZ is significantly lower than expected for the observed gas densities.
The complex interplay of energetic processes in the GC allows for different potential answers to this problem of low SFE, but none of those processes alone can explain the discrepancy \citep{Kruijssen2014}.
A general picture of episodic starbursts in gas rings in galactic centers introduced by \citet{Krumholz2015} and \citet{Krumholz2017} might solve the SFE problem in the particular case of the CMZ and set the framework for another intriguing feature of the GC: a ring-like structure of dust and molecular gas.
This structure (see Fig.~\ref{figure: GC overview} for an overview) is projected onto an infinity ($\infty$) shape which follows several arcs, the most prominent being the so-called ``dust ridge'' stretching from the massive molecular cloud G0.253+0.016 (``the Brick'') to the star forming region Sgr B2 \citep{Lis1999}.
It might be continued via an [$l^+, b^-$] arc and another loop at negative longitude ([$l^-, b^+$], [$l^-, b^-$]) passing through the star forming region Sgr C.
The syntax [$l^{+/-}, b^{+/-}$] denotes the four quadrants in Galactic coordinates at positive/negative Galactic longitude (l) and latitude (b).
Clouds at [$l^-, b^-$] and [$l^+, b^+$] are thought to be located in front of Sgr~A*, i.e. the near side of the GC that is visible in silhouette against the background, whereas most of the [$l^-, b^+$] and [$l^+, b^-$] gas is more likely to be on the back side, i.e. behind Sgr~A* \citep{Bally2010,Molinari2011}.

Various models of the gas and dust distribution in the GC exist:
a simple bar model \citep{Morris&Serabyn1996}; variations of a spiral arm model in which the apparent ring is formed by the inner part of two spiral arms \citep{Sofue1995,Sawada2004,RF2008,RF2011,Ridley2017}; a closed, twisted elliptical ring detected in dust emission \citep{Molinari2011} and a sequence of open-ended gas streams \citep[][hereafter K15]{Kruijssen2015}.
\citetalias{Kruijssen2015} highlight the impossibility of closed orbits in extended gravitational potentials and provide a better fit to single-dish ammonia emission in position-position-velocity (PPV) space as was confirmed by \citet{Henshaw2016a} for three molecular species in single-dish observations.
Thus, we will focus on this model and, henceforth, call it the ``stream model'' in contrast to the ``ring model'' and the ``spiral arm'' model.
According to the stream model, the stream of GC molecular clouds oscillates radially and vertically, and can be traced for $\sim 1.5$ revolutions around the GC.
The radial oscillation periodically brings dense molecular gas closer ($r \sim 60$\,pc at pericenter) to the gravitational center (traced by Sgr~A*) and deeper into the potential whereas apocenter occurs at $r \sim 120$\,pc.
As the CMZ clouds slowly evolve towards low virial ratios as the turbulent energy dissipates \citep{Walker2015,Krumholz2015,Henshaw2016b}, it is statistically more likely that cloud collapse occurs at the pericenter where the compressive tidal forces are strongest along the orbit and thus a cloud receives the final nudge to transition to self-gravitation.
Subsequent SF stages will then happen downstream from the pericenter passages and could potentially be observed as a SF sequence if the orbit is sufficiently sampled with molecular clouds that start to collapse at a similar point of their orbits.
This model of triggered star formation was first proposed by \citet{Longmore2013b} due to the observation of different evolutionary SF stages along the dust ridge.
The two young stellar clusters in the GC, Arches and Quintuplet, may also support this model as their orbits and ages are consistent with formation at a common point after collapse of their respective parent clouds at a pericenter passage \citep[Fig.~\ref{figure: GC overview},][]{Stolte2014, Kruijssen2015}.
Beside stars and star formation tracers, cloud properties are also expected to show evolutionary behavior along the gas streams which has not been thoroughly tested yet.
Weak hints towards rising gas temperatures in the dust ridge are suggested by \citet{Ginsburg2016}, while a recent paper by \citet{Kauffmann2016b} based on N$_2$H$^+$ data from the Galactic Center Molecular Cloud Survey cannot confirm, nor exclude, the possibility of triggered evolution when examining mass-size relation and SF suppression.
Both analyses were based on a low number of measurements ($\sim 35$ measurements in dust ridge clouds in \citealt{Ginsburg2016} and six clouds in \citealt{Kauffmann2016b}) and lack the statistical power to detect potential evolution in the presence of scatter.

In this paper, we aim to test the prediction of the stream model by searching for coherent temporal evolution of molecular gas properties on a broad statistical basis using the ``Survey of Water and Ammonia in the Galactic Center'' (SWAG).
SWAG targets 42 molecular and atomic species at $\sim 21 - 25$\,GHz of which at least 20 are detected and thus offers an ideal database for such an analysis.
SWAG maps the CMZ in the region of $\sim -1^\circ$ to $\sim +2^\circ$ in Galactic longitude at latitudes $|b| < 0.4^\circ$ using the Australia Telescope Compact Array (ATCA) and achieves a spectral resolution of 0.4\,\kms at $\sim 20"$ spatial resolution.
One of the survey's signature targets is the ammonia molecule, the properties of which are well suited to explore the thermal and kinematic structure of the ISM \cite{Ho1983}.
In an ammonia (NH$_3$) molecule, the nitrogen atom can tunnel through the plane of hydrogen atoms (inversion) at a rotational-state-dependent frequency which can be related to temperature.
These inversion frequencies are closely spaced ($\Delta \nu_{(1,1)...(6,6)} \sim 1.4$\,GHz) while covering a large energy range (excitation energy $22<$ T$_{l}$\,[K] $ \lesssim 400$) that traces at moderately dense gas of $10^4 - 10^6$\,cm$^{-3}$ due to the critical density being $n_{crit} \sim10^3$\,cm$^{-3}$ \citep[e.g.][]{Shirley2015}.
Detections of ammonia span a large range of sources from circumstellar envelopes \citep[e.g.][]{Betz1979} over galactic dark clouds \citep[e.g.][]{Ho1978} and nearby galaxies such as NGC 253 \citep[e.g.][]{Ott2005,Lebron2011,Mangum2013,Gorski2017} to heavily star-forming galaxies like ULIRGS \citep[e.g. Arp 220,][]{Ott2011,Zschaechner2016b} and even at high redshift ($z = 0.9$) as absorption against continuum emission \citep{Henkel2008}.
In the GC, SWAG offers high resolution ($\sim 0.9$\,pc) interferometric ammonia maps that show the intermediate density gas structure most relevant to the stream model because large scale diffuse emission $\gtrsim 1'.4$ ($\gtrsim 3.3$\,pc at 8.3\,kpc\footnote{We adopt this distance measurement by \citet{Reid2014} throughout the paper. The IAU recommends to use 8.5\,kpc while a recent study by \citet{Boehle2016} found 7.86\,kpc.}) is filtered out by the ATCA interferometer in this setup.
Spectral fitting of the ammonia hyperfine structure allows us to construct maps of line-of-sight velocity, line width, opacity and kinetic gas temperature which can be tested for evolutionary behavior under the assumption of a kinematic model.
Applying this method to SWAG data and the stream model allows us to check gas properties for absolute time dependencies that can be expected if a sequential star formation sequence is present in one or more of the stream segments.

As the structures in the GC are complex, simple names can be misleading.
Throughout this paper we will refer to the ``ring'' as the general $\infty$-like structure of CMZ gas without being interested in the detailed structure or kinematics.
``Stream'', however, refers to the \citetalias{Kruijssen2015} model that introduced a stream as a set of four segments of an orbit wrapping around the GC whose subdivision does not have physical meaning but simplifies descriptions.

In this paper, we summarize the observational setup and data reduction of the SWAG survey in §\ref{section: observations and data reduction} on the basis of observations of one third of the total area, covering the region between Sgr B2 and Sgr C.
§\ref{section: results} shows and discusses the primary data products (channel and moment maps, spectra, position-velocity diagrams) and §\ref{section: hyperfine structure fitting} describes the fitting of the ammonia hyperfine structure inversion lines.
In section 5, we use the derived thermal and kinetic properties of GC molecular clouds to discuss their properties in the context of the orbital model of \citetalias{Kruijssen2015} by reconstructing potential time dependencies of these properties.
The resulting relations are discussed in the context of the star formation sequence as proposed by \citet{Longmore2013b}.
A summary in §\ref{section: summary and conclusions} concludes this work.
Appendices show further SWAG data products (\ref{appendix: SWAG data products}), details on imaging (\ref{appendix: mosmem fails}, the ammonia thermometer (\ref{appendix: ammonia thermometer}) and hyperfine structure line fitting (\ref{appendix: hyperfine fitting}), as well as further fit derived data products (\ref{appendix: temperature maps} \& \ref{appendix: further temperature comparisons}) and further time evolution plots (\ref{appendix: further sequence plots}).

\section{Observations and Data Reduction}\label{section: observations and data reduction}

The ``Survey of Water and Ammonia in the Galactic Center'' (SWAG) is a large survey aiming to map the entire Central Molecular Zone (CMZ) of the Milky Way from $\sim-1^\circ$ to $\sim+2^\circ$ in Galactic longitude at latitudes $|b| <0.4^\circ$.
The boundaries correspond to an integrated surface brightness level of 0.1\,K\,\kms of \nh333 from Mopra single-dish observations \citep[beam size $\sim 2.4$'][]{Ott2014b}.
The achieved spatial resolution is $\sim 22.8"$ ($20.6" - 25.0"$ in the range of 25.4\,GHz - 21.2\,GHz) which, at the Galactic Center distance, corresponds to $\sim 0.9$\,pc.
The velocity resolution is $\sim 0.4$\,\kms (0.37\,\kms to 0.45\,\kms for 25.4\,GHz - 21.2\,GHz).
The observations are performed in a mosaic of $\sim 6500$ pointings with the Australia Telescope Compact Array (ATCA) interferometer during $\sim 525$ hours spread over three years (2014 - 2016).
At a wavelength of around 1.3\,cm, 42 spectral lines were targeted in the range of 21.2 - 25.4\,GHz; amongst these are six metastable and three non-metastable ammonia inversion lines, water maser and atomic radio recombination lines and complex organic molecules.
At least 20 of the targeted lines are unambiguously detected.

\subsection{Observations}

The observations for SWAG were carried out with the ATCA\footnote{The Australia Telescope Compact Array is part of the Australia Telescope National Facility which is funded by the Australian Government for operation as a National Facility managed by CSIRO.} interferometer.
Achieving the maximum sensitivity to extended emission requires the most compact array configuration, H75, with baselines of $31-89$\,m between five dishes.
A sixth antenna which is not considered in this work provides baselines of $\sim 4.4$\,km to the inner antennas.
The primary beam FWHM of the 22\,m dishes at $\nu = 23$\,GHz is $\sim 2.4'$.
Every position in the targeted area is covered by the primary beam of at least three pointings.
In the radio K band regime, the H75 array configuration is not sensitive to emission extending over more than $\sim 1'.4$ (3.3\,pc) due to interferometric spatial filtering.

\begin{figure*}
	\centering
	\includegraphics[width=\linewidth]{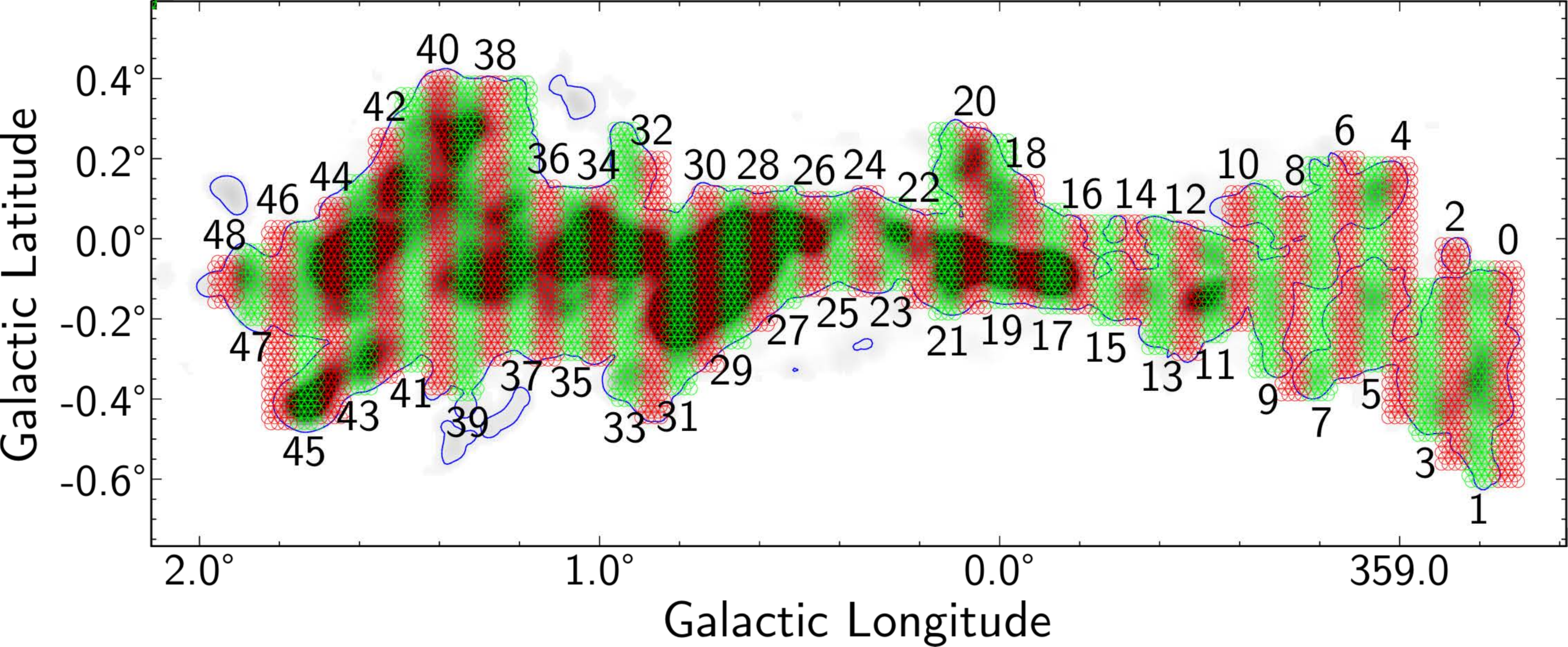}
	\caption{SWAG observing layout. The spatial coverage of SWAG is derived from the 0.1\,K\,\kms contour (blue) of the Mopra single dish CMZ survey by \citet[][background gray scale image]{Ott2014a}. 40 to 220 pointings are included in an observation strip (numbered from west to east and alternately colored). This work includes data of the inner CMZ covered by strips \#10 to \#30.}
	\label{figure: SWAG strips}
\end{figure*}

\begin{figure}
	\centering
	\includegraphics[width=\linewidth]{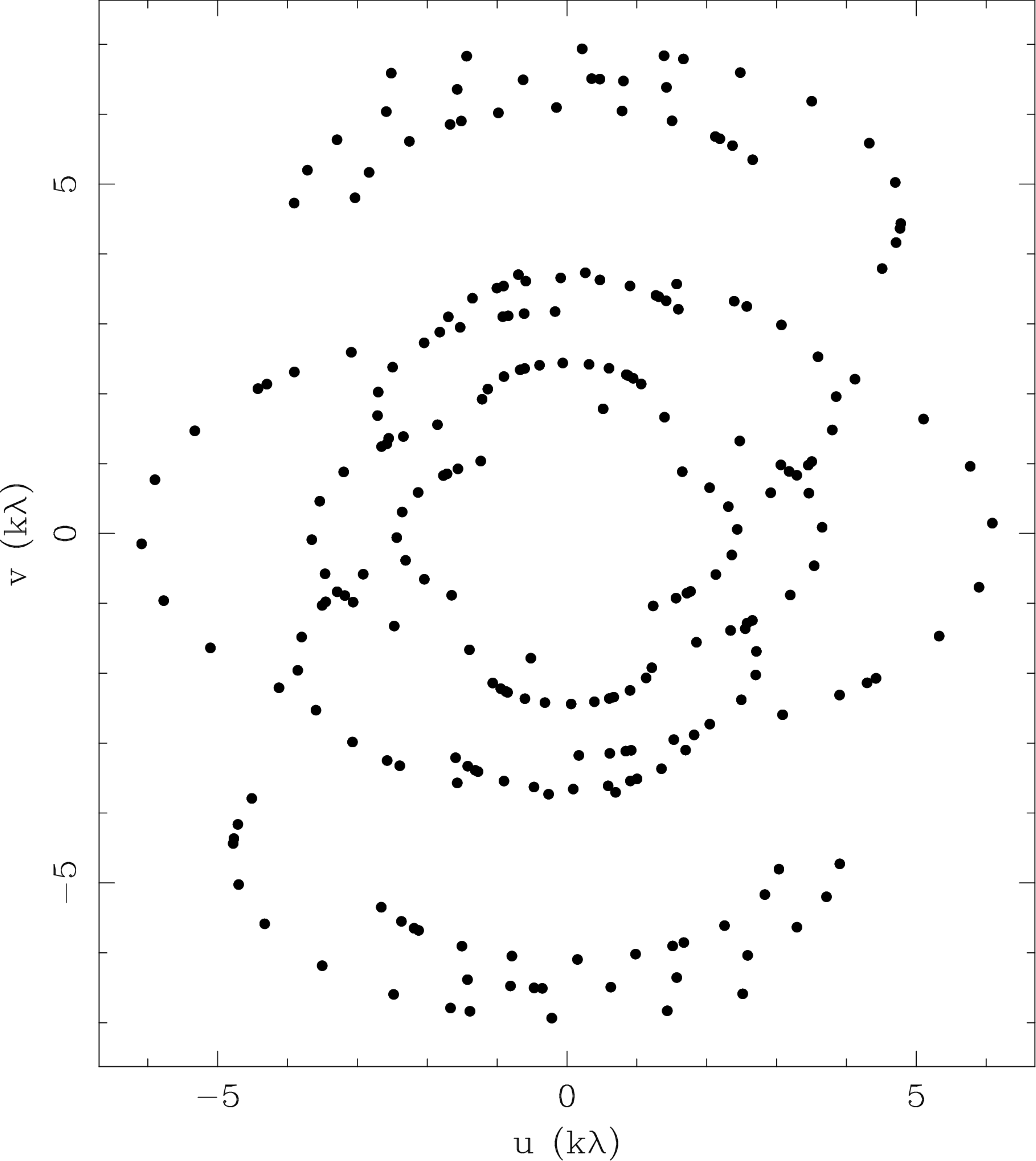}
	\caption{Typical $u,v$ coverage of a single pointing of SWAG using the inner array of five antennas (excluding the far out antenna ca06). Note that overlapping pointings increase the $u,v$ coverage by a factor of $\sqrt{3}$ ($\sqrt{2}$ for edge pointings) which is not included in this plot. The $u,v$ coverage was optimized by the scanning technique (``l-scan'') shown in fig. 1 of \citet{Ott2014a}.}
	\label{figure: UV coverage}
\end{figure}

Between 40 and 220 pointings are combined into strips of $\sim 4.5'$ width, as shown in Fig.~\ref{figure: SWAG strips}, that are observed between subsequent observations of the phase calibrator.
Long strips are split into Galactic northern and southern parts to ensure frequent phase calibrations.
In order to maximize the $u,v$ coverage, pointings are not observed one after the other, but in rows of even and odd numbers as explained in \citet{Ott2014a}.
The rows are aligned parallel to Galactic longitude (``l-scan'' in \citealt{Ott2014a}) while the first element of each row determines its parity (odd, even).
Rows of odd parity are observed consecutively, followed by the left-out rows of even parity at somewhat later local sidereal times (LST) and therefore other projected baselines, which increases the $u,v$ coverage.
The resulting typical $u,v$ coverage of a single pointing is shown in  Fig.~\ref{figure: UV coverage}.

The Compact Array Broadband Backend \citep[CABB,][]{Wilson2011} provides two IF bands, each of 2\,GHz bandwidth. In the CFB 64M-32k mode adopted for these observations, each 2\,GHz band was divided into $32 \times 64$\,MHz ``continuum'' channels with 16 ``zoombands'' of $2048 \times 32$\,kHz channels selected in each IF band for spectral line observations.
The integration time of each pointing is set to $8 \times 30\,\mathrm{sec} =4$\,min as a trade-off between low noise and total project time but varies due to re-observations and flagging.
The eight individual integrations were scheduled for best $u,v$ coverage to be observed over the whole available range of LST from about 15:00 to 22:00\,h as much as possible.
Due to pointing overlap, the actual sensitivity is higher by a factor of $\sqrt{3}$ ($\sqrt{2}$ for edge pointings) than can be expected from the integration time of a single pointing.
The expected noise calculated from the radiometer equation $\sigma \propto \frac{T_{sys}}{\sqrt{\Delta\nu \tau}}$ as a function of system temperature T$_{sys}$, bandwidth $\Delta \nu$ and integration time $\tau$ is $\sigma = 28$\,mK for a typical $T_{sys} = 80$\,K in one channel of $\Delta\nu = 32$\,kHz ($\sim 0.4$\,\kms at 23\,GHz) and $t = 240$\,s.
This value corresponds to $\sigma \sim 6.0$\,m\jybeam for \nh333 ($\nu = 23.87$\,GHz, beam $26.03" \times 17.71"$) which is about half of the measured noise of $\sim 13.1$\,m\jybeam (§ \ref{section: deconvolution}) because of imperfect deconvolution due to the sparsely sampled $u,v$ plane.

Observations were scheduled in 3 week blocks in the southern hemisphere winter months of July/August in each year from 2014 to 2016.
The total project time including calibration data sums up to $\sim 720$\,h ($\sim 204$\,h for the data presented in this paper) including additional time needed to re-observe some regions whose observations were affected by bad weather, resulting in increased noise.
The total on-source time was $\sim 525$\,h ($\sim 150$\,h shown in this paper).

Bandpasses and delay calibrations for each antenna are derived from daily 10\,min integrations on PKS 1253-055 (3C 279).
Its flux of $\sim 16-20$\,Jy at frequencies around 23\,GHz is bright enough to derive accurate bandpass solutions.
The gain/phase (complex gain) calibrator PKS 1710-269 was observed for 2\,min every ${20-30}$\,min, depending on the field, after each l-scan and in between l-scans if necessary.
Daily flux calibrations of 5\,min integration time are performed on the radio galaxy PKS 1934–638, which is the standard primary flux density calibrator for the ATCA below 50\,GHz.

\subsection{Calibration}
All steps from data import to imaging are performed with the ATCA specific version of the package \textsc{miriad}\footnote{\url{http://www.atnf.csiro.au/computing/software/miriad}} \citep{Sault1995}.
The necessary steps were developed into a data reduction pipeline that provides the base for reducing the entire survey.

\subsubsection{Flagging}
Raw data are imported via \texttt{atlod} with options \texttt{birdie}, \texttt{nocacal}, \texttt{noif} and \texttt{opcorr} to automatically flag resonant instrument modes (birdies), initial array calibration data (cacal), set correct intermediate frequency mapping behavior and to apply atmospheric opacity corrections.
Some of the array setup data were not detected by the automatic routine and had to be flagged manually with \texttt{uvflag}.

Resonant modes in ATCA's backend system cause additional ``birdies'' in channels $n \times 1024 + 1$ $(n = 1,2,...)$ that are not identified during data import.
The missing channels are interpolated later when averaging and Doppler-correcting the visibilities to a common resolution of 2\,\kms.
To remove additional correlator artifacts and extreme RFI, calibrator data were clipped at 200\,Jy and CMZ observations at 10\,Jy.
Additional flagging of bad data was done by eye in the interactive task \texttt{blflag}.
Polarization measurements were not required for the presented analysis and all crosshands/cross-polarizations (XY and YX for ATCA's linear feeds) were flagged.
Antenna ca06 offers high spatial resolution at baselines of $>4300$\,m but was also flagged due to lack of intermediate $u,v$ points which drastically down-weights these long baselines.
Apart from aforementioned instrumental birdies, little radio frequency interference (RFI) is present at frequencies of $\sim 23$\,GHz.

\subsubsection{Complex Gain Calibration}
Bandpass amplitudes and phases are calculated on the designated bandpass calibrator PKS 1253-055.
The resulting correction is then applied to the gain/phase calibrator PKS 1710-269 and flux density calibrator PKS 1934-638, and complex antenna gains are derived with \texttt{mfcal}.

Calibration solutions were inspected visually to identify remaining problems that were subsequently flagged.
Re-calibrations were performed until well-behaved solutions were found. 

\subsubsection{Flux Density Calibration}

\begin{figure*}
	\centering
	\includegraphics[width=\linewidth]{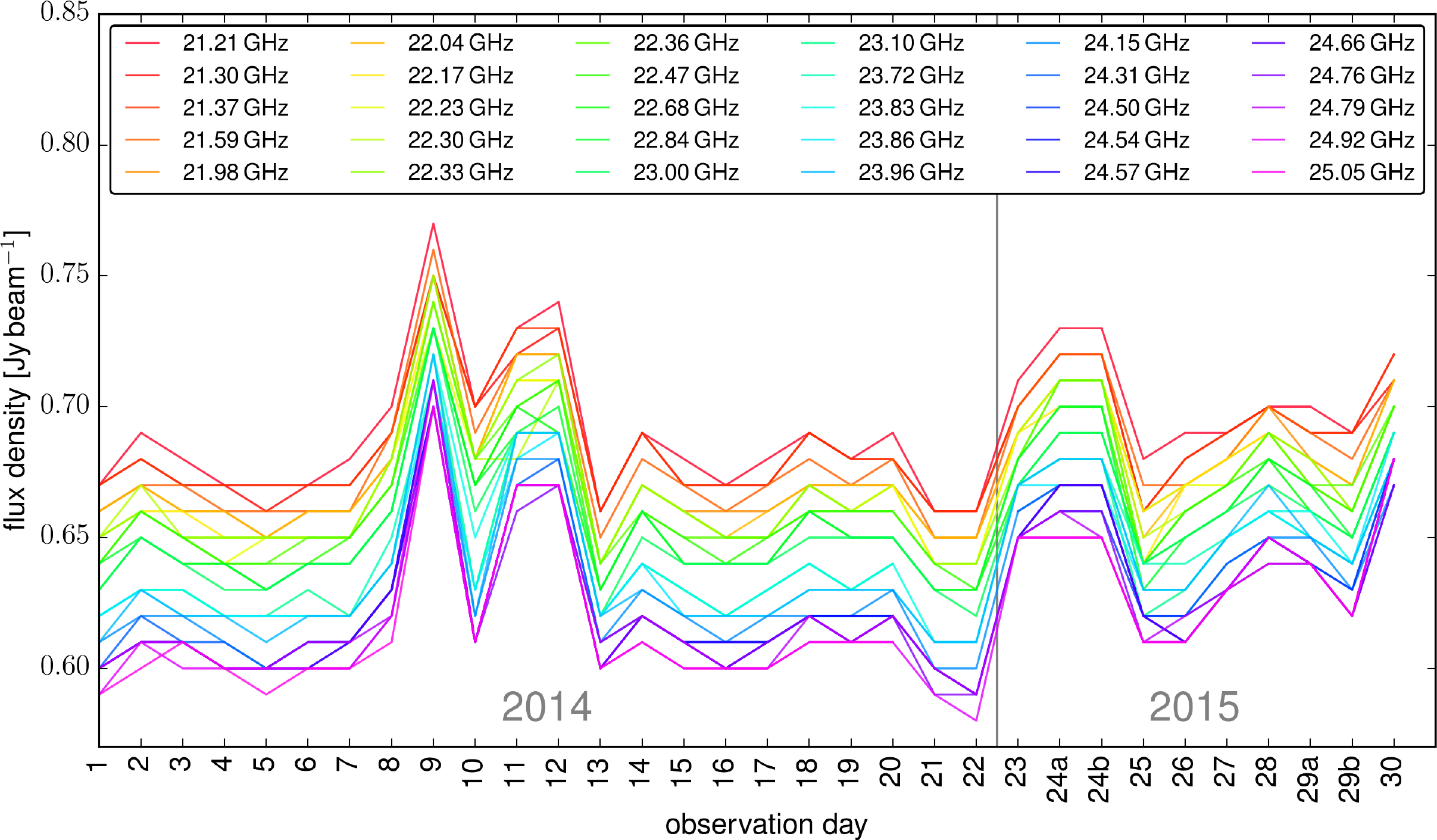}
	\caption{Flux density of the phase calibrator PKS 1710-269 per observation day after all calibration steps for the 30 frequency bands of CABB. On observation day 24 and 29, correlator crashes required new calibrations which are labeled a and b corresponding to flux densities before and after the crash, respectively.  Flux densities are constant across observation days within $\sim 15$\%. The increasing values towards lower frequency is an intrinsic property of the source: a quasar with spectral index $\sim-0.5$. The precision of the \textsc{miriad} task \texttt{imstat} is two decimal places which leads to quantized data.}
	\label{figure: flux calibration stability}
\end{figure*}

The calibration quality is assessed with Fig.~\ref{figure: flux calibration stability}.
The phase calibrator PKS 1710-269 was imaged including all calibration corrections for all observation days and all zoombands and the peak flux density listed.
Measured flux densities per zoomband are constant within $\sim 15$\%, which is typical for radio observations at this frequency.
The origin of this variability cannot be identified without absolute measurements and can be due to changing atmospheric conditions or intrinsic luminosity variation in the QSO PKS 1710-269.
Thus, 15\% is an upper limit on the flux density variation across observation days which is similar to the absolute flux uncertainty.

Spectral lines observed in the same zoomband, however, share a common flux density uncertainty and thus flux ratios and derived quantities are more accurate.
This provides a robust basis for the derivation of the kinetic gas temperature, which is primarily determined from ammonia line ratios.

\subsubsection{Continuum Subtraction}

Over the relatively small spectral range of a zoomband of $\sim 64$\,MHz, a linear approximation captures the spectral index of any continuum emission well enough.
Line-free channels could not be identified in visibility spectra due to the low brightness of most of the spectral lines; instead, they are calculated from the line's rest frequency and a window of $\pm 400$ channels (corresponding to $\sim 178-150$\,\kms at $21.0-25.0$\,GHz) which is large enough to exclude the typical velocity range of molecular gas in the CMZ of $\pm 150$\,\kms.
Any possible emission outside these line-free windows is weak and does not affect the continuum fit.
Fit and subtraction was done using the \texttt{uvlin} task forcing a first-order polynomial.

Rest frequencies are set according to table~\ref{table: ammonia info} that lists values from Splatalogue\footnote{\url{www.splatalogue.net}}.

Merging five channels to obtain a spectral resolution of 2\,\kms increases the signal-to-noise ratio by a factor of $\sqrt{5} \sim2.2$, while the spectral resolution is still adequate for the expected typical line widths of $5-25$\,\kms in ammonia emission.

\floattable
\begin{deluxetable}{lcccccc}
	\tablewidth{\textwidth}
	\tablecaption{For each ammonia transition (1), we detail the (2) rest frequency, (3) excitation energy of the upper transition state, (4) natural weighted beam sizes ($\mathrm{b}_{maj} \times \mathrm{b}_{min}$), (5) beam position angle, RMS noise for a channel width of 2\,\kms in the two regions of (6) normal ($\sigma_{normal}$) and  (7) elevated noise ($\sigma_{high}$). $\sigma_{normal}$ is relevant for most of the maps beside the two outermost strips (see Fig.~\ref{figure: SWAG strips}) on each side that are described by $\sigma_{high}$. Given values are an average over 20 ($\sigma_{normal}$) and 12 ($\sigma_{high}$) measurements in two channels at $\pm 150$\,\kms relative to the rest frequency.
		\label{table: ammonia info}}
	\tablehead{\colhead{transition} & \colhead{frequency} & \colhead{E$_u / k_B$} & \colhead{beam size} & \colhead{PA} & $\sigma_{normal}$ & $\sigma_{high}$\\ & [GHz] & [K] & ["] & [$^\circ$] & \multicolumn{2}{c}{[m\jybeam]}\\ \colhead{(1)} & \colhead{(2)} & \colhead{(3)} & \colhead{(4)} & \colhead{(5)} & \colhead{(6)} & \colhead{(7)}}
	\startdata
	\nh311 & 23.69451 & 24.4 & $26.2 \times 17.8$ & $89.3^\circ$ & 13.4 & 19.4\\
	\nh322 & 23.72263 & 65.6 & $26.2 \times 17.8$ & $89.3^\circ$ & 13.1 & 19.3\\
	\nh333 & 23.87013 & 124.7 & $26.0 \times 17.7$ & $89.3^\circ$ & 13.1 & 18.6\\
	\nh344 & 24.13942 & 201.7 & $25.7 \times 17.5$ & $89.3^\circ$ & 12.9 & 16.1\\
	\nh355 & 24.53299 & 296.5 & $25.3 \times 17.2$ & $89.3^\circ$ & 12.0 & 16.2\\
	\nh366 & 25.05603 & 409.2 & $24.8 \times 16.9$ & $89.3^\circ$ & 13.2 & 16.2\\
	mean value & & & & & 13.0 & 17.6\\
	\enddata
\end{deluxetable}

\subsection{Imaging}

\subsubsection{Deconvolution}\label{section: deconvolution}

Fourier transformation from the visibility to the image domain is performed with the task \texttt{invert} that includes primary beam correction and combines the visibilities of single pointings into a mosaic.
The Briggs weighting parameter is set to \texttt{+2}, which is close to natural weighting.
The cell size (pixel size) is set to $5"$ in RA and DEC, which oversamples the minor axis of the synthesized beam ($24.8" \times 16.9"$, 89.3$^\circ$ at 25.1\,GHz) by a factor >3.
At lower frequency, the beam size grows up to $26.2" \times 17.8"$ at 23.7\,GHz which results in a variation of linear size of $\sim 5\%$ between \nh311 and (6,6).
These beam size are $\sim 1.7$ and $\sim 1.2$ times the naively expected size of $\sim 15" \times 15"$ due to weighting.
Table~\ref{table: ammonia info} lists the obtained beam sizes for \nh311 to (6,6).

Each pointing is imaged over 512 pixel (FOV $42'.7$) and then integrated linearly into the mosaic with another weighting function that smooths noise across regions of differing pointing coverage \citep[\textsc{miriad} default method,][]{Sault1996}.

The extended emission of most spectral lines is generally better deconvolved by an algorithm that also uses extended sources for modelling.
The only possibility of extended modeling that \textsc{miriad} offers is the maximum entropy method (MEM).
Our tests with the mosaic versions of MEM (\texttt{mosmem}) and clean (\texttt{mossdi}), on SWAG data confirmed that more extended emission was reconstructed by MEM over regular clean.

A first run with up to 10 iterations deconvolves the dirty image well enough to construct a mask that prevents treating regions without significant emission.
Image restoration is done with \texttt{restor}.
A mask is calculated to contain all pixels with SNR above $\sim 5$ which corresponds to 65.0\,m\jybeam in the semi-cleaned image.

Deconvolution is then repeated with 50 iterations in \texttt{mosmem} only inside the pixels that were set as relevant by the mask.
The restored images still contain sidelobe structures, especially around strong sources and in strong lines.
The difficulty to obtain better images with \texttt{mosmem} is described in more detail in Appendix~\ref{appendix: mosmem fails}.

The root mean square (RMS) noise is not constant within and across the data cubes but increases slightly with decreasing frequency due to the proximity to an atmospheric water line that increases the telescope's system temperature.
At a level of $\sim 13.0$\,m\jybeam, the difference between the channels at -150\,\kms and +150\,\kms relative to rest frequency is $0.1-0.6$\,m\jybeam for the ammonia lines listed in Table~\ref{table: ammonia info}.
Thus, a common value is calculated as the mean of minimal and maximal noise measurements.
Spatial variation of noise across pointings is negligible except for the two outermost strips of the data considered in this work (\#10/11 and \#30/31, see Fig.~\ref{figure: SWAG strips}) that show increased noise levels in the 2014/2015 data due to bad weather.
Hence, two values $\sigma_{normal}$ and $\sigma_{high}$ are needed to describe the noise properties in the inner map and the $l^+/l^-$ edges, respectively.
Table \ref{table: ammonia info} lists these noise levels for the six metastable ammonia lines.
Any further mention of noise will refer to $\sigma_{normal}$ as only two major molecular clouds (Sgr~B2 and Sgr~C) are located in the region of elevated noise, and they are strong sources with high SNR.

\subsubsection{Masking of the convolved data cubes}\label{section: map masking}

The outer edges of all data cubes exhibit a $\sqrt{3}$ higher noise than the rest of the maps because of non-overlapping pointings.
These edges are blanked from the edge inwards by $\sim 50"$ and $\sim 100"$ ($\sim 1/3$ and $2/3$ of the primary beam, respectively) as the smaller radius of $50"$ proved to be enough to exclude any visible higher noise edges in ammonia distribution maps whereas spectral fitting is affected by increased noise to $\sim 90"$ from the edge.

\subsubsection{Moment Maps}\label{section: moment map production}

Data cubes were collapsed along the spectral axes with the task \texttt{moment} in \textsc{miriad} to obtain moment maps.
Map edges are blanked with the $50"$ edge mask and noise is excluded at levels of $3 \sigma$ and $5 \sigma$ for intensity maps (peak intensity; moment 0, integrated intensity) and velocity maps (moment 1, intensity weighted mean velocity; moment 2, velocity dispersion), respectively.

\subsubsection{Position-Velocity Diagrams}

Position-velocity cuts through the restored data cube edge masked at 100" were calculated with the task \texttt{impv} in the \textsc{CASA} software package\footnote{Common Astronomy Software Applications, \textit{https://casa.nrao.edu/}} \citep{McMullin2007} along Galactic longitude, averaging over all emission in Galactic latitude.

\section{Results}\label{section: results}

Figures in this section show \nh333 as representative of all six observed metastable ammonia inversion lines as it is typically the brightest ammonia line observed with SWAG.
The corresponding figures for \nh311, (2,2), (4,4), (5,5) and (6,6) can be found in appendix \ref{appendix: SWAG data products}.

\subsection{Data cubes}

\begin{figure*}
	\centering
	\includegraphics[height=0.9\textheight]{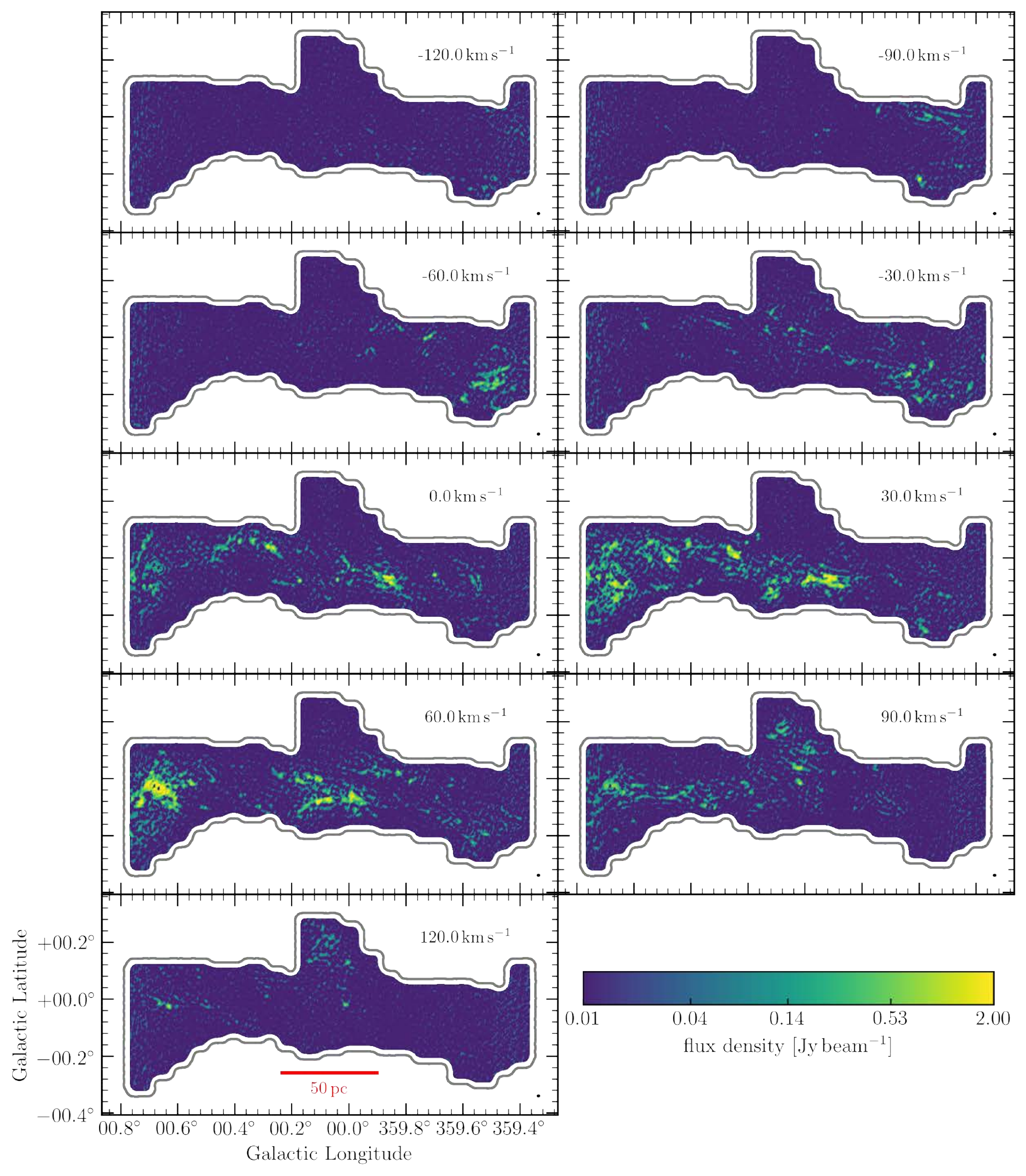}
	\caption{Channel maps of \nh333. Every 15$^{\mathrm{th}}$ channel of 2\,\kms (separated by 30\,\kms) in the range of -120\,\kms to +120\,\kms is shown. Velocity and beam ($26.03" \times 17.71"$, $89.3^\circ$) are indicated in the top and bottom right corner, respectively. The conversion factor from flux density to brightness temperature for this beam size at 23.87\,GHz is $\mathrm{T\ [K]}/\mathrm{S\ [Jy\,beam^{-1}]} = 4.65$.}
	\label{figure: nh333 channel map}
\end{figure*}

Channel maps of 2\,\kms width in steps of 15 channels (30\,\kms) are shown in Fig.~\ref{figure: nh333 channel map}.
All velocities are specified in the LSRK frame using the radio approximation.
Emission can be detected in the range of $\sim-180$\,\kms to $\sim 150$\,\kms with emission related to the ring-like feature between $\sim -150$\,\kms and $\sim 120$\,\kms.
Extended gas clumps at $l \sim 0^\circ, b \sim 0.15^\circ$ are offset in velocity from the surrounding gas and are detected among the highest velocity gas in the field of view.
Typical flux density values for molecular clouds in the \nh333 line are on the order of a few hundred m\jybeam with peaks of 4.23\,\jybeam (19.6\,K) in the 20\,\kms cloud and 5.28\,\jybeam (24.8\,K) in Sgr B2.
The spatial distribution is very similar among the ammonia lines despite the varying flux density levels (§\ref{figure: nh333 channel map} and \ref{appendix: SWAG channel maps}).

\subsection{Sample Spectra}

\begin{figure}
	\centering
	\includegraphics[width=\linewidth]{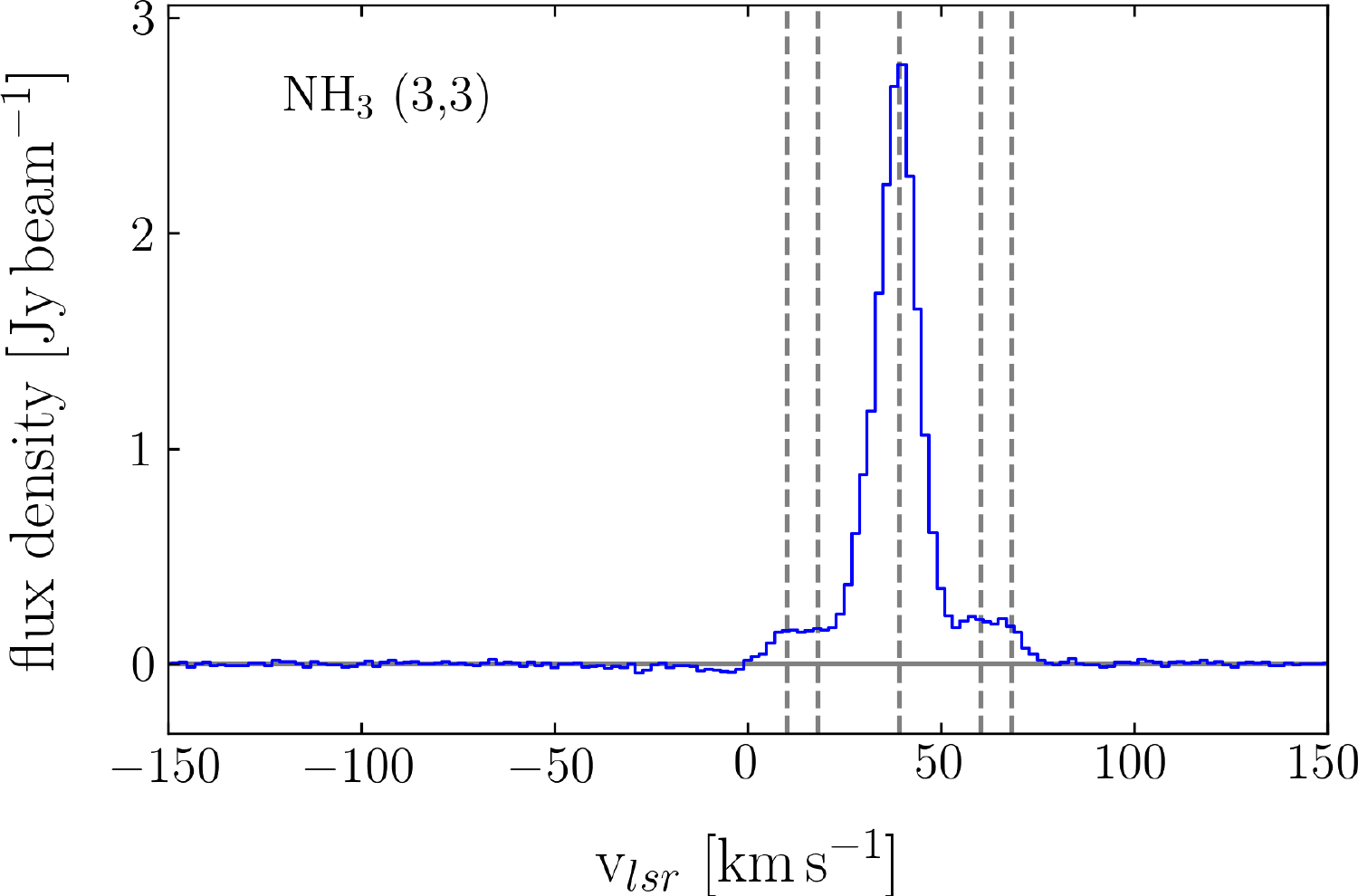}
	\caption{Sample spectrum toward the Brick ($l=0.2328^\circ$, $b=0.0107^\circ$)  extracted from the \nh333 cube. RMS noise is 13.1\,m\jybeam (0.061\,K). The positions of the hyperfine satellites according to Table~\ref{table: ammonia hyperfine components} are indicated by dashed vertical lines.  Note the broad line width that blends the outer hyperfine components (having separations of $\pm 21.1$\,\kms and $\pm 29.1$\,\kms from the main line) into a pedestal.}
	\label{figure: spectrum nh333 cloud e/f}
\end{figure}

Depending on the structure of the source and the spectral line, hyperfine satellite components and multiple components along the line-of-sight are detected.
A bright single component example spectrum of \nh333 is shown in Fig.~\ref{figure: spectrum nh333 cloud e/f} toward the Brick (G0.253+0.016).
Absorption against continuum supposedly from free-free emission is detected towards the 50\,\kms cloud, the Brick, cloud c and f, and some smaller clouds in the $l^+, b^-$ and $l^-, b^+$ portions of the ``ring'' (cf. Fig.~\ref{figure: GC overview}).

Further spectra at the same position in the Brick for \nh311 to (6,6) are shown in Appendix~\ref{appendix: SWAG example spectra}

\subsection{Image Moments}

\begin{figure*}
	\centering
	\includegraphics[height=0.9\textheight]{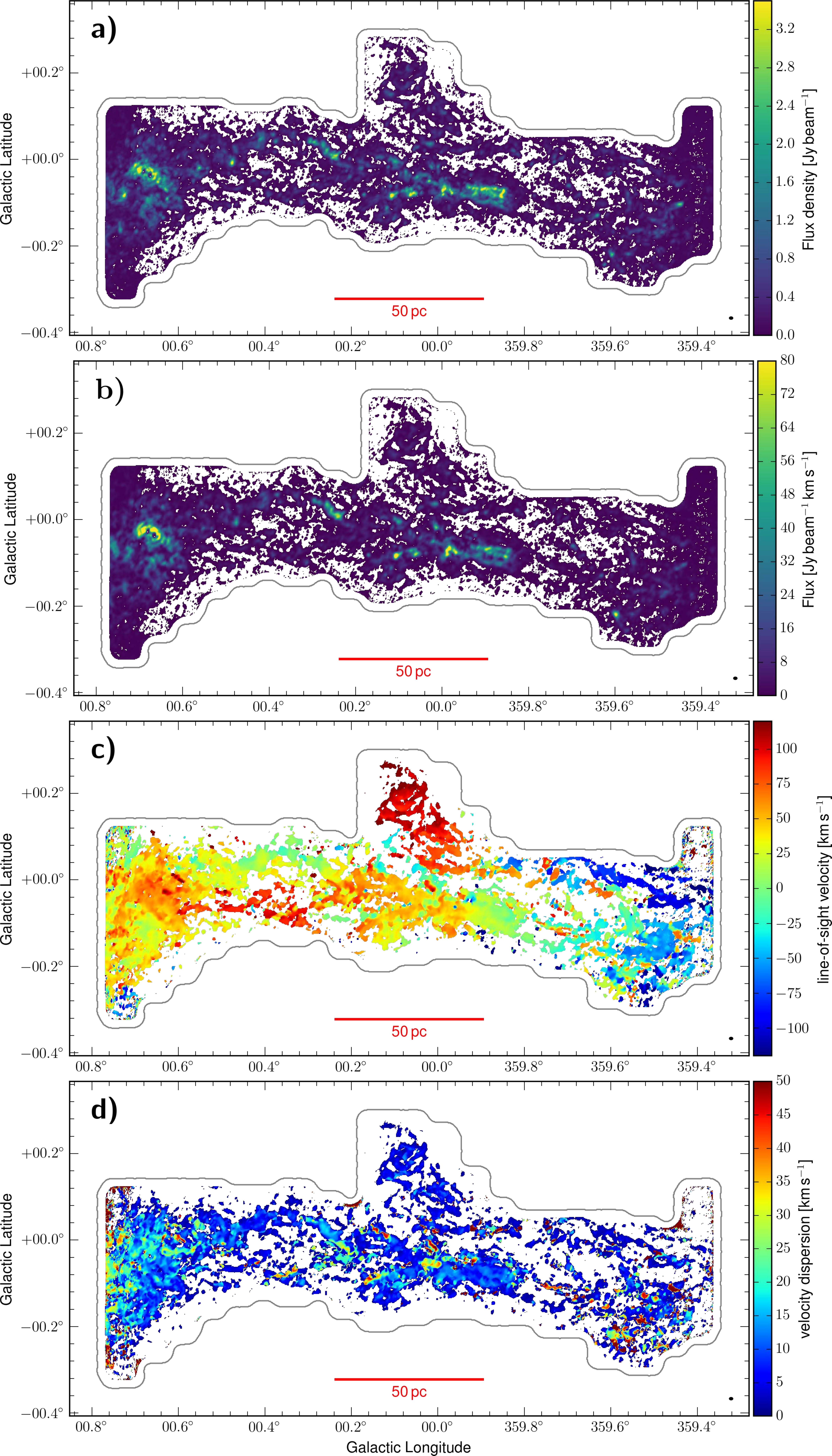}
	\caption{\nh333 moment maps: \emph{a}) peak intensity, \emph{b}) integrated intensity (moment 0), \emph{c}) intensity weighted mean velocity (moment 1), \emph{d}) intensity-weighted velocity dispersion (moment 2). The intensity maps a) and b) are masked at $3 \sigma$, velocity maps c) and d) are masked at $5 \sigma$ with an rms noise of $\sigma = 13.1$\,mJy. The beam of $26.0" \times 17.7"$ (1.05\,pc $\times$ 0.71\,pc) is indicated in the lower right hand corner of each panel.}
	\label{figure: nh333 moment map}
\end{figure*}

Image moment maps are shown in Fig.~\ref{figure: nh333 moment map} for orders 0 (integrated intensity), 1 (velocity centroid) and 2 (velocity dispersion), the top panel shows peak intensity.

Peak intensity (Fig.~\ref{figure: nh333 moment map} a) and integrated intensity (Fig.~\ref{figure: nh333 moment map} b) show an almost identical distribution of emission, with Sgr~C being a notable exception of strong peak but low integrated intensity indicating narrow line widths.
Sgr~B2, the arching structure of the dust ridge ($0.2^\circ < l < 0.6^\circ$) and the sequence of massive clouds at $-0.2^\circ < l < 0.15^\circ$ with a bifurcation\footnote{This term is adopted from \citetalias{Kruijssen2015} and denotes the apparent splitting after the 50\,\kms cloud when moving towards positive longitudes, best seen in Fig.~\ref{figure: nh333 channel map} at 60\,\kms.} at $l \sim 0.1^\circ$ are most prominent in both intensity maps.
Line-of-sight velocity (Fig.~\ref{figure: nh333 moment map} c) shows a complex pattern at $0.0^\circ < l < 0.2^\circ$ of multiple partially overlapping emission components along the line-of-sight.
At negative longitudes, the coherent structure of negative velocities is intermingled with clouds at positive velocities.
In total, the molecular arcs trace an apparently ring-like structure.
As the interferometer filters out diffuse emission, the molecular clouds in this ring-like structure must be rather clumpy which was already reported previously \citep{Bally2010,Molinari2011,Ott2014b,Ginsburg2016}.

Using the definition $l^{+/-}$ and $b^{+/-}$ denoting positive/negative values of Galactic longitude and latitude, the general sense of rotation in the CMZ is consistent with the scenario $[l^-,0]$ $\rightarrow$ $[l^-,b^-]$ $\rightarrow$ $[0,0]$ $\rightarrow$ $[l^+,b^+]$ $\rightarrow$ $[l^+,0]$ $\rightarrow$ $[l^+,b^-]$ $\rightarrow$ $[0,0]$ $\rightarrow$ $[l^-,b^+]$ found in previous studies \citep[e.g.][]{Molinari2011,Kruijssen2015,Ginsburg2016}.
This, however, does not necessarily mean that arcs are physically connected, e.g. note the 50\,\kms jump between $[l^-,b^-]$ and $[l^+,b^+]$, but indicates that the bulk of the gas is following the same sense of rotation.
Typical velocity dispersions in the moment~2 map are found to be $5-25$\,\kms but velocity dispersions for individual clouds are expected to be lower because of  multiple components along the line-of-sight and, for the lower excitation states, blending of hyperfine components.

Moment maps of \nh311 to (6,6) are shown in Appendix~\ref{appendix: SWAG moment maps}.

\subsection{Structure in Position-Velocity Space}

\begin{figure}
	\centering
	\includegraphics[width=\linewidth]{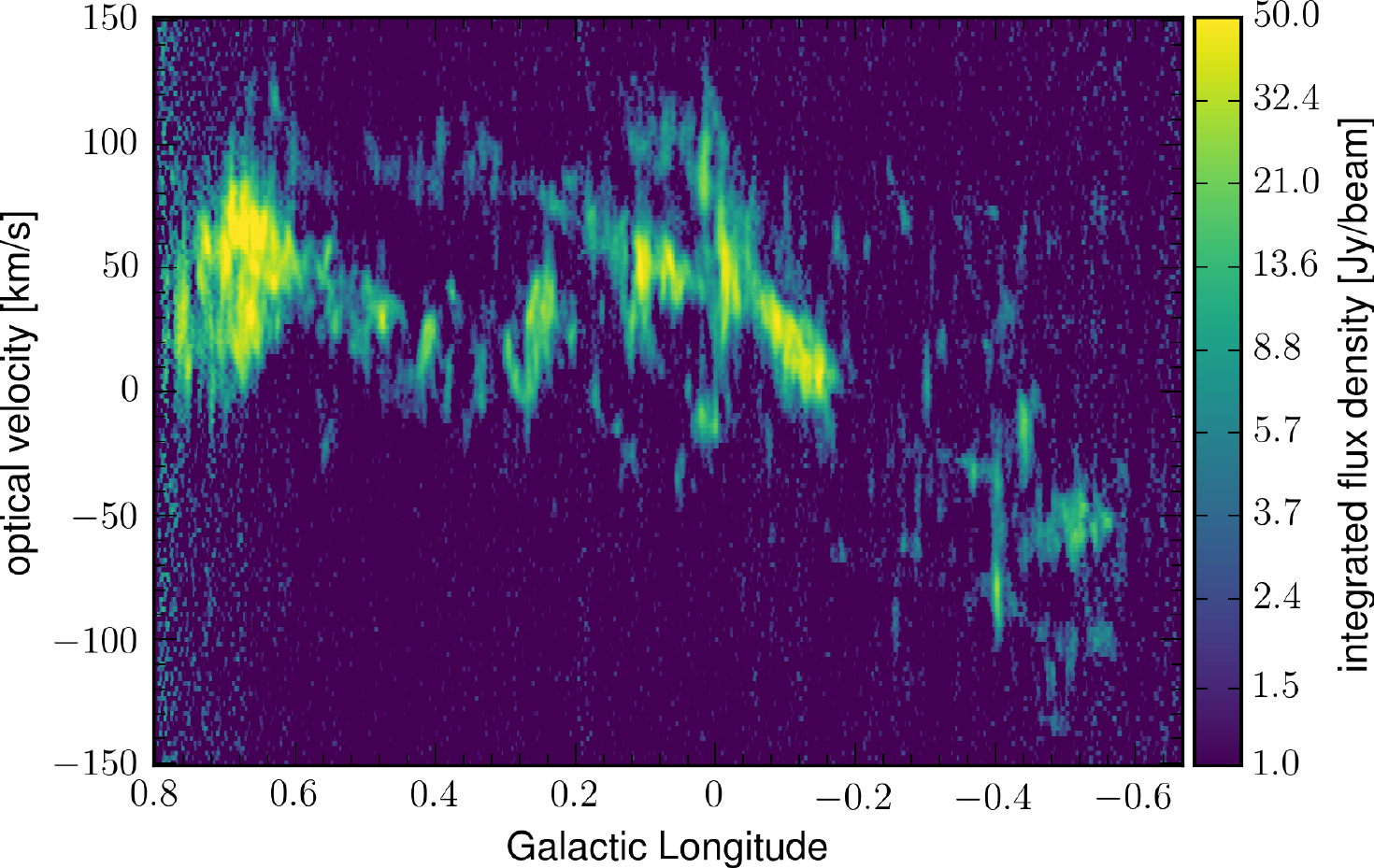}
	\caption{Position-velocity diagram along Galactic longitude, integrated over Galactic latitude.}
	\label{figure: nh333 pV along l}
\end{figure}

A position-velocity diagram along Galactic longitude integrated over Galactic latitude traces the ring-like structure in the CMZ (Fig.~\ref{figure: nh333 pV along l}) which can be seen as roughly parallel structures traversing the diagram diagonally.
A detailed analysis of the spatial and kinematic structure is given in section \ref{section: fit results kinematics} based on fit-derived kinematics.

\section{Analysis: Pixel by Pixel Hyperfine Structure Fitting}\label{section: hyperfine structure fitting}

Several basic cloud properties can be derived from the metastable ammonia inversion lines \cite{Ho1983}.
The tunneling frequency of the nitrogen atom through the plane of three hydrogen atoms (inversion) depends on the rotational state (J,K) of the molecule.
The population of these states is temperature dependent according to a Boltzmann law.
For each rotational state J (angular momentum), there exists another ladder of states with quantum number K (angular momentum projected onto the symmetry axis).
Each state decays quickly down the K-ladder to the lowest energy state (J,K=J), e.g. \nh311.
Those states are metastable as they can only de-excite through forbidden transitions if the density is too low for collisional de-excitation and thus live long enough to be commonly observable in the ISM.
The inversion lines are further split into hyperfine structure lines by the interaction of the electric quadrupole moment of the nitrogen atom with the electric field of the molecule's electrons which allows the derivation of gas opacity from the ratio of hyperfine components.
Fits of the entire line structure of at least two metastable ammonia lines thus yield opacity-corrected column density, line-of-sight velocity, line width and rotational temperature.
The rotational temperature is a lower limit and can be used to estimate the kinetic gas temperature up to approximately the line's excitation energy via model derived conversion functions.

Details on the ammonia thermometer, the derivation of opacity corrected column density, rotational temperature and its conversion to kinetic gas temperature are described in Appendix~\ref{appendix: ammonia thermometer}.

\subsection{Hyperfine Structure Fitting in \texorpdfstring{\textsc{class}}{CLASS}}\label{section: hyperfine fitting CLASS}

In order to derive gas temperatures without having to assume optically thin emission, line temperatures, line widths and opacities need to be known.
We derive these quantities by fitting the ammonia hyperfine structure with the \textsc{class}\footnote{\url{http://iram.fr/IRAMFR/GILDAS/doc/html/class-html/}} package in \textsc{gildas}\footnote{\url{https://www.iram.fr/IRAMFR/GILDAS/}}.
The 100" edge-masked data cubes were masked at $3 \sigma$ per channel and 5\,\jybeam\kms in integrated flux density to ensure sufficient SNR for successful fits.
Fitting is done with the \textsc{class} functions \texttt{method nh3(j,j)} and \texttt{minimize} for $\mathrm{J} = 1,2,3$.
The higher transitions $\mathrm{J} = 4,5,6$ are not implemented but can be fitted with \texttt{method hfs} using the relative positions and strengths of the ammonia hyperfine structure components of \citet{TownesSchawlow1975} (listed in Table~\ref{table: ammonia hyperfine components}).
The fitting algorithm is constrained to line opacities in the range $0.1 \leq \tau \leq 30$ which introduces a factor of 

\begin{equation}
\lim_{\tau \rightarrow 0} \frac{0.1}{\tau}\frac{1-e^{-\tau}}{1-e^{-0.1}} = 1.051
\end{equation}

relative to the line temperature of optically thin emission ($\tau \rightarrow 0$).
This deviation by up to 5.1\% is small compared to the general flux density uncertainties of $\sim 15\%$ (Fig.~\ref{figure: flux calibration stability}) and can be neglected.
The fit is further constrained to the strongest emission component along the line-of-sight (Fig~\ref{figure: good bad fit}, a and b).
Occasionally, unsuccessful fits occur due to blending of closely spaced emission components (Fig.~\ref{figure: good bad fit}, c and d).
These are discarded based on reduced $\chi^2$ (\textsc{class} parameter $\mathtt{lineRMS} > 0.1$), physically implausible values ($\Delta v (FWHM) > 50.0$\,\kms) and very large error of the fitted parameter ($\Delta (v_{los}) > 10.0$\,\kms, $\Delta (\Delta v) > 10.0$\,\kms).
Table~\ref{table: fit success rates} lists the statistics of pixel selection and fitting.

\subsubsection{Sample Boltzmann Plot}\label{section: fit results boltzmann plot}

\begin{figure}
	\centering
	\includegraphics[width=\linewidth]{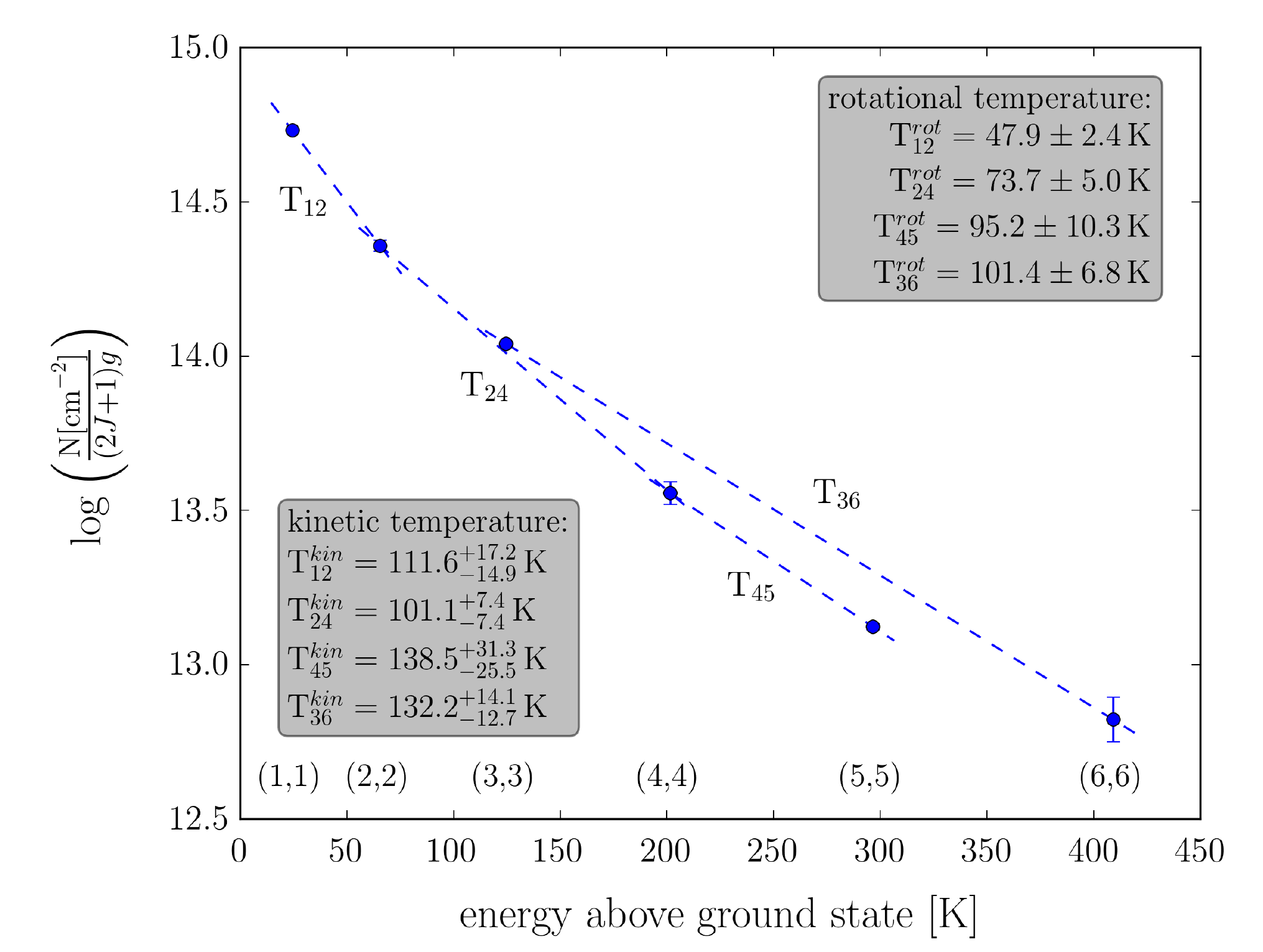}
	\caption{Typical Boltzmann plot (excitation state-scaled column density as a function of excitation temperature or energy above ground state) of ammonia emission in the Brick. Four temperature estimates (proportional to inverse slope) are sketched and listed in the two boxes. Typical errors are $5-10\%$ in rotational and $10-25\%$ in kinetic temperature with increasing uncertainty for higher kinetic temperatures due to flattening of the $\mathrm{T}_{kin} - \mathrm{T}_{rot}$ conversion (Table~\ref{table: Trot Tkin}, \citealt{Morris1973}, \citealt{Ott2011}, \citealt{Gorski2017}).}
	\label{figure: boltzmann plot}
\end{figure}

A Boltzmann plot (appendix \ref{appendix: boltzmann plot}, cf. \citealt{Goldsmith1999}) typical for the ammonia emission in the Galactic Center is shown in Fig.~\ref{figure: boltzmann plot}.
The observed shapes do not follow a single linear relation but become more shallow around \nh333 to \nh344 which can be due to the non-linear T$_{kin}$-T$_{rot}$ conversion or be indicative of a multiple temperature medium as is already known in the literature \citep[e.g.][]{Huettemeister1993, Mills2013}.
Errors in derived column density are typically small ($\Delta$N$_u$/N$_u$ of order few percent) but larger (few tens of percent) at the edges of clouds due to lower SNR which allows the derivation of  accurate temperatures of typically $5-10\%$ relative error (T$_{rot}$) and $10-25\%$ after conversion to kinetic temperature.
Additionally, the $\sim 15\%$ flux density error (Fig.~\ref{figure: flux calibration stability}) contributes to column density errors.

\subsection{Results of Hyperfine Structure Fitting}

As in section \ref{section: results}, only \nh333 is shown to illustrate features common in all observed ammonia lines.
As representative temperature measure, we choose to show \temp24 calculated from \nh322 and \nh344. The strongest line, \nh333, yields the temperature \temp36 which suffers larger relative errors than \temp24 due to the low intensity of \nh366 emission.
We also calculate \temp12 and \temp45.
Other combinations, like \temp23, cannot be calculated due to \nh322 being a para state (one hydrogen nuclear spin anti-parallel) while \nh333 is an ortho state (all hydrogen nuclear spins parallel).
Generally, \nh{3}{i}{i} with $i = 3n$ (n integer) are ortho, the others are para states.
As the relative abundance of ortho vs. para ammonia is not known, temperatures can only be estimated from two para states or two ortho states.
For \temp12, the $\mathrm{T}_{rot} - \mathrm{T}_{kin}$ conversion (Table~\ref{table: Trot Tkin} and fig.~5 in \citealt{Ott2011}, see also \citealt{Morris1973}) flattens which does not permit a reliable derivation of temperatures above $\sim 50-60$\,K.
\temp24, \temp45 and \temp36 do not suffer such problems \citep{Gorski2017} and yield very similar results.
The corresponding maps for \temp12, \temp45 and \temp36 can be found in appendix \ref{appendix: temperature maps}.

\subsubsection{Column Density}\label{section: fit results column density}

\nh333 column density N$_l$ (Fig.~\ref{figure: nh333 column density}) traces much of the ring-like gas stream by construction of the mask applied before hyperfine structure fitting.
A peak of $\mathrm{N}_{33} = 7.7 \times 10^{15}$\,\sqcm is reached in Sgr~B2.
Massive clouds like the Brick, G+0.10-0.08 and the 20\,\kms cloud are also found to have large column densities ($\mathrm{N}_{33} \gtrsim 1.5 \times 10^{15}$\,\sqcm) while \nh333 is not largely excited and detected at column densities $\lesssim 1.5 \times 10^{15}$\,\sqcm.
\nh311 column density reaches a maximum of $1.2 \times 10^{16}$\,\sqcm in Sgr~B2 where \nh366 is still detected at up to $2.6 \times 10^{15}$\,\sqcm.

\begin{figure*}
	\centering
	\includegraphics[width=\linewidth]{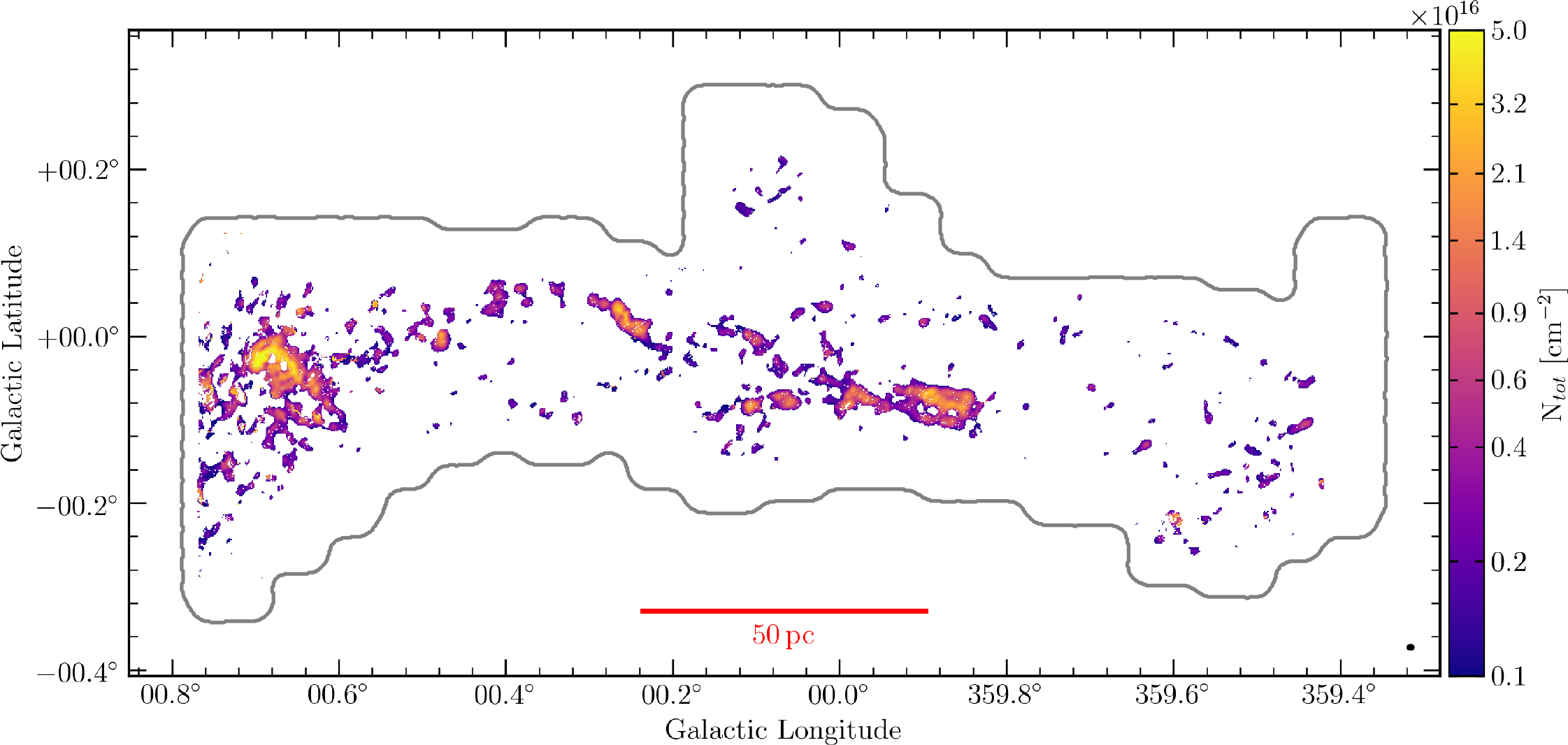}
	\includegraphics[width=\linewidth]{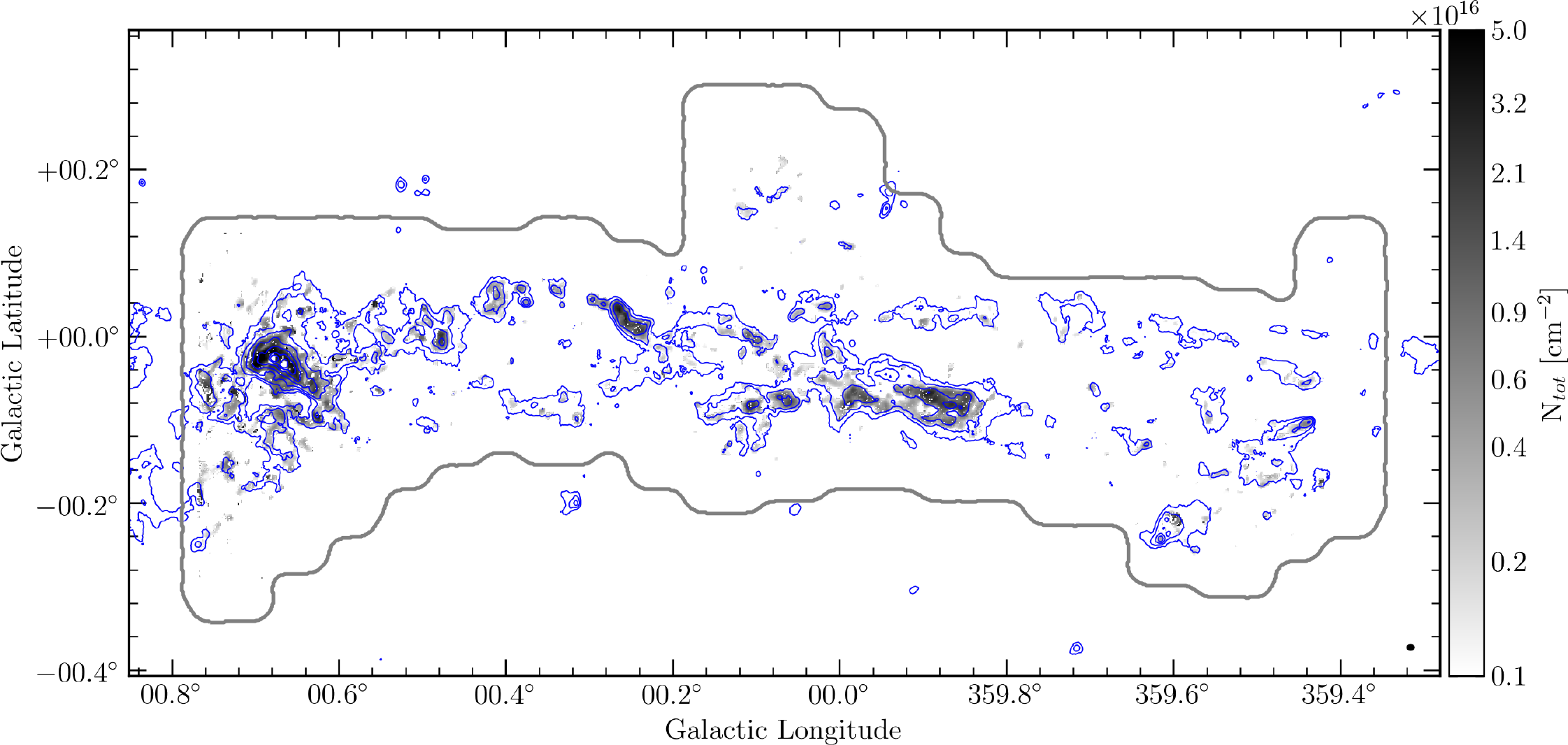}
	\caption{\emph{top}: Total ammonia column density according to eq. \ref{equation: total column density}. The dynamical range is $\sim 5 \times 10^{15}$\,\sqcm to $\sim 6.8 \times 10^{16}$\,\sqcm. In warm clouds like Sgr~B2, this figure represents a lower limit as the unobserved states $\mathrm{J}>6$ can be populated in non-negligible fractions \citep{Mills2013}. \emph{bottom}: The same map of total column density overlaid with 870\,$\mu$m dust emission from ATLASGAL \citep{Schuller2009} at contours of powers of two (1, 2, ..., 64) \jybeam. The ammonia distribution closely follows dust emission in regions of high ammonia column density ($\gtrsim 1 \times 10^{16}$\,\sqcm).}
	\label{figure: total column density}
\end{figure*}

The total ammonia column density as derived from eq.~\ref{equation: total column density} (Fig.~\ref{figure: total column density}) shows a similar picture as does the \nh333 column density.
The highest column densities are reached in Sgr~B2 ($6.8 \times 10^{16}$\,\sqcm), the Brick ($3.9 \times 10^{16}$\,\sqcm) and the 20\,\kms cloud ($3.0 \times 10^{16}$\,\sqcm) whereas most smaller clouds along the ``ring'' have $\mathrm{N}_{tot} \lesssim 1.0 \times 10^{16}$\,\sqcm.
The clouds at $l \sim 0.1^\circ, b \sim 0.2^\circ$ show the lowest detected column densities ($1.0 \times 10^{15} \lesssim \mathrm{N}_{tot}\ \mathrm{cm^{-2}} \lesssim 2.5 \times 10^{15}$) of clouds that do not fall below the fitting thresholds. 
Other clouds of this size typically exhibit higher column densities by a factor of 2-3.
A small cloud of high column density is located at $l=-0.40^\circ, b=-0.22^\circ$ with $\mathrm{N}_{tot}$ up to $2.6 \times 10^{16}$\,\sqcm ($\mathrm{N}_{33} = 6.4 \times 10^{15}$\,\sqcm) and surrounded by gas of $\mathrm{N}_{tot} \sim 10^{16}$\,\sqcm.
This cloud is assumed to lie in the foreground \citep{Longmore2013b} and is discussed to be influenced by an intermediate mass black-hole \citep{Oka2016} due to unusually large CO velocity dispersion.

The ammonia distribution as traced by the NH$_3$ column density map closely follows 870\,$\mu$m dust emission obtained by ATLASGAL \citep{Schuller2009} as can be seen in Fig.~\ref{figure: total column density} (bottom).
The correlation is most pronounced for gas at higher column density than $\sim 1 \times 10^{16}$\,\sqcm.
Lower column density gas is mostly found in regions of weak dust emission but the presence of dust does not imply the detection of ammonia at $\mathrm{N}_{tot} \gtrsim 2 \times 10^{15}$\,\sqcm.
Almost all (dense) ammonia gas is accompanied by dust emission.
Notable exceptions from the overall good correlation are found in Sgr~B2 and Sgr~A*.
Sgr~B2 (N) and Sgr~B2 (M) are detected in absorption in ammonia whereas the dust emission peaks at those locations.
All of Sgr~B2 that is detected in emission, however, does correlate very well with the 870\,$\mu$m contours.
At Sgr~A*, the $870\,\mu$m emission locally peaks but the ammonia column density is $\ll 1 \times 10^{15}$\,\sqcm as this region falls below the selection thresholds for hyperfine structure fitting.
The emission from Sgr A* at submm is contributed by optically thick synchrotron emission rather than thermal dust emission and no correlation is observed.

\subsubsection{Kinematics}\label{section: fit results kinematics}

\begin{figure*}
	\centering
	\includegraphics[width=\linewidth]{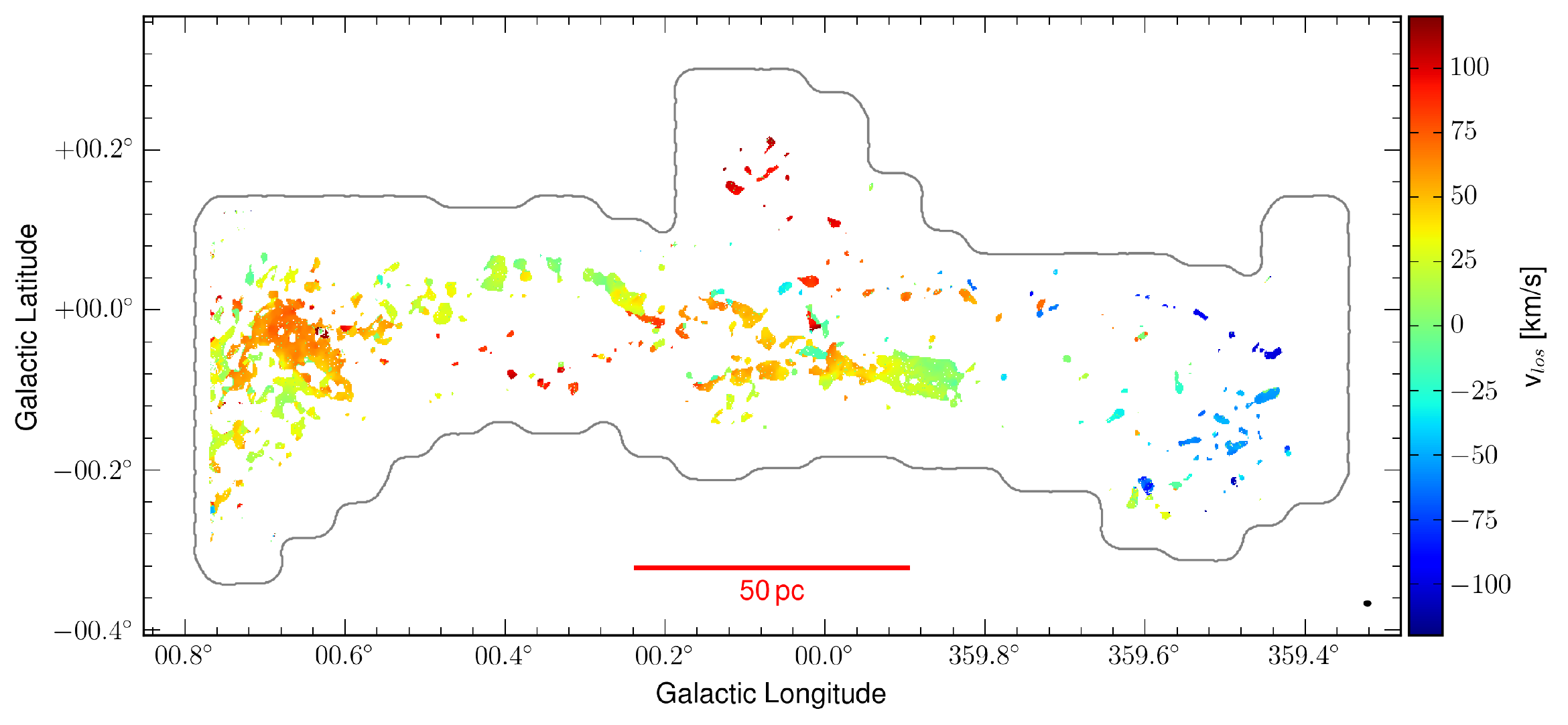}
	\caption{\nh333 line-of-sight velocity derived from fitting the hyperfine structure.}
	\label{figure: nh333 los velocity}
\end{figure*}

The fit-derived line-of-sight velocity of \nh333 (Fig.~\ref{figure: nh333 los velocity}) generally agrees well with the moment 1 map (Fig.~\ref{figure: nh333 moment map}c).
Due to the constraints for successful fitting (§\ref{section: hyperfine fitting CLASS}), the mapped area is considerably smaller.
Significant differences between both maps are not detected which confirms that ``successful'' fits are indeed correct and potential ``bad'' fits were excluded.

\begin{figure*}
	\centering
	\includegraphics[width=\linewidth]{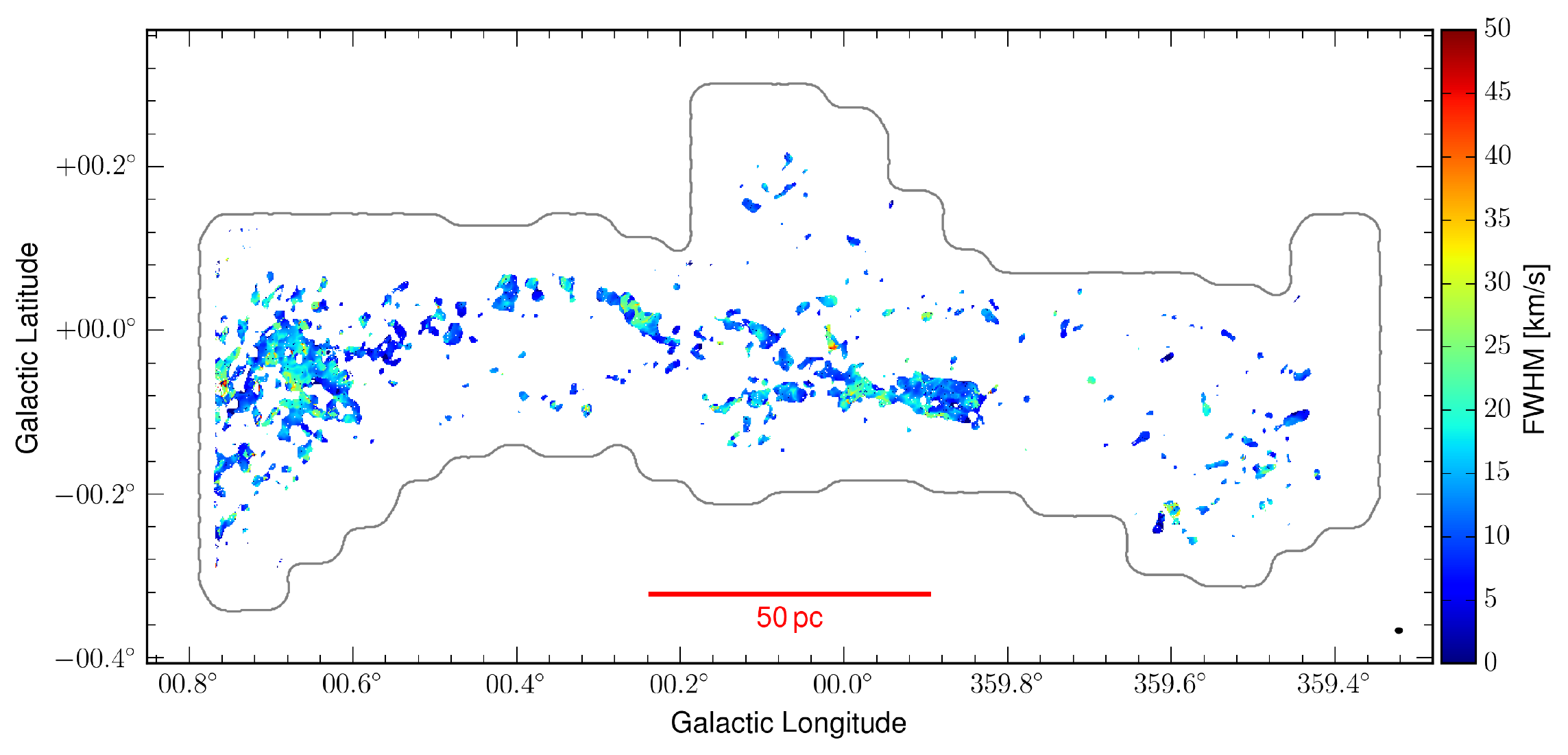}
	\caption{\nh333 line width (FWHM) derived from fitting the hyperfine structure.}
	\label{figure: nh333 line width}
\end{figure*}

\begin{figure}
	\centering
	\includegraphics[width=\linewidth]{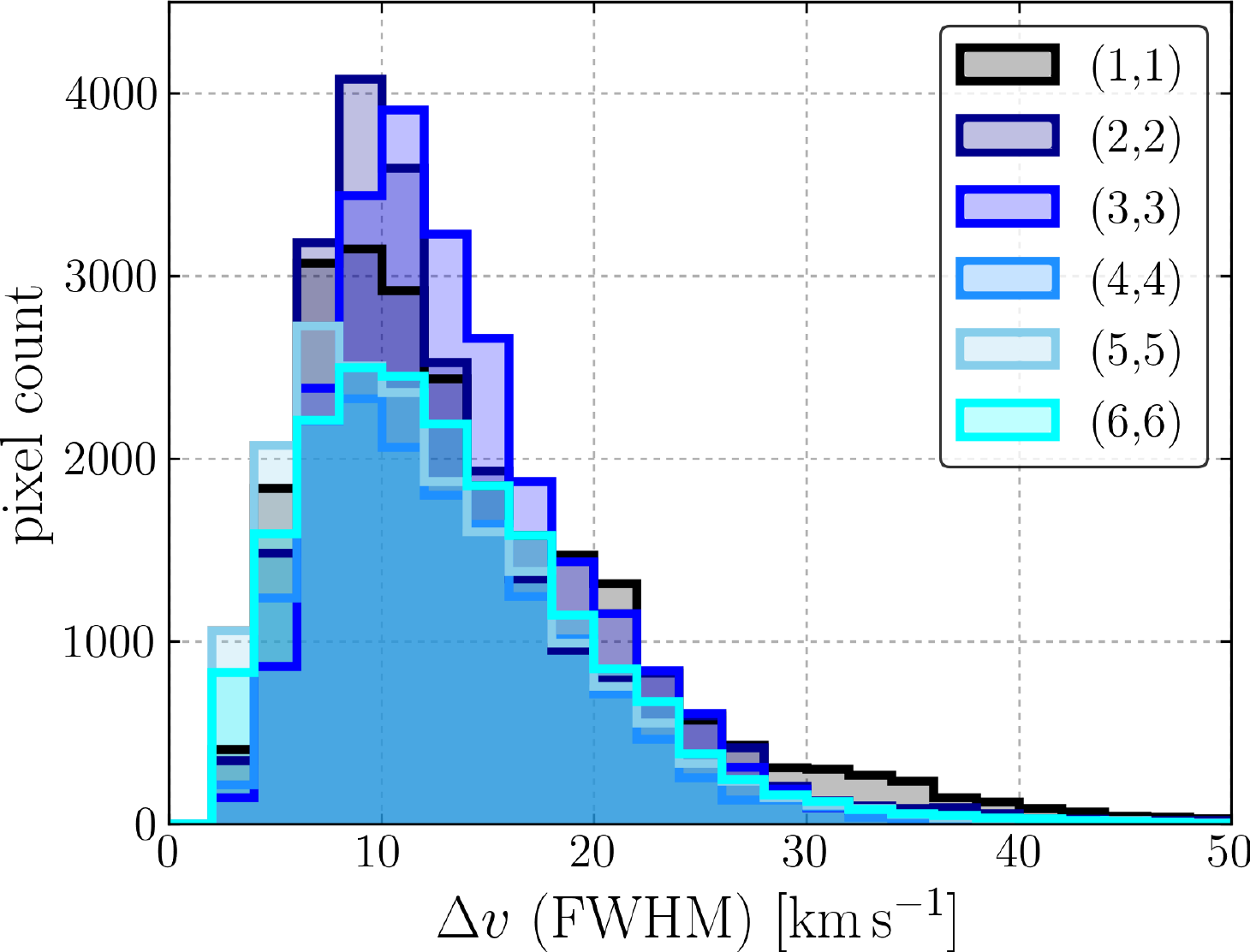}
	\caption{Distribution of FWHM of metastable ammonia hyperfine lines in the Galactic Center. The bin size of 2\,\kms corresponds to the spectral resolution of the data cubes.}
	\label{figure: histogram line widths}
\end{figure}

The line widths of the fitted map (Fig.~\ref{figure: nh333 line width}) and the moment 2 map (Fig.~\ref{figure: nh333 moment map} d) qualitatively agree about regions of higher or lower line width but show very different quantitative results.
After correcting for the conversion between FWHM (fitted map) and velocity dispersion $\sigma$ (moment map), a typical factor of $\sim 1.5-2$ (coresponding to differences of $< 10$\,\kms) remains but extended regions all across the CMZ display factors of $3-5$ (differences of 20\,\kms up to $> 50$\,\kms).
The larger line widths are derived from the moment map in all cases, meaning the image moment analysis overestimates the line width with respect to line fitting.
This is partly due to unaccounted blending of hyperfine structure and mostly caused by multiple line-of-sight components in the moment map.
\cite{Henshaw2016a} find similar differences between these two line width estimates in the CMZ based on HNCO data.
Depending on the number of detected emission components along the line-of-sight, their median difference varies between 3.4\,\kms (single component) and 18.8\,\kms when four components are present.
The maps derived from hyperfine structure fitting are thus considered to be more reliable than the simple moment analysis when investigating the kinematics although fitting discards all but the strongest component along the line-of-sight, introducing a bias towards strong sources.

The distribution of fitted line widths for all ammonia transitions is shown in Fig.~\ref{figure: histogram line widths} and shows a typical line width of $8 - 16$\,\kms consistent for all six observed ammonia lines.
The peak of the distribution is located at $8-10$\,\kms with only \nh333 having the peak at slightly higher velocity of $10-12$\,\kms.

\subsubsection{Opacity}\label{section: fit results opacity}

\begin{figure*}
	\centering
	\includegraphics[width=\linewidth]{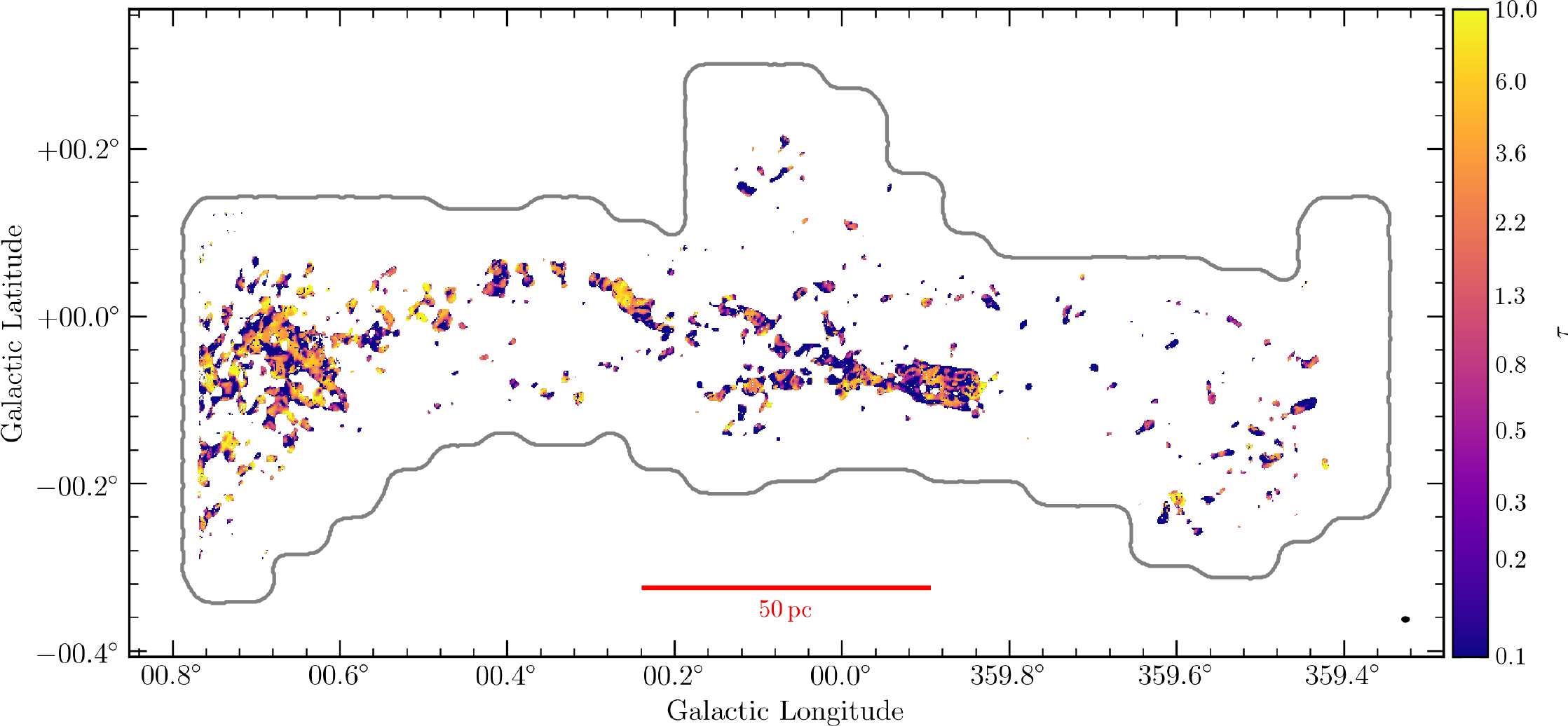}
	\caption{\nh333 opacity derived from fitting the hyperfine structure. Large areas in CMZ clouds are not optically thick ($\tau \lesssim 1$) and either optically thin ($\tau \sim 0$) or marginally thick ($\tau < 1$) Several extreme exceptions of $\tau \gtrsim 10$ exist in the Brick, an extension to the 20\,\kms cloud and the surrounding of Sgr~B2.}
	\label{figure: nh333 opacity}
\end{figure*}

The opacity map of \nh333 (Fig.~\ref{figure: nh333 opacity}) shows that a significant amount of fitted pixels of the six ammonia species exhibit opacities $\tau \gtrsim 1$ where the often used assumption of optically thin emission does not hold anymore.
This is the case not only in the very high column density clouds (like Sgr B2 and the Brick) but also in other dust ridge clouds and the inner regions of smaller clouds at negative longitudes and the backside portion of the ``ring'' ($l^+, b^-$).
A weak correlation between high column density (Fig.~\ref{figure: nh333 column density}) and high opacity (Fig.~\ref{figure: nh333 opacity}) is found in the Brick, G+0.10+0.00, 20 and 50\,\kms clouds, and the foreground cloud G-0.40-0.20.
Other clouds are optically thin at low column density or do show high opacity despite having low  column density (N$_{u,33} <2.5 \times 10^{15}$\,\sqcm, e.g. G0.50+0.05 or G-0.60-0.09), indicative of a smaller filling fraction of clumpy gas.
High opacities, especially in very small regions, must be taken with care as the edge of a region corresponds to low signal-to-noise ratio ($\gtrsim 3\sigma$, see §\ref{section: hyperfine fitting CLASS} for details) and thus random noise peaks may be mistaken as hyperfine structure satellite components by the fitting routine.
Due to the low flux of these pixels, derived column density is only marginally affected but opacity which depends on the flux ratio of hyperfine structure components can be artificially elevated.
This opacity distribution is consistent with \citet{Huettemeister1993} who observed \nh311, (2,2), (4,4) and (5,5) in selected clouds in the GC.
In the (1,1) line, they find ``strong'' sources at $\tau_{11} \sim 4$ and an average $\bar{\tau}_{11} = 2.3$ for ``weak'' sources which matches our observations of $\bar{\tau}_{11} = 2.0$ (median 1.2) and $\tau_{11} \gg 1$ in clouds when considering that SWAG also observed the less dense ISM inbetween clouds.
We can also confirm the trend to lower opacity in \citet{Huettemeister1993} for higher-J lines as the median opacity decreases to the lower fitting limit ($\tau = 0.1$) for \nh344 and higher.

\subsubsection{Temperature}\label{section: fit results temperature}

\begin{figure*}
	\centering
	\includegraphics[width=\linewidth]{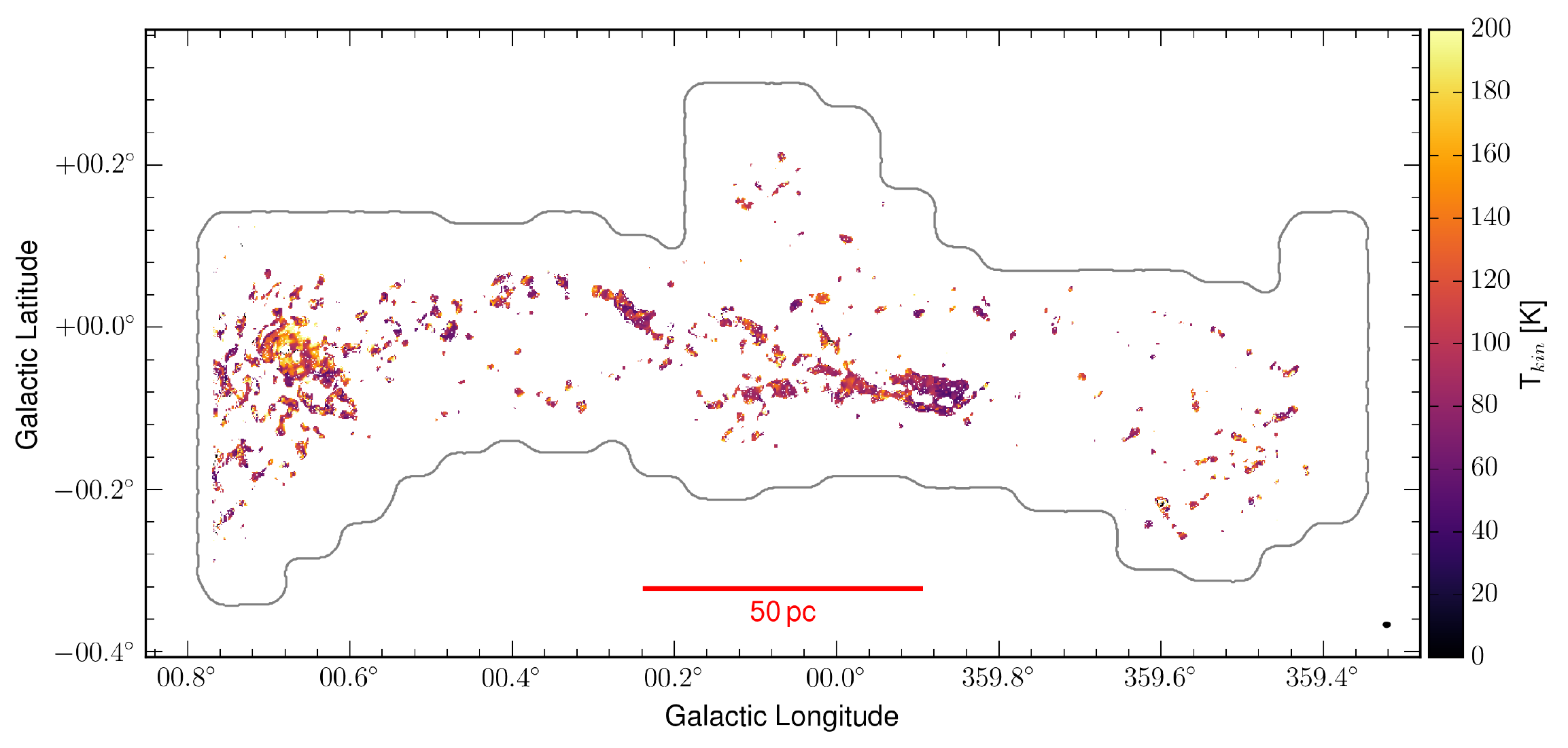}
	\caption{Kinetic ammonia gas temperature map (\temp24) derived from \nh322 and \nh344 hyperfine structure fits.}
	\label{figure: T24 kin map}
\end{figure*}

\begin{figure}
	\centering
	\includegraphics[width=\linewidth]{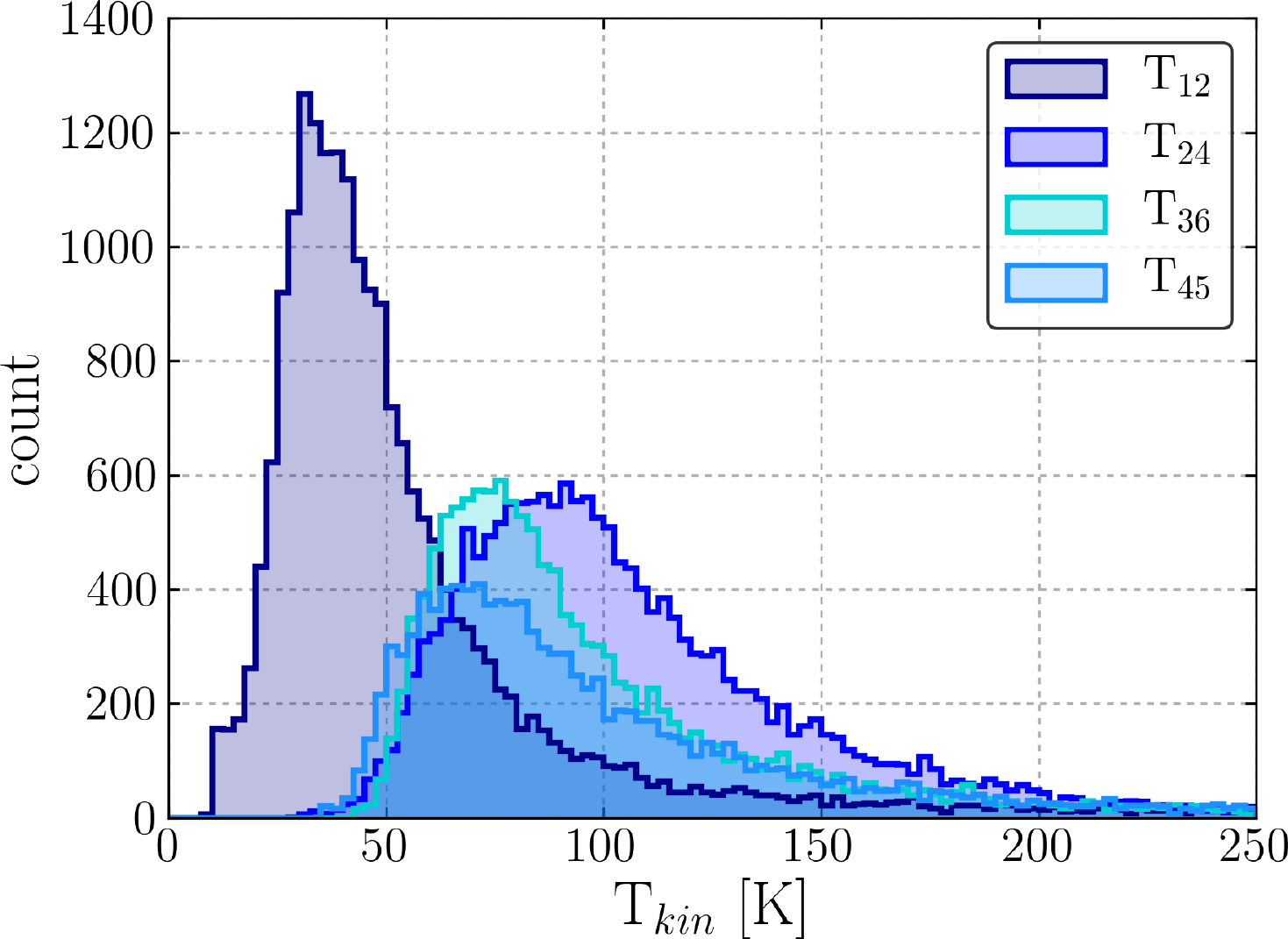}
	\caption{Distribution of temperatures in the Galactic Center as derived for four kinetic ammonia gas temperatures. The conversion of rotational to kinetic temperature is taken from \citet{Ott2011} as described in §\ref{appendix: ammonia thermometer}. T$_{ij}$ denotes the temperature measure derived from the lines \nh3ii and \nh3jj. \temp12 is sensitive to cold gas whereas higher lines need warmer conditions to be excited and are thus succeptible only to increasingly warm gas. \temp24, \temp45 and \temp36 trace similarly warm gas of $> 50$\,K.}
	\label{figure: histogram temperature}
\end{figure}

Ammonia gas temperature as traced by kinetic temperature T$_{24}$ which is derived from \nh322 and \nh344 is shown in  Fig.~\ref{figure: T24 kin map}.
The typical temperature range observed in the Galactic Center is $50 - 120$\,K, only Sgr B2 is considerably hotter with $\sim 200$\,K.

A histogram of temperature measures \temp12, \temp24, \temp45 and \temp36 is shown in Fig.~\ref{figure: histogram temperature} to illustrate the temperature distribution of the CMZ.
\temp12 cannot reliably trace temperatures $\gtrsim 50-60$\,K (see excitation energies listed in Table~\ref{table: ammonia info}) and thus its distribution peaks at $\sim 35$\,K tracing cooler molecular gas.
The other three temperature measures of higher lines are reliable to at least $\sim 200$\,K before the rotational-kinetic conversion starts to introduce large uncertainties and are therefore easier to interpret.
Their distributions are similar in peak temperature and spread with \temp24 yielding higher values by $\sim 15$\,K.
Regarding the spatial distribution, most of the gas is found at intermediate temperatures at $60-100$\,K and few hot regions $\mathrm{T} > 150$\,K.

The detection of two temperature components in the molecular gas is consistent with \citet{Huettemeister1993} who found gas at rotational temperatures $\mathrm{T}_{rot} \sim 25$\,K and $\mathrm{T}_{rot} \sim 200$\,K.
Our cold ($\mathrm{T}_{kin} = 25-50$\,K) component is thus compatible with \citet{Huettemeister1993} considering the rotational-kinetic temperature conversion.
However, the temperature of the warm components differ significantly by a factor of $\sim 2-2.5$ in rotational temperature.
\citet{Ao2013} and \citet{Ginsburg2016} derived kinetic temperatures of dense molecular gas over large areas of the CMZ based on APEX observations of H$_2$CO (formaldehyde).
They find the gas to be warm at 50\,K to $>100$\,K \citep{Ao2013} and $\sim 60$\,K to $>100$\,K \citep{Ginsburg2016} which is consistent with our detection of a warm gas component.
Regarding the spatial distribution, the temperatures known in the literature match our results:
Sgr~B2 is hot at $\sim 200$\,K (this work), $>150$\,K \citep{Ginsburg2016}; the 20\,\kms and 50\,\kms cloud and the Brick show temperatures of $80-100$\,K (this work, \citealt{Guesten1981},  \citealt{Mauersberger1986}, \citet{Ao2013}, \citet{Ginsburg2016}).

We estimate the fraction of cold ($\sim 35$\,K) and warm ($\sim 80$\,K) gas of the total detected ammonia column density by fitting the Boltzmann plot (Fig.~\ref{figure: boltzmann plot}) with two components.
This allows the calculation of the column density of cold and warm ammonia independently and hence the relative abundances.
Over the whole CMZ, we find the mean column density fractions to be $\sim 55$\,\% and $\sim 45$\,\% for the cold and warm component, respectively.
In the 20\,\kms and 50\,\kms clouds, the cold component dominates at $50\% - 70\%$ (mean: $\sim 68\%$) relative contribution to the observed ammonia column density.
In Sgr~B2, a larger fraction of column density is in the warm component which results in both components being almost equally strong ($\sim 50\%$ cold and warm).
Mean values of $\sim 60\%$ cold gas fraction are found in the dust ridge and $\sim 53\%$ in clouds at negative longitudes (excluding the 20\,\kms and 50\,\kms clouds).
The ``Three Little Pigs'' (G+0.05-0.07 ``sticks'', G+0.10-0.08 ``stones'',  G+0.15-0.09 ``straw'') are a group of clouds of similar global properties but very different substructure.
They are also striking in cold gas fraction $\mathrm{f}_{cold}$ as G+0.10-0.08 (``sticks'') shows the highest $\mathrm{f}_{cold}$ over an extended region at $\sim 81\%$ whereas its neighbor G+0.05-0.07 (``stones'') ($\mathrm{f}_{cold} \sim 65\%$) seems inconspicuous and much closer to the CMZ mean.
G+0.15-0.09 ``straw'' does not show up as its ammonia emission was to low to be hyperfine structure fitted.
These results show that the typical gas temperature is not the same over the whole CMZ but is either $\sim 35$\,K or $\sim 80$\,K depending on the considered region.

\begin{figure}
	\centering
	\includegraphics[width=\linewidth]{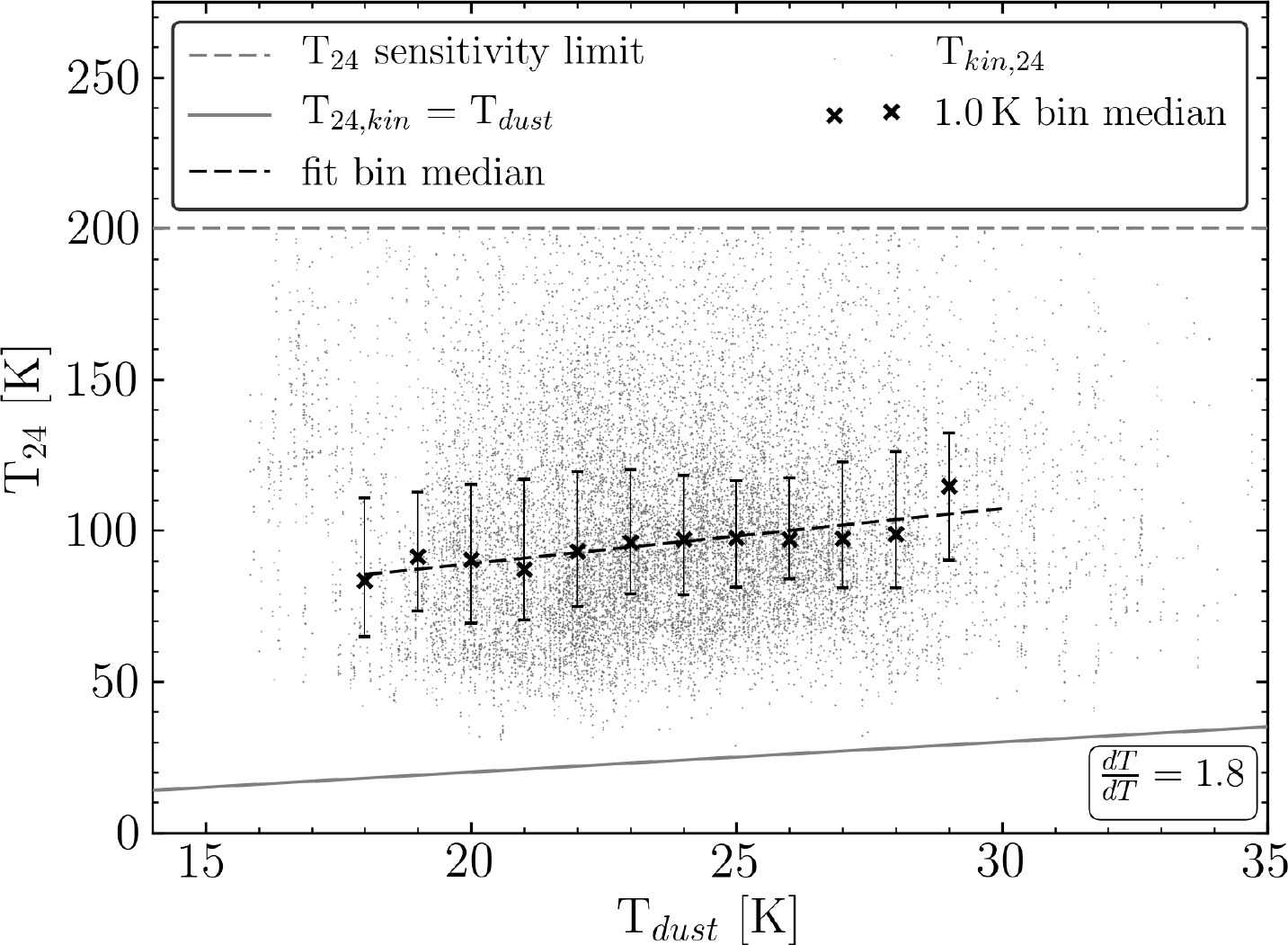}
	\caption{Correlation of Hi-Gal dust temperature \citep{Molinari2011} and kinetic ammonia gas temperature \temp24. Black crosses denote median $\mathrm{T}_{24}$ temperature in bins of 1.0\,K dust temperature which are linearly fitted (black dashed line). The errors are the range that includes 50\% of measurements in the respective bin. Median temperatures are correlated despite the offset relative to $\mathrm{T}_{24} = \mathrm{T}_{dust}$.}
	\label{figure: Tdust vs T24}
\end{figure}

The relation between Hi-GAL dust temperature \citep{Molinari2011} and ammonia gas temperature is shown in Fig.~\ref{figure: Tdust vs T24} for \temp24.
Further tracers are shown in Fig.~\ref{figure: Tdust vs TNH3}.
\temp24 is offset well above the one-to-one relation by a factor of $\sim 3$ to $>5$.
A linear best-fit, calculated in bins of 1\,K for $18\,\mathrm{K} \leq \mathrm{T}_{dust} \leq 30$\,K, yields a gradient of $\mathrm{d}T_{24} / \mathrm{d}T_{dust} = 1.8$ and fitting without binning results in no correlation (gradient $\sim 0.0$) due to cut-off effects of the limited temperature reliability range.
This result implies that, in the CMZ, the same heating mechanism is likely to be responsible for both dust and gas heating despite their temperatures not being coupled to the same value.
This hint towards a common heating source contradicts the prediction by \citet{Clark2013} who explain the temperature structure of the Brick (G0.253+0.016) by independent heating of gas and dust by cosmic rays and interstellar radiation field, respectively.
The observed correlation is compatible with heating by turbulence on large scales and compression on cloud scales (cf. §5) as was also suggested by \citet{Ginsburg2016}.
The fact that dust acts as a coolant in the CMZ is already known in the literature for a long time \citep[e.g.][]{Guesten1981,Molinari2011,Ginsburg2016} and can be confirmed by our new data.
Similar results are obtained for warm molecular gas traced by T$_{45}$ and T$_{36}$ (dT$_{45}/\mathrm{dT}_{dust} = 2.4$, dT$_{36}/\mathrm{dT}_{dust} = 1.4$).
The cold molecular gas component (T$_{12}$) only weakly correlates with dust temperatures (dT$_{12}/\mathrm{dT}_{dust} = 0.5$) which implies a temperature component that is not reflected by dust measurements.

\subsubsection{Estimation of the beam filling factor}

A simple estimation of the beam filling factor $\eta_f$ can be done based on the comparison of T$_{kin}$ and brightness temperature T$_{b}$.
For $\eta_f = 1$, both temperatures must be equal, but if the beam is partly covered by emission of temperature T$_{kin}$ and partly by weak (i.e. cold) background, the observed T$_{b}$ is lower.
A first order estimate is given by $\eta_f = \mathrm{T}_{b} / \mathrm{T}_{kin}$ under the assumptions of $\eta_f$ being identical for the two lines \nh{3}{i}{i} and \nh{3}{j}{j} that define temperature \temp{i}{j}.
This furthermore assumes high optical depth and the lines' excitation temperature being approximately given by $\mathrm{T}_{kin}$.
Typical values of dense clouds (20\,\kms cloud, Brick, cloud d, Sgr B2) are in the range $0.10 - 0.15$, Sgr C and less dense clouds are covered at $\eta_f = 0.05 - 0.10$ (e.g. the ``Three Little Pigs'' G0.05-0.07, G0.10-0.08 G0.15-0.09).
G-0.40-0.25 sticks out from all other clouds at negative longitudes with $\eta_f \sim 0.10$ which the third highest $\eta_f$ observed after Sgr B2 and the 20\,\kms cloud.
All other clouds show beam filling factors below 0.05.
The average size of cloud substructure (clump size) for $\eta_f = 0.10$ (0.05 0.15) is then 0.27\,pc (0.19\,pc, 0.33\,pc).
This is consistent with recent interferometer measurements with SMA and ALMA \citep{Rathborne2014b,Kauffmann2013,Kauffmann2016a,Lu2016}.

\subsubsection{Correlation of Fitted Parameters}\label{section: fit results parameter correlations}

\begin{figure}
	\centering
	\includegraphics[width=\linewidth]{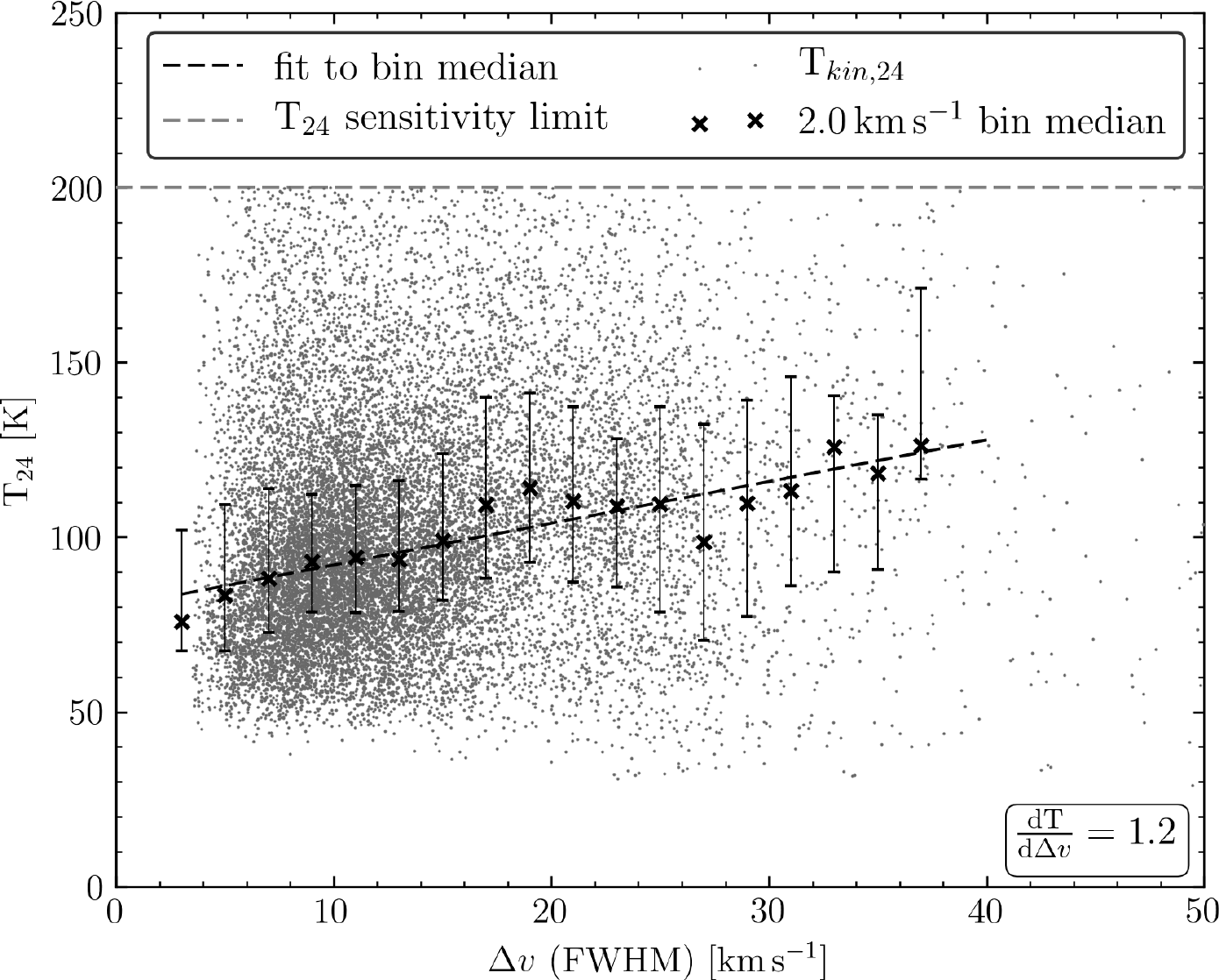}
	\caption{Kinetic ammonia temperature \temp24 as a function of line width of \nh322. Median temperatures in bins of 2.0\,\kms width are marked by black crosses and fitted linearly. The errors are the range that includes 50\% of measurements in the respective bin. A cut-off at 200\,K is applied above which the conversion of rotational to kinetic temperature becomes increasingly unreliable (Table~\ref{table: Trot Tkin}, \citealt{Morris1973}, \citealt{Ott2011}, \citealt{Gorski2017}).}
	\label{figure: FWHM22 vs T24 kin}
\end{figure}

Fig.~\ref{figure: FWHM22 vs T24 kin} shows the relation between line width and temperature for \temp24 and the corresponding line \nh322.
As before, temperatures scatter between 50\,K and the sensitivity cut-off at 200\,K for line widths of $\sim 5$\,\kms to $30-40$\,\kms with no obvious correlation.
Temperature medians in bins of 2\,\kms width, however, follow a linear relation at $d\mathrm{T} / d\mathrm{(\Delta v)} = 1.2$.
The increase in line width due to thermal broadening scales sub-linearly as $\mathrm{\Delta v} \propto \sqrt{\mathrm{T}}$ and is about one order of magnitude ($\Delta v_{thermal} \sim 2$\,\kms) weaker than the observed linear relation.
Assuming that the line width is not contaminated by blending of clouds along the line of sight, this implies that on a statistical basis, the more turbulent gas is warmer but large variations between clouds do exist as also previously detected by \citet{Immer2016} and \citet{Ginsburg2016}.

\section{Discussion}\label{section: discussion}

\begin{figure}
	\centering
	\includegraphics[width=\linewidth]{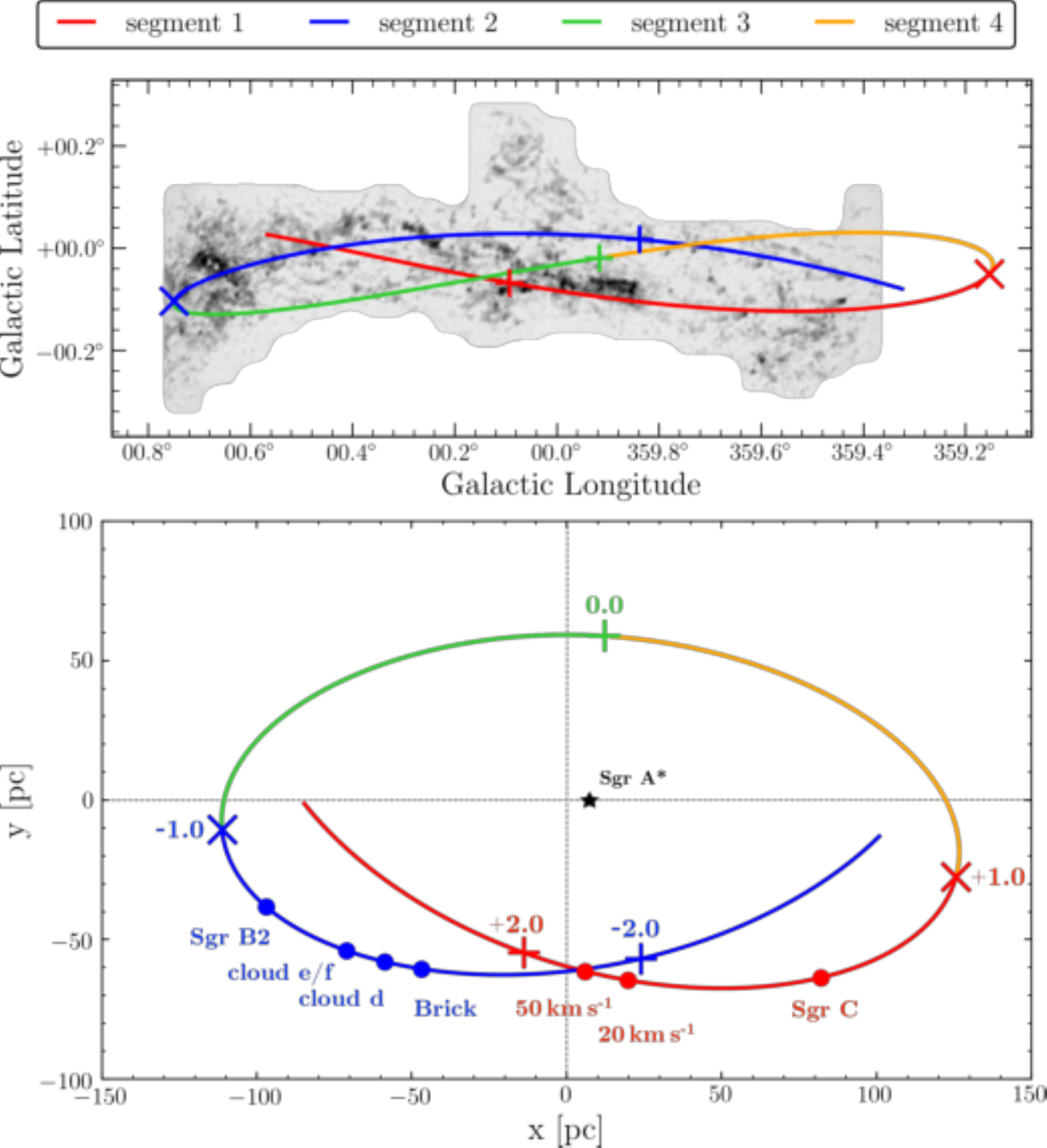}
	\caption{Overview of the stream orbit according to \citetalias{Kruijssen2015}. Pericenters are marked by plus symbols, crosses denote apocenters. The bottom of the gravitational potential at Sgr~A* is marked by a star. \emph{Top:} Projected view of the stream orbit in Galactic coordinates. Background image: SWAG \nh333 integrated intensity. \emph{Bottom:} Top-down view in physical coordinates. The observer is situated in negative $y$-direction. Major molecular clouds are labeled. Time since far side pericenter passage is given for peri- and apocenters in Myr.}
	\label{figure: stream model}
\end{figure}

We now compare the observations to models proposed in the literature with a focus to the stream model by \citetalias{Kruijssen2015} as introduced in §\ref{section: introduction}.
In practice, this means positions on the sky are mapped to times and examined for consistent sequential behavior.
Fig.~\ref{figure: stream model} offers an overview of the orbit's 3D structure, relation to major molecular clouds and evolutionary time-scales.

\subsection{Orbital Fitting and Statistical Analysis}

The mapping of cloud positions to times in the \citetalias{Kruijssen2015} model is done in position-position-velocity (PPV) space.
Within $\pm 20$\,\kms in line-of-sight velocity and $0.15^\circ$ projected distance, a cloud is matched to the nearest point along the orbit for which time and other orbital parameters are known.
The line-of-sight velocity of clouds is given by the mean velocity derived from the NH$_3$ hyperfine structure fits (§\ref{section: fit results kinematics}).
Clouds further from the stream than $0.15^\circ$ in the Galactic longitude/latitude plane are considered to not belong to the stream as this is approximately the maximum spatial separation found by \citetalias{Kruijssen2015} (see their fig.~4).
Out of the 24383 fitted pixels, we match 19453 pixels to the stream (79.8\%) while 4930 pixels (20.2\%) contain emission from clouds not related to the gas stream\footnote{This gas is mostly located on the polar spur at $l \sim 0$ and in the surrounding of Sgr~B2 and thus too far ($> 0.15^\circ$) from the stream.}.

We examine the kinetic ammonia temperatures \temp24, \temp45 and \temp36, as well as column density and line width of the six metastable inversion lines covered by SWAG along the stream.
Medians are calculated in bins of 0.01\,Myr, if the bin contains more than 10 members and therefore the sample size is large enough for the median to be considered robust against outliers.
Errors to the medians are given by the range that includes the central 50\% of all measurements in the respective bin.
The resulting time dependencies of physical parameters are assessed by fitting linear functions to the medians if a trend is apparent.

\subsection{Time Dependencies}\label{section: time dependency plots}

The time dependencies of kinetic temperature \temp24, column density and line width of \nh333 are shown in the following sections to represent results that appear in all of the ammonia temperature measures (\temp24, \temp45, \temp36) or ammonia inversion lines ($\mathrm{J} = 1, ..., 6$) if not explicitly noted otherwise.
Plots of the other temperature measures and ammonia lines can be found in Appendix~\ref{appendix: further sequence plots}.
The origin of the time scale in plots versus time is chosen to match \citetalias{Kruijssen2015} and as such identifies 0.0\,Myr ($\pm 2.03$\,Myr) with far (near) side pericenter passages, respectively (as labeled in Fig.~\ref{figure: stream model}).

Although ammonia hyperfine structure fits were derived at the projections of all four stream segments, several time ranges are not well sampled.
Especially, $0.3 \lesssim t\,\mathrm{[Myr]} \lesssim 1.5$ is not covered.
This time range corresponds to stream segment 4 where little gas is found due to the large-scale l$^+$/l$^-$ asymmetry of the CMZ.
The time ranges most interesting to this study, the dust ridge and around the pericenter passages at 0\,Myr and +2\,Myr, are well sampled.

Fitting of potential evolution can be done in up to four regions of significant data density: in the dust ridge, $-1.90 < \mathrm{t\ [Myr]} < -1.1$, at the far side pericenter passage, $-0.30 < \mathrm{t\ [Myr]} < 0.05$, near side pericenter passage, $1.75 < \mathrm{t\ [Myr]} < 2.25$ and in clouds near Sgr~C, $1.50 < \mathrm{t\ [Myr]} < 1.75$.

\begin{figure*}
	\centering
	\includegraphics[height=0.9\textheight]{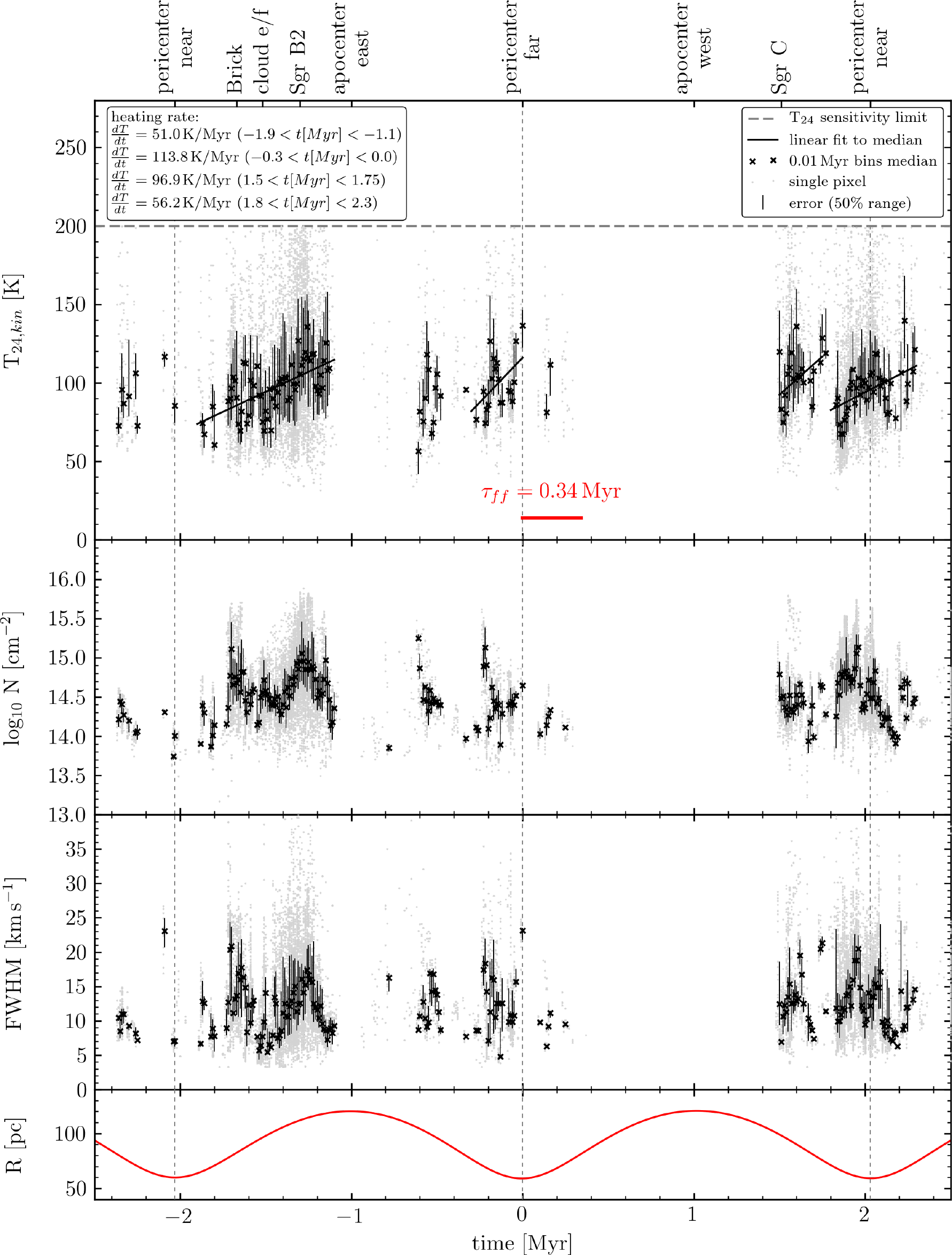}
	\caption{Kinetic ammonia temperature \temp24 (top), \nh333 column density (middle top) and \nh333 line width (middle bottom) as a function of time since far side pericenter passage (bottom x-axis, top x-axis shows important dynamical points in the K15 model and massive molecular clouds). The bottom panel shows radial distance from the center of the Galactic potential. Measurements for individual pixels (light grey points) are overlaid with black crosses denoting medians in bins of 0.01\,Myr and error intervals that include 50\% of the data in each bin. Linear fits to the median temperature and corresponding heating rates are shown for four regions of consistent time evolution.}
	\label{figure: time vs all}
\end{figure*}

\subsubsection{Gas Temperature}\label{section: discussion gas temperature}

The top panel of Fig.~\ref{figure: time vs all} shows the kinetic ammonia temperature \temp24 as a function of time.
Median kinetic temperatures of $\sim 55$\,K to $\sim 135$\,K are detected with typical error margins of [-11\,K,+15\,K].
Four sequences of consistently increasing temperature are found in the time intervals [-1.9\,Myr, -1.1\,Myr], [-0.3\,Myr, 0.05\,Myr], [1.5\,Myr, 1.8\,Myr] and [1.8\,Myr, 2.3\,Myr].
Positive linear temperature gradients of 51\,\KMyr and 56\,\KMyr are derived in the dust ridge and around near side pericenter whereas steeper gradients of 114\,\KMyr at pericenter far and 97\,\KMyr at pericenter near are fitted.
Due to the short sampled time range and few temperature measurements, the sequence at pericenter far must be taken with caution.
The error intervals are large due to the chosen conservative error estimation (§\ref{section: time dependency plots}) which causes the heating rate errors derived from $\chi^2$-fitting to also show a substantial error of typically $\sim 100\%$ which is likely overestimated given the consistent trend.
The mean heating rate of the three individual measurements (\temp24, \temp45, \temp36) thus has a relative error of approximately 58\%.

In the dust ridge (Brick to Sgr~B2, $-1.9 < \mathrm{t\,[Myr]} < -1.1$), the covered time range (0.8\,Myr) and statistics ($\sim 9200$ measurements) yield reliable heating rates of 47.1\,\KMyr to 57.9\,\KMyr (mean of $52.0$\,\KMyr) in the three partly independent temperature measures \temp24, \temp45 and \temp36 (Table.~\ref{table: temperature gradients}).
The absolute increase is from 75\,K to 110\,K (\temp24), 60\,K to 105\,K (\temp45) and 60\,K to 100\,K (\temp36) or $\sim 40$\,K on average.

The short sequence at far side pericenter passage ($-0.30 < \mathrm{t\ [Myr]} < 0.00$) is affected by low number statistics and has heating rate errors larger than 100\%.
Any conclusion for this time range should thus be treated merely as a qualitative but not as a quantitative result.

Between western apocenter of the orbit and pericenter on the near side, two independent sequences of increasing temperature are found, separated by a discontinuity at 1.8\,Myr where the median temperature sharply drops from $\sim 120$\,K to $\sim 85$\,K.
This is again seen at a very similar level in \temp24, \temp45 and \temp36.
Heating rates differ between the sequence at Sgr~C and pericenter, the former being consistently higher at $\sim 102$\,\KMyr than the latter ($\sim 52$\,\KMyr, mean heating rates).

An overview of derived heating rates in the four sequences is given in Table~\ref{table: temperature gradients}.

\floattable
\begin{deluxetable}{ccccc}
	\tablecaption{Ammonia kinetic temperature gradients (heating rates) of the four recovered temporal sequences (Fig.~\ref{figure: time vs all}) and orbital phase without and with adjusting the zeropoint (Fig.~\ref{figure: phase vs T24}). Errors are $\sim 100\%$ for individual heating rates and $\sim 58\%$ for the mean.
		\label{table: temperature gradients}}
	\tablehead{time range & $\displaystyle \frac{T_{24}}{dt}$ & $\displaystyle \frac{T_{45}}{dt}$ & $\displaystyle \frac{T_{36}}{dt}$ & mean\\ & [\KMyr] & [\KMyr] & [\KMyr] & [\KMyr]}
	\startdata
	$-1.90<t \,\mathrm{[Myr]}<-1.10$ & \multirow{2}{*}{51.0} & \multirow{2}{*}{47.1} & \multirow{2}{*}{57.9} & \multirow{2}{*}{52.0}\\
	(dust ridge)\\[3pt]
	$-0.30<t \,\mathrm{[Myr]}< 0.00$ & \multirow{2}{*}{113.8} & \multirow{2}{*}{77.3} & \multirow{2}{*}{67.7} & \multirow{2}{*}{86.3}\\
	(pericenter far)\\[3pt]
	$1.50 <t \,\mathrm{[Myr]}< 1.75$ & \multirow{2}{*}{96.9} & \multirow{2}{*}{122.8} & \multirow{2}{*}{84.8} & \multirow{2}{*}{101.5}\\
	(Sgr~C)\\[3pt]
	$1.80<t \,\mathrm{[Myr]}<2.30$ & \multirow{2}{*}{56.0} & \multirow{2}{*}{57.2} & \multirow{2}{*}{41.6} & \multirow{2}{*}{51.6}\\
	(pericenter near)\\[3pt]
	combined (phase) & 14.6 & 14.6 & 17.7 & 15.6\\
	combined (phase, adjusted) & 47.7 & 58.4 & 46.3 & 50.8\\
	\enddata
\end{deluxetable}

\subsubsection{Column Density}\label{section: discussion column density}

Median column densities show no evidence of time dependence (Fig.~\ref{figure: time vs all}, middle top, Fig.~\ref{figure: time vs column density}) in the dust ridge.
No consistent trend is apparent as column densities increase strongly in the Brick and Sgr~B2 above a flat or, at most, very weakly rising base level.
These clouds are already known to be denser than other dust ridge clouds and thus expected to be more massive outliers to any potential underlying relation \citep{Lis1994a,Lis1994b,Molinari2011,Longmore2012,Pillai2015}.
Given that the column densities in between the Brick and Sgr~B2 show no rise and also drop to the same level after Sgr~B2, we conclude that there is no evolutionary column density sequence in the dust ridge.

The clouds around near side pericenter passage show weak evidence for decreasing column densities where the median column density drops for J = 1, 2, 3 but can be considered constant within the scatter for J = 4, 5, 6.
This relation is likely also influenced by two individual objects, the 20\,\kms and 50\,\kms clouds although density bumps are not as prominent as in Sgr~B2.

At far side pericenter, medians scatter by up to $\sim 1$\,dex and no conclusion can be drawn.
The column densities in the Sgr~C region are constant for all six ammonia tracers with very low scatter compared to the rest of the covered time ranges.

Qualitatively and quantitatively very similar results are derived for opacity due to the correlation of opacity and column density and hence not discussed further.

\subsubsection{Line Width}\label{section: discussion line width}

The evolution of median line width (FWHM) is similar to column density but shows even less consistent behavior as can be seen in Fig.~\ref{figure: time vs all} (middle bottom) for \nh333 (J = 1, ..., 6 in Fig.~\ref{figure: time vs linewidth}).
In the dust ridge, line widths vary by a factor of $\sim 2.5$ with peaks in the Brick and Sgr~B2.
There is no sign that these could be just outliers to an underlying trend as line widths are at similar levels of $6-10$\,\kms before the Brick, in between Brick and Sgr~B2 and after Sgr~B2.
Any potential time dependency of line width thus cannot be stronger than a few \kms\,Myr$^{-1}$ to be hidden within the observed scatter.
Results are very similar for the other ammonia lines (J=1,2,4,5,6) at slightly varying absolute level as could already be seen in the line width histogram (Fig.~\ref{figure: histogram line widths}).

In the other time ranges that show temperature evolution, no sign of consistent line width trend can be found.
As for column density, 20\,\kms and 50\,\kms cloud are weak outliers towards higher values.

The increase in line width at the position of large and massive clouds can be caused by scaling relations \citep[``Larson laws'', ][]{Larson1981}, be an effect of hyperfine structure fitting or projection.
CMZ clouds are known to follow a steep scaling relation of line width $\sigma$ with cloud radius R: $\sigma \propto \mathrm{R}^{0.6-0.7}$ \citep{Shetty2012,KruijssenLongmore2013,Kauffmann2016b} and thus line widths are expected to increase by a factor of $\sim 1.5$ between the small clouds b/c and larger clouds like the Brick and cloud e/f \citep{Immer2012}.
The observed variation in line width of a factor $\sim 2.5$ is thus only partly due to variations in cloud size and other sources of scatter must be present, such as unresolved blending of multiple emission components along the line-of-sight as shown in Fig.~\ref{figure: good bad fit} d).
Spectra with two close components with separations of a few to a few tens of \kms are more likely to be present in larger clouds and thus cause the same signature as the scaling relation does, increasing the overall scatter.
Projection effects may also increase the apparent line width due to pile up along the line-of-sight at the tangent points of the orbit. This can play a role in Sgr~B2 but is not expected to affect the other clouds (cf. Fig.~\ref{figure: stream model}).

\subsection{Discussion in the Context of of the Stream Model and Tidal Triggering}\label{section: discussion triggered collapse}

As shown above, sequences of increasing temperature are found but no such sign is recovered for line width and column density.
At near side pericenter passage an anti-correlation of increasing temperature and potentially decreasing column density is found whereas the positive temperature gradients in the dust ridge, at far side pericenter and around Sgr~C do not have a detectable counterpart in column density or line width.
Based on the concept of tidally triggered SF \citep{Longmore2013b}, we discuss the evolution at and after pericenter passage separately in §\ref{section: discussion compression} and §\ref{section: discussion dust ridge}.
By the time-tagging method used in this work, the interpretation is furthermore linked to the stream model of \citetalias{Kruijssen2015}, but general conclusions can be transfered to other orbital models.
If a model includes radial oscillation, it can trigger cloud collapse and may give rise to a sequence of star formation.
If the clouds were on circular orbits, our time-tagging method would still result in accurate measurements which means the basic conclusions are not sensitive to the used orbital model.

\subsubsection{Sequences at Pericenter Passage}\label{section: discussion compression}

Qualitatively, increasing temperatures during pericenter passage can occur by converting the gravitational energy released by cloud compression or collapse to heat through a turbulence cascade if cooling is not efficient enough to keep the cloud at a constant temperature \citep[e.g.][]{Ginsburg2016}.
The effect on line width by such a process is not immediately clear and likely needs detailed simulations.
In general, turbulence should increase the observed line width which would result in a rising sequence of line width following the rise in temperature.
However, as the energy makes its way through the cascade to smaller scales R, the dissipation rate $\epsilon \propto \sigma / \mathrm{R}$ also increases allowing for little or even no change in the effective line width $\sigma$.
If the turbulent line width does increase, it will still be difficult to detect due to variations among clouds (§\ref{section: discussion line width}).

Radiative cooling generally is an efficient process \citep[e.g.][]{Juvela2001} and it is not obvious why it could be so inefficient for temperatures to rise by $> 50$\,\KMyr.
As shown by Fig.~\ref{figure: nh333 opacity}, a significant fraction of clouds in the CMZ are optically thick for line emission at $\sim 25$\,GHz which can trap cooling photons inside clouds and diminish cooling efficiency.
Ammonia is not a dominant cooling line, but other lines which are more important for cooling (CO, C$^+$ and, at least in the Brick, neutral oxygen \citet{Clark2013}) can be expected to be even more optically thick than NH$_3$ and the same argument applies.
If true, the hint towards decreasing column density might also be an effect of opacity as in optically thick media, the observed emission originates from an optical depth of $\tau \sim 1$ which is the envelope rather than core of the cloud.
The outer layers of a cloud moving through pericenter along the stream orbit are sheared and can even become unbound while the inner cloud is pushed into self-gravity and collapse as shown observationally by \citet{Rathborne2014,Federath2016} and modeled by \cite{Kruijssen2017a} and Kruijssen et al. (in prep.).
Hence, a trend of decreasing column density can be observed.

The observed mean heating rates at pericenter passage of $\sim 52$\,\KMyr and $\sim 86.3$\,\KMyr are large, but cannot be easily compared to theoretic predictions because the complex interplay of dynamic and hydrodynamic processes require simulations including radiative transfer.
Without these simulations, it is also not known at which time before pericenter passage tidal effects start to become non-negligible and affect temperature.
The same applies to the time after pericenter passage when compression and shear are reduced which could in principle also lead to cooling which is not observed.
Detailed simulations can be expected in the near future as the general feasibility was already demonstrated \citep{Clark2013,Kruijssen2017a}.
Without simulations, we can estimate the time range of strong tidal influence on the clouds from Fig.~\ref{figure: time vs all}.
The indication of increasing temperature in Fig.~\ref{figure: time vs all} suggests that \emph{at least} 0.25\,Myr before reaching pericenter a temperature change is detectable.
A change in temperature might occur earlier than 0.25\,Myr before pericenter passage, but unfortunately our data do not cover that time range to check if temperatures are flat before heating sets in.
At 0\,Myr and 2\,Myr, the deepest point in the potential is reached at R$_{peri} \sim 59$\,pc while 0.25\,Myr corresponds to R $\sim 1.22$ R$_{peri} = 72$\,pc.
Apocenter is reached at R$\sim 120$\,pc, so tidal effects becoming relevant at $\sim 75 -80$\,pc from the center of the potential is plausible.
In the spherically symmetric enclosed mass distribution in the GC \citep{Launhardt2002}, the gravitational potential at $\lesssim 0.25$\,Myr before reaching pericenter\footnote{or $\sim50$\,pc before reaching pericenter at $\sim200$\,\kms orbital velocity which is appropriate near pericenter \citet{Kruijssen2015}} is symmetric to the respective range after pericenter, i.e. tidal effects are expected to play a role until at least 0.25\,Myr after pericenter passage.
After pericenter passage, some processes reverse, i.e. tidal pressure is released which can allow cooling but cloud collapse is ongoing and early star formation may set in.
This complex interplay of physical processes make it very difficult to define an end point of strong tidal influence from the observational data.
Generally, observations show that the temperatures are consistent with turbulent energy dissipation \citep{Ginsburg2016} as would be expected from triggered cloud collapse.

\subsubsection{Sequences after Pericenter Passage}\label{section: discussion dust ridge}

This observed sequence in the dust ridge starts about one free-fall time $\tau_{ff} = 0.34\,Myr$ after pericenter passage whereas the other ones start before.
This means that if cloud collapse was triggered, a cloud's core is collapsed whereas previously sheared but still bound gas is reaccreted.
As the collapse of outer cloud layers proceeds, the release of gravitational energy declines which implies flattening temperature gradients.
However, no obvious decrease in heating rate is found in the dust ridge meaning cloud collapse is still ongoing at similar intensity as at pericenter passage or a second heating source starts to contribute.
In the framework of triggered SF, star formation could be a second source of heating. 
As an estimate, star formation can set in as early as one free-fall time after the cloud itself started to collapse, which is 0.34\,Myr for a typical CMZ cloud, and coincides with the time since the Brick passed pericenter \citepalias{Kruijssen2015}.
Accordingly, early signs of embedded star formation have been found in one region of the Brick \citep{Longmore2012,Johnston2014,Rathborne2014}, i.e. heating by star formation is just about to set in.

These effects increase the temperature if the gas cannot cool efficiently enough which apparently is the case as gas temperatures are known to be significantly offset from dust temperatures \citep[Fig.~\ref{figure: Tdust vs T24} and e.g.][]{Guesten1981,Molinari2011,Ginsburg2016}.
Therefore, the increase in temperature cannot simply be attributed to a single of these processes but a combination at varying relative strengths, which is not known at the moment for Galactic Center conditions and cannot be inferred from our observations.
In order to disentangle the contribution of star formation from the kinematic and gravitational effects, a complex hydrodynamic gas simulation including star formation feedback is needed.

Collapsing clouds are expected to show a signature of increasing column density that is not observed which might be a consequence of shearing outer cloud layers during pericenter passage.
It might also be an observational effect of too low resolution to trace the collapse. 
Collapse happens quickest on the highest density, lowest virial ratio scales. 
It is known in the literature that the column density profiles of the Brick and bricklet clouds are centrally concentrated and that the volume density of the Brick increases towards the center \citep{Walker2015,Rathborne2016}.
The collapse will therefore happen quickest at the center, on scales smaller than the SWAG beam which will wash out potential small scale effects.
In higher resolution observations, \citet{Walker2015} found evidence for increasing gas and (proto-) stellar density along the stream orbit from the Brick to Sgr B2.

As discussed in §\ref{section: discussion gas temperature}, the sequence of increasing temperature around Sgr~C is likely not connected to the sequence at near side pericenter due to the sharp discontinuity and differing heating rate.
The observation of 24\,$\mu$m emission in this region and the higher temperature compared to the pericenter sequence suggest that these clouds already underwent tidal triggering of collapse at far side pericenter (see also \citealt{Kruijssen2015} regarding the position of Sgr~C).
This places Sgr~C as a successor to the dust ridge sequence in terms of evolution, but not in time.
Finding no column density trend in Sgr~C matches the picture as this cloud is supposedly collapsed already and actively forming stars \citep{Kendrew2013,Lu2016} which powers the rise in temperature.

\subsubsection{Connection of Sequences at and after Pericenter Passage}\label{section: discussion link phases}

According to the \citetalias{Kruijssen2015} model, the empirically found sequences are directly associated in the sense that the star formation dominated phase (dust ridge and Sgr~C) follows the tidal compression dominated phase (near and far pericenter passage).
Orbital phase, however, cannot be the only parameter controlling the physical state of gas along the stream as discussed in §\ref{section: discussion gas temperature}.
It seems likely that the clouds at and after pericenter passage are similar but not identical in properties, i.e. all clouds may follow the same evolution but the gas in the dust ridge was initially cooler or could cool more efficiently when cloud collapse was triggered.
The required offset in temperature to match the sequences is small ($\sim 15-20$\,K) compared to the observed heating rates ($\sim 52$\,\KMyr).
As we do not know the initial cloud properties, this remains speculation and we can only state that the observed sequences can be the result of tidally triggered collapse assuming a scatter of initial cloud temperature of $\sim 20$\,K.
On this note, it should be kept in mind that individual cloud properties as well as collapse triggering are statistically distributed and are not identical for subsequent occurrences.
Estimates for the variation of physical parameters from cloud to cloud can be drawn from the analysis of velocity corrugations in the [$l^-$, $b^-$] segment of the stream that is expected to be a precursor to the dust ridge \citep{Henshaw2016b}.
For this analysis, \citet{Kruijssen2017a} report standard deviations of initial cloud parameters of 0.19\,dex (mass), 0.09\,dex (radius), 0.16\,dex (density) and 0.08\,dex (free-fall time scale) among the potential precursors to the dust ridge only.
The standard deviations would naturally be larger when also including the progenitors to the other temperature sequences identified in this paper.
If the variations of initial cloud temperature are at a similar level as these properties, the expected temperature scatter is $\gtrsim 20$\,K, larger than the observed offsets of $15-20$\,K.
These numbers show that the observed temperature sequences at and after pericenter passage may be causally related as the temperature deviation is within the expected scatter.

\subsubsection{Orbital Phase and Variation of the Trigger Point}

\begin{figure*}
	\centering
	\includegraphics[height=0.8\textheight]{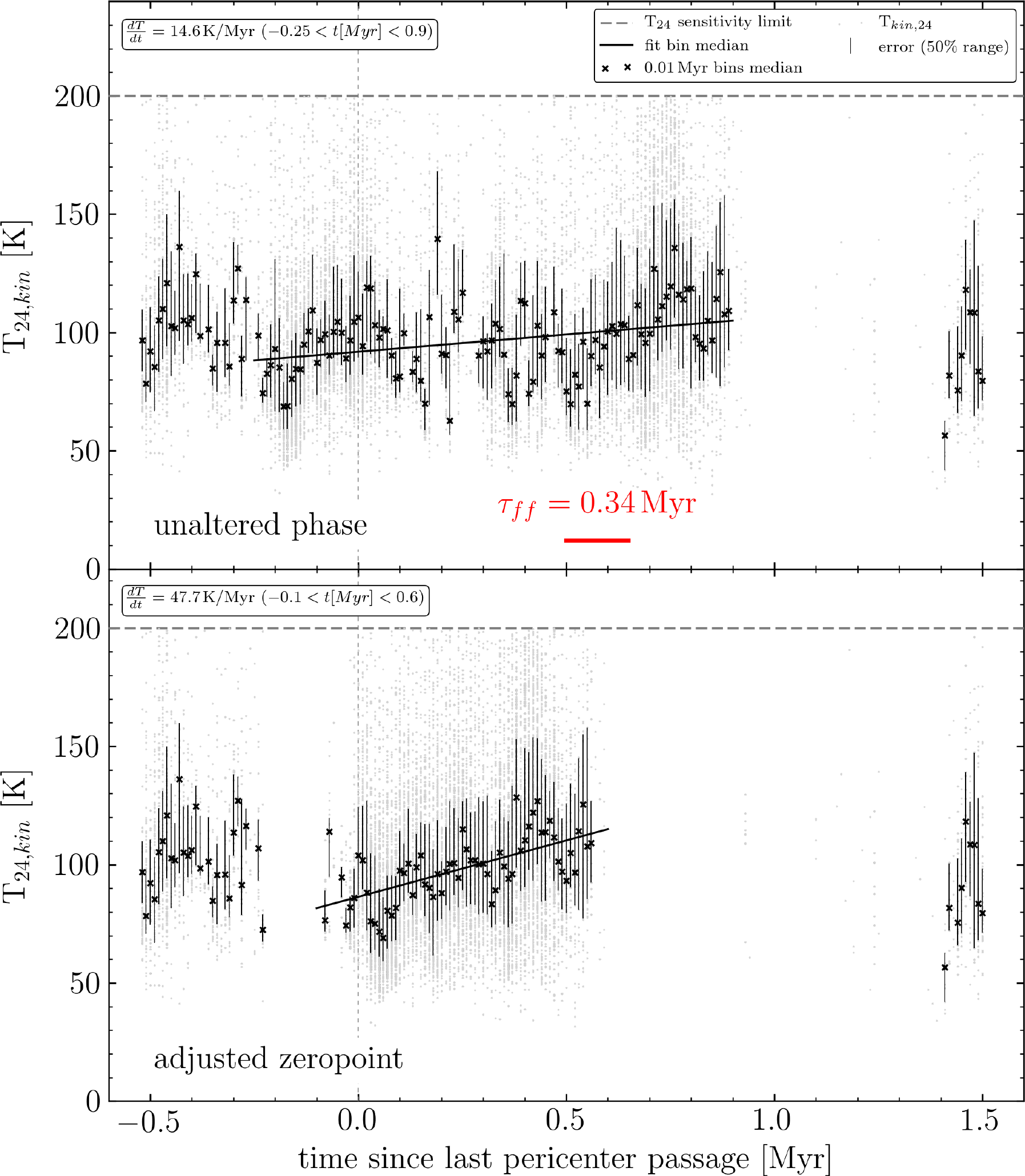}
	\caption{Kinetic ammonia temperature \temp24 as a function of time since the last occurrence of pericenter passage (orbital phase). Symbols and error bars are as in Fig.~\ref{figure: time vs all}.
		The top panel shows the unaltered data as in Fig.~\ref{figure: time vs all} (top panel) whereas in the bottom panel the zeropoint of the sequences in the dust ridge, pericenter near and pericenter far are adjusted to maximize the heating rate. Within a range of reasonable offsets in zeropoint of $\pm 0.3$\,Myr, the best fit is achieved for -0.3\,Myr (dust ridge), +0.2\,Myr (pericenter near) and +0.2\,Myr (pericenter far).
		The fitted time ranges of -0.25\,Myr to +0.9\,Myr and -0.1\,Myr to 0.6\,Myr result from the sum of the fits in the time domain.}
	\label{figure: phase vs T24}
\end{figure*}

The periodic occurrence of pericenter passages every $\sim 2.0$\,Myr suggest that temperatures can also be plotted against orbital phase which is shown in Fig.~\ref{figure: phase vs T24} (top panel).
Although four temperature sequences could be defined as a function of absolute time, no global sequence is found.
Only in the time range 0.5\,Myr to 0.9\,Myr shows a consistent trend of rising temperatures as the data in this region originates almost exclusively from the dust ridge.
The other sequences are lost due to superposition of differing temperature levels that also smears the discontinuity at $\mathrm{t}_{peri} = 1.75$\,Myr which occurs at a phase of $\mathrm{t}_{last} = -0.25$\,Myr.
However, fitting a linear relation in the time range -0.25\,Myr to 0.90\,Myr (sum of the four ranges identified in §\ref{section: time dependency plots}) still results in a positive heating rate of 14.6\,\KMyr.
Almost identical results are obtained for \temp45 and \temp36 with relative errors of $\sim 100\%$ as listed in Table~\ref{table: temperature gradients}.
These results are in line with \citet{Kauffmann2016b} who find a trend in density gradients of six CMZ clouds but no clear trend in star formation activity per unit dense gas as a function of orbital phase.

The signature of rising temperature could be expected to become stronger by transforming to orbital phase if the four sequences are samples of a single consistent evolution.
However, it is highly unlikely that all clouds are triggered to collapse exactly at pericenter but rather within some time range around pericenter.
By allowing the zeropoint $\mathrm{t}_{start}$ of each sequence to vary by up to $\pm 0.3$\,Myr, the sequential behavior can be recovered (Fig.~\ref{figure: phase vs T24} bottom panel).
The heating rate is maximized at a value of $\sim 48$\,\KMyr for zeropoints of $\mathrm{t}_{start} = -0.3$\,Myr (dust ridge), $\mathrm{t}_{start} = +0.2$\,Myr (pericenter near) and $\mathrm{t}_{start} = +0.2$\,Myr (pericenter far).
At 0.3\,Myr before or after pericenter, the radial distance from the bottom of the GC potential is $\mathrm{R} = 77$\,pc ($\mathrm{R} = 60$\,pc at pericenter) and thus the clouds are still deep in the potential where tidal forces can be expected to be sufficiently strong to trigger cloud collapse.
The fact that heating rates similar to the individual values can be recovered by only adjusting the exact trigger point without taking potential variations of initial cloud conditions or variations of the gravitational potential into account, suggests that $\mathrm{t}_{start}$ is one of the major sources of scatter of a downstream SF sequence.
The stacked and $\mathrm{t}_{start}$-adjusted mean heating rates are smaller than the per-segment heating rates implying that another factor still erases part of the trend. This is the vertical (temperature) offset due to slightly different initial conditions which has been discussed in §\ref{section: discussion link phases}.
This observation confirms the warning by \citet{Kruijssen2017a} that simple stacking of pericenter passages is not advisable due to scatter.
However, problems can be diminished if it is possible to estimate the scatter in $\mathrm{t}_{start}$.
Additional scatter can result from the local environment (e.g. radiation field) that does not have to be identical for two clouds of same evolutionary stage at different position.
For instance, pericenters on near and far side are identified with each other in terms of evolutionary stage despite the spatial separation of $\sim 120$\,pc.
Hence, influences of the environment can be very different at a given time since \emph{last} pericenter passage.

\subsection{Sequential Evolution in other CMZ Models}\label{section: discussion conclusion}

Although this analysis is based on the kinematic stream model of \citetalias{Kruijssen2015}, similar results would be obtained assuming a different orbital model\footnote{Of course, this also assumes that a consensus about the 3D position of CMZ clouds is reached. Results differ greatly if a cloud is placed on the far side of the GC instead of in front.}.
However, the interpretation of the found sequences in §\ref{section: discussion triggered collapse} relies on a varying tidal field strength along the orbit.
In the context of a dense gas torus \citep{Molinari2011} instead of streams, the basic requirement for tidal triggering of cloud collapse is still present \citep[as argued by][]{Longmore2013b}: gas passes close to the bottom of the Galactic Center gravitational potential.
It is irrelevant if this is due to a displaced orbit (ring model), elliptical orbits (x$_1$/x$_2$ orbits in the bar model) or a radially oscillating stream.
Differing dynamics (e.g. orbital speed) imply that the time axis of Fig.~\ref{figure: time vs all} is stretched or squished but the progression fundamentally stays the same.
The major difference in the context of this work is the question if another trigger point on the far side of the orbit is present or not which does not strongly affect this analysis as most of the gas is detected in the dust ridge or near side pericenter.
Thus, a definite distinction between models is not possible but the hints of a third trigger point at far side pericenter passage favors the \citetalias{Kruijssen2015} model.
As not all the models agree on the clouds' positions along the line-of-sight, any connection between the found sequences rely on the adopted model.
Some versions of the two spiral arms model place the 20 and 50\,\kms clouds or Sgr~C and the dust ridge on different arms \citep{Sofue1995,Sawada2004}, thus no connection is given whereas in a recent spiral arm model \citep{Ridley2017} and the elliptical ring model, the dust ridge immediately follows the 20 and 50\,\kms clouds.
The \citetalias{Kruijssen2015} model is different from all other published orbital models of the CMZ in that it predicts two different pericenters on the near side. This is consistent with our observations but the absence of dense gas on segment 2 before the dust ridge does not allow a definite distinction between models.

\subsection{Relation to other recent GC SF research}

Finding a gas temperature sequence in the CMZ stream segments confirms implications that followed from the star formation triggering model by \citet{Longmore2013b}, but does not directly confirm that the clouds themselves evolve as expected.
Heating or cooling of molecular clouds does not necessarily have to be associated with star formation or collapse in the theoretically expected way, but can be related to other mechanisms such as cosmic ray heating as well \citep[e.g.][]{Yusef-Zadeh2007,Ao2013,Clark2013}.
A sequence of rising temperature could be observed as a function of distance to the potential heating sources Sgr~A* and Sgr~B2.
For these two sources and the distribution of measurements, the fitted time since pericenter passage can be related to physical distance.
The observed temperature sequence in the dust ridge rises towards Sgr~B2 which is qualitatively consistent with it being the locally dominating heating source.
The other sequences, however, do not match this picture because the molecular clouds on the far side show an inverse trend and gas at negative longitudes requires a second heating source to the east that does not influence the close-by western end of the dust ridge.
Similar arguments apply to Sgr~A*.
Clouds are observed to heat up beyond pericenter (closest distance to Sgr~A*) and the 20 and 50\,\kms clouds are colder than Sgr~C which is $\sim 1.5$ times further from Sgr~A*.
However, these arguments also rely on the 3D distribution of gas in the GC that is still debated \citep[e.g.][]{McGary2001,Clavel2013,Takekawa2015,Kruijssen2015,Churazov2017}.

It is therefore essential to gain more insight in cloud evolution in the stream and the GC in general in order to link it to observations of star formation tracers (in the dust ridge) and stars (Arches and Quintuplet clusters).
\citet{Barnes2017} extend the SF sequence downstream from Sgr~B2 by finding evidence of a gradual increase in stellar feedback activity as measured from the size of the hot gas bubbles driven by young massive stars.
The Arches and Quintuplet clusters are found to be consistent in kinematics and age to have formed by tidally triggered cloud collapse and thus seem to fit in the Galactic Center star formation model (\citetalias{Kruijssen2015}).
If this can be confirmed by future observations, the time range covered by the model would greatly increase from pre-collapse clouds ($\mathrm{t_{peri}} \sim -1\ \mathrm{to} -2$\,Myr) to evolved stars (Quintuplet, $\mathrm{t_{peri}} \sim 5$\,Myr) and span the whole time range of star formation.
Immediate insights for cloud evolution could be gained from a study of internal cloud kinematic, size, shape and density.

Beside the ISM ingredients gas and dust, the key to a better understanding of (proto-) star formation in extreme environments such as the CMZ are the stars themselves.
Evolution in gas and possibly also in dust is accompanied by increasing signs of star formation along the dust ridge that we suspect to partially power the rise in temperature.
Starting from an increased base level after initial heating by tidal effects, the Brick is still cold and shows little star formation \citep{Longmore2012,Kauffmann2013,Rodriguez&Zapata2013} while cloud c contains H$_2$CO and SiO masers \citep{Ginsburg2015a}, cloud d contains a H$_2$CO maser suggesting massive star formation \citep{Immer2012} and Sgr B2 is heavily forming stars heating the surrounding medium.
From the qualitative picture, gas and stars seem to go in hand providing a rough picture that needs to be tested in more detail in order to reach the goal of a well-calibrated absolute timeline of star formation in the CMZ.

\section{Summary and Conclusions}\label{section: summary and conclusions}

We describe the observations and data reduction of the Survey of Water and Ammonia in the Galactic Center (SWAG) and present the first data products.
This paper covers the central $\sim 250$\,pc of the Central Molecular Zone in the Galactic Center in the six metastable ammonia inversion lines \nh311 to (6,6) at a resolution of $\sim 22"$ ($\sim 0.9$\,pc).
The gas distribution is asymmetric with respect to the Galactic Center (Sgr~A*) as most gas is detected at positive longitudes.
We find ammonia gas in the known arching structures, especially in the dust ridge spanning from the Brick (G0.253+0.016) to Sgr~B2.
Typical \nh333 main beam brightness is found to be of order a few hundred m\jybeam (few K) but reaches up to 5.28\,\jybeam (24.8\,K) in Sgr~B2.
Ammonia gas at $[0,b>0]$ is oriented approximately perpendicular to the Galactic plane and shows significantly higher velocities than the rotating arc structures.
Typical velocity dispersions are $5-25$\,\kms but are affected by blending of ammonia hyperfine components and multiple emission components along the line-of-sight.

Fitting the hyperfine structure allows us to derive maps of line width, opacity, column density and kinetic gas temperature.
We find the line widths to be $\Delta v \mathrm{(FWHM)} = 6-18$\,\kms with a typical value of $\sim 10$\,\kms, consistent for the six ammonia species.
Large areas of the CMZ are optically thin with the more massive clouds being somewhat optically thick at $\tau < 5$.
Exceptions of high opacity ($\tau \gtrsim 10$) are the Brick, several clouds surrounding Sgr~B2, an extension to the 20\,\kms cloud and G-0.40-0.20.
Ammonia column densities reach up to $\mathrm{N} \sim 7 \times 10^{16}$\,\sqcm in Sgr~B2 and several $10^{16}$ in the massive CMZ clouds.
Smaller clouds on the gas ``ring'' show lower column densities of $\mathrm{N} < 10^{16}$\,\sqcm.

The kinetic gas temperatures exhibit two components, a cold component of $25 - 50$\,K traced by \temp12 and a warm component of $50-120$\,K traced by \temp24, \temp45 and \temp36 where \temp{i}{j} denotes the temperature derived from ammonia states \nh{3}{i}{i} and \nh{3}{j}{j}.
Sgr~B2 stands out as the warm component is detected at up to $\sim 200$\,K.
On average over the CMZ, the cold component contributes $\sim 55\%$ to the ammonia column density, the warm component $\sim 45\%$.
In the 20\,\kms and 50\,\kms clouds, the cold gas fraction is higher ($\sim 68\%$), as is in the dust ridge ($\sim 60\%$).
The warm gas component is more pronounced in Sgr~B2 where both, cold and warm gas, contribute equally ($\sim 50\%$ each) to the total column density.
We confirm the kinetic gas temperature to be offset from dust temperature for which we derive offsets of $\sim 10-15$\,K for the cold component and $\sim 75-100$\,K for the warm component.
Despite the offset, median gas temperatures of both components are linearly correlated with the dust temperature.

One possibility to exploit this rich data set is given in §\ref{section: discussion} by statistically analyzing CMZ temperature, column density and line width as a function of time based on the dynamical model of \citetalias{Kruijssen2015}.
We find evidence for a consistent, gradual heating of $52$\,\KMyr in two sequences in the dust ridge and at near side pericenter passage in the three partially independent ammonia temperature tracers \temp24, \temp45 and \temp36.
Near Sgr~C a steeper trend of $\sim 100$\,\KMyr is detected whereas at far side pericenter passage, we identify a qualitative increase of gas temperature.
A hint on time dependent evolution is also found as a decreasing sequence of column density at near side pericenter passage while other time ranges show no or no consistent evolution.
No evolution is found in line width but rather variations between clouds dominate over potential weaker evolutionary trends.
This time dependence can be explained in the context of \citet{Longmore2013b} and \citetalias{Kruijssen2015} as being the result of tidal triggering of cloud collapse when passing pericenter.
Gravitational energy released through a turbulence cascade triggered by tidal compression is potentially the source of heating near pericenter.
The observed sequences of rising temperatures can result if the heat cannot be radiated away efficiently in the dense molecular clouds.
Clouds downstream the triggering point continue to collapse and begin to form stars in their cores which introduces an increasing source of heat that likely powers the second sequence of rising gas temperatures.
Stacking of the identified sequences results in a weaker correlation due to scatter of the exact collapse triggering point, i.e. clouds do not collapse at identical orbital phase.
By allowing the starting point of each sequence to vary by up to $\pm 0.3$\,Myr, we recover a heating rate of $\sim 48$\,\KMyr and estimate the triggering point to scatter by -0.2\,Myr to +0.3\,Myr.

Our work demonstrates the power of ammonia temperature measurements in inferring physical conditions.
Usually a tool for investigations on small scales, we greatly expanded the spatial scale as was never done before to derive temperatures, column density and opacity in the inner CMZ and, finally, the whole CMZ in the full SWAG survey.
Despite the large area, SWAG offers high resolution which allows statistical conclusions like the observed rising temperature sequences.
The other data of SWAG, spatial as well as spectral coverage will be presented in forthcoming papers.

\acknowledgments

HB acknowledges support from the European Research Council under the Horizon 2020 Framework Program via the ERC Consolidator Grant CSF-648505.
JMDK gratefully acknowledges funding from the German Research Foundation (DFG) in the form of an Emmy Noether Research Group (grant number KR4801/1-1, PI Kruijssen), from the European Research Council (ERC) under the European Union's Horizon 2020 research and innovation programme via the ERC Starting Grant MUSTANG (grant agreement number 714907, PI Kruijssen), and from Sonderforschungsbereich SFB 881 ``The Milky Way System'' (subproject P1) of the DFG.
ER is supported by a Discovery Grant from NSERC of Canada.
T.P. acknowledges support from the Deutsche Forschungsgemeinschaft, DFG via the SPP (priority program) 1573 “Physics of the ISM”.

\software{CASA \citep{McMullin2007}, CLASS (\url{http://www.iram.fr/IRAMFR/GILDAS}), miriad \citep{Sault1995}, APLpy \citep{aplpy}}

\facility{ATCA(CABB)}

\newpage
\appendix

\section{SWAG data products}\label{appendix: SWAG data products}

\subsection{Channel Maps}\label{appendix: SWAG channel maps}

\begin{figure}[H]
	\centering
	\includegraphics[height=0.8\textheight]{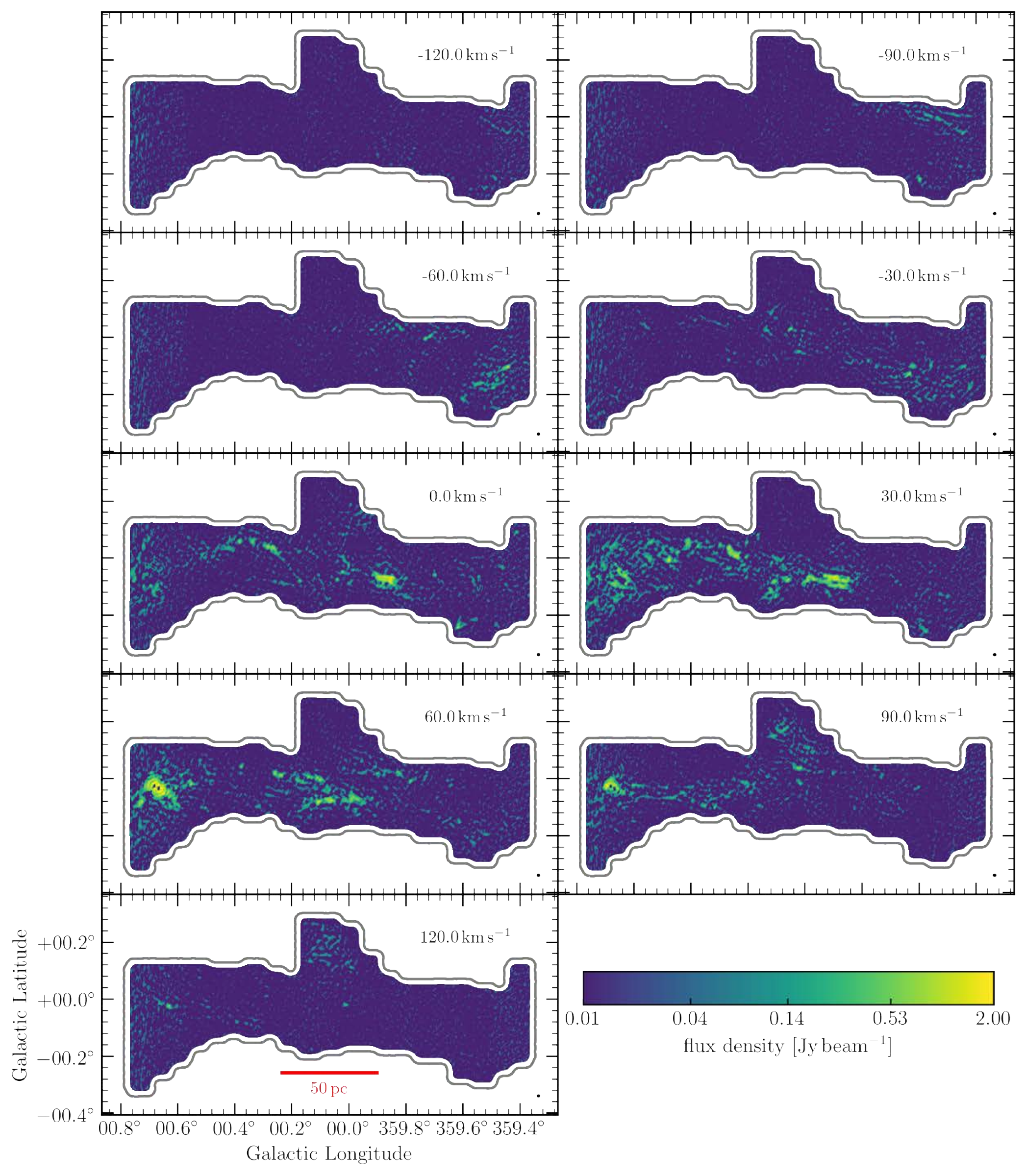}
	\caption{Channel maps of \nh311. Every 15$^{\mathrm{th}}$ channel of 2\,\kms (separated by 30\,\kms) in the range of -120\,\kms to +120\,\kms is shown. Velocity and beam area ($26.22" \times 17.84"$, $89.3^\circ$) are indicated in the top and bottom left corner, respectively. The conversion factor from flux density in \jybeam to brightness temperature in K for this beam size at 23.69\,GHz is $\mathrm{T\ [K]}/\mathrm{S\ [Jy\,beam^{-1}]} = 4.65$.}
	\label{figure: nh311 channel map}
\end{figure}

\begin{figure}[H]
	\centering
	\includegraphics[height=0.8\textheight]{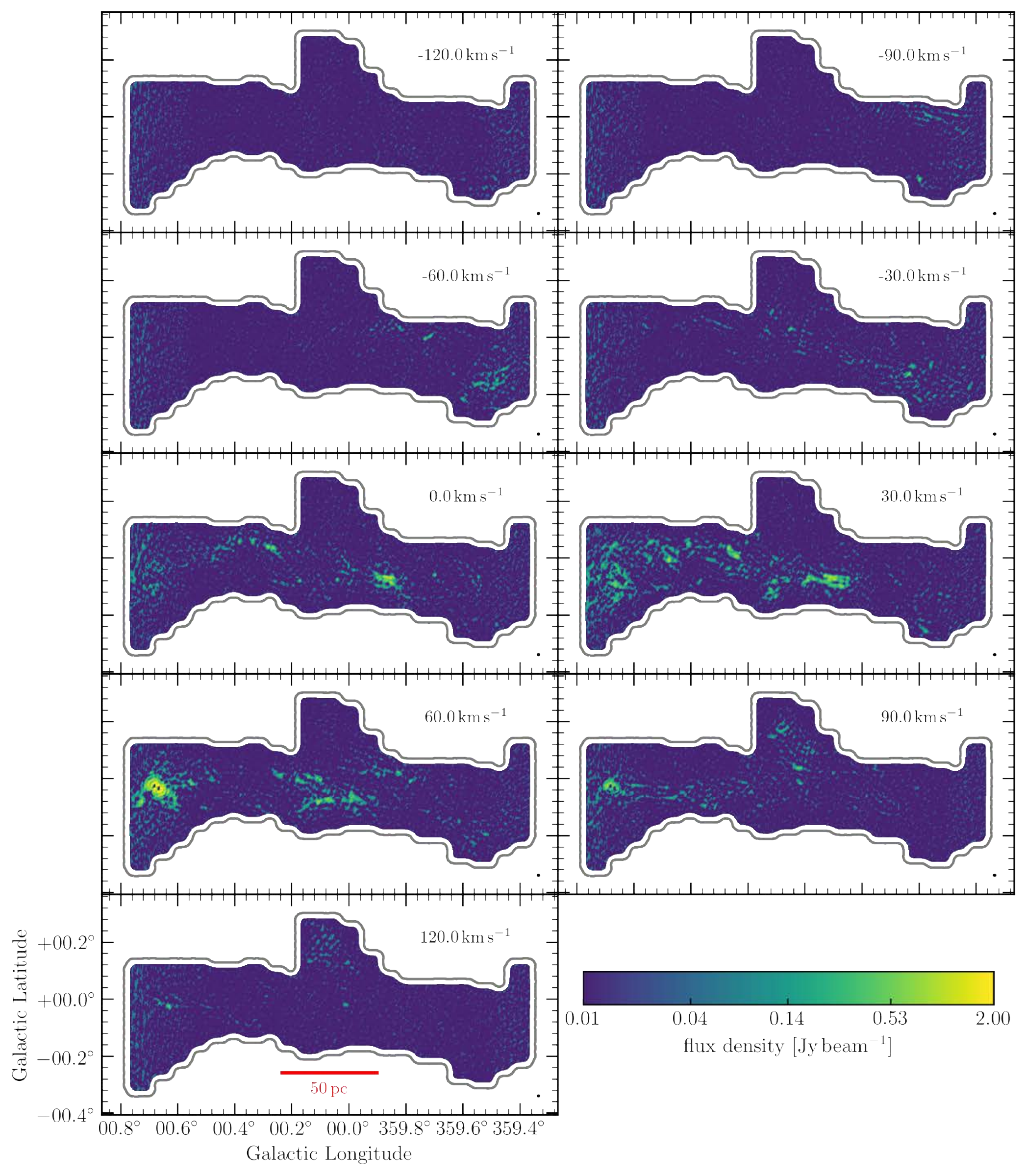}
	\caption{Channel maps of \nh322. Every 15$^{\mathrm{th}}$ channel of 2\,\kms (separated by 30\,\kms) in the range of -120\,\kms to +120\,\kms is shown. Velocity and beam ($26.19" \times 17.82"$, $89.3^\circ$) are indicated in the top and bottom left corner, respectively. The conversion factor from flux density in \jybeam to brightness temperature in K for this beam size at 23.72\,GHz is $\mathrm{T\ [K]}/\mathrm{S\ [Jy\,beam^{-1}]} = 4.65$.}
	\label{figure: nh322 channel map}
\end{figure}

\begin{figure}[H]
	\centering
	\includegraphics[height=0.8\textheight]{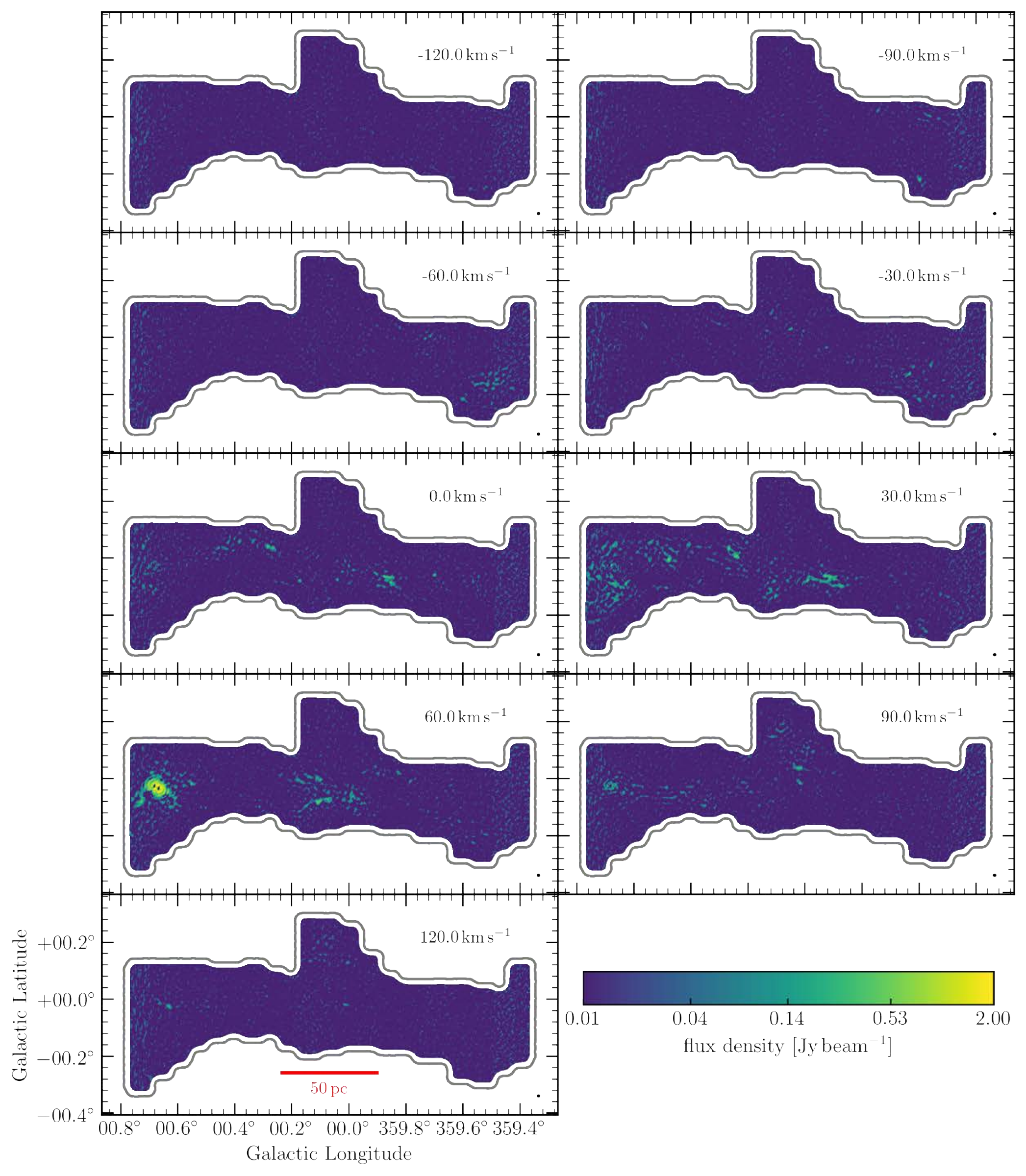}
	\caption{Channel maps of \nh344. Every 15$^{\mathrm{th}}$ channel of 2\,\kms (separated by 30\,\kms) in the range of -120\,\kms to +120\,\kms is shown. Velocity and beam ($25.74" \times 17.51"$, $89.3^\circ$) are indicated in the top and bottom left corner, respectively. The conversion factor from flux density in \jybeam to brightness temperature in K for this beam size at 24.14\,GHz is $\mathrm{T\ [K]}/\mathrm{S\ [Jy\,beam^{-1}]} = 4.65$.}
	\label{figure: nh344 channel map}
\end{figure}

\begin{figure}[H]
	\centering
	\includegraphics[height=0.8\textheight]{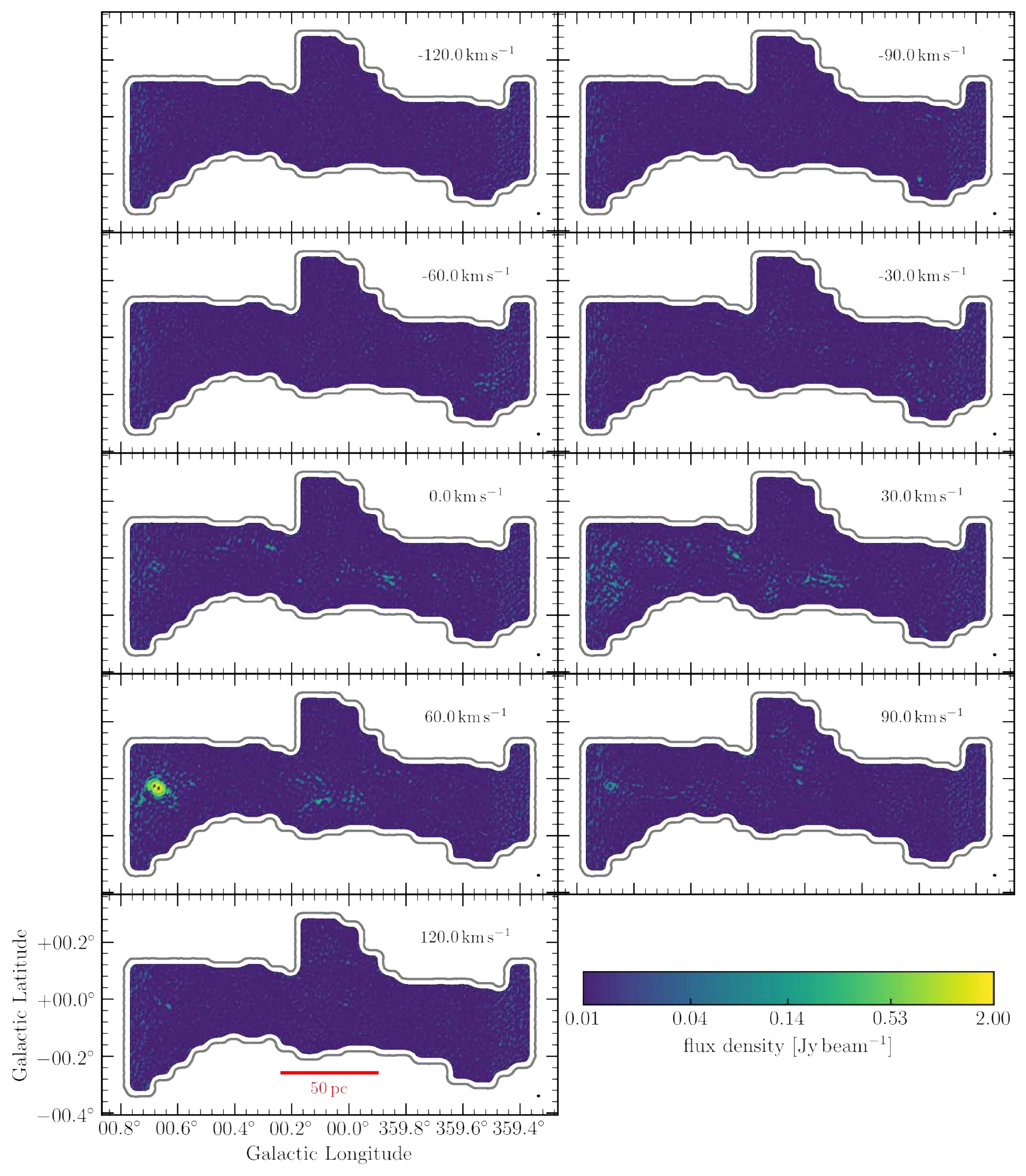}
	\caption{Channel maps of \nh355. Every 15$^{\mathrm{th}}$ channel of 2\,\kms (separated by 30\,\kms) in the range of -120\,\kms to +120\,\kms is shown. Velocity and beam ($25.32" \times 17.23"$, $89.3^\circ$) are indicated in the top and bottom left corner, respectively. The conversion factor from flux density in \jybeam to brightness temperature in K for this beam size at 24.53\,GHz is $\mathrm{T\ [K]}/\mathrm{S\ [Jy\,beam^{-1}]} = 4.65$.}
	\label{figure: nh355 channel map}
\end{figure}

\begin{figure}[H]
	\centering
	\includegraphics[height=0.8\textheight]{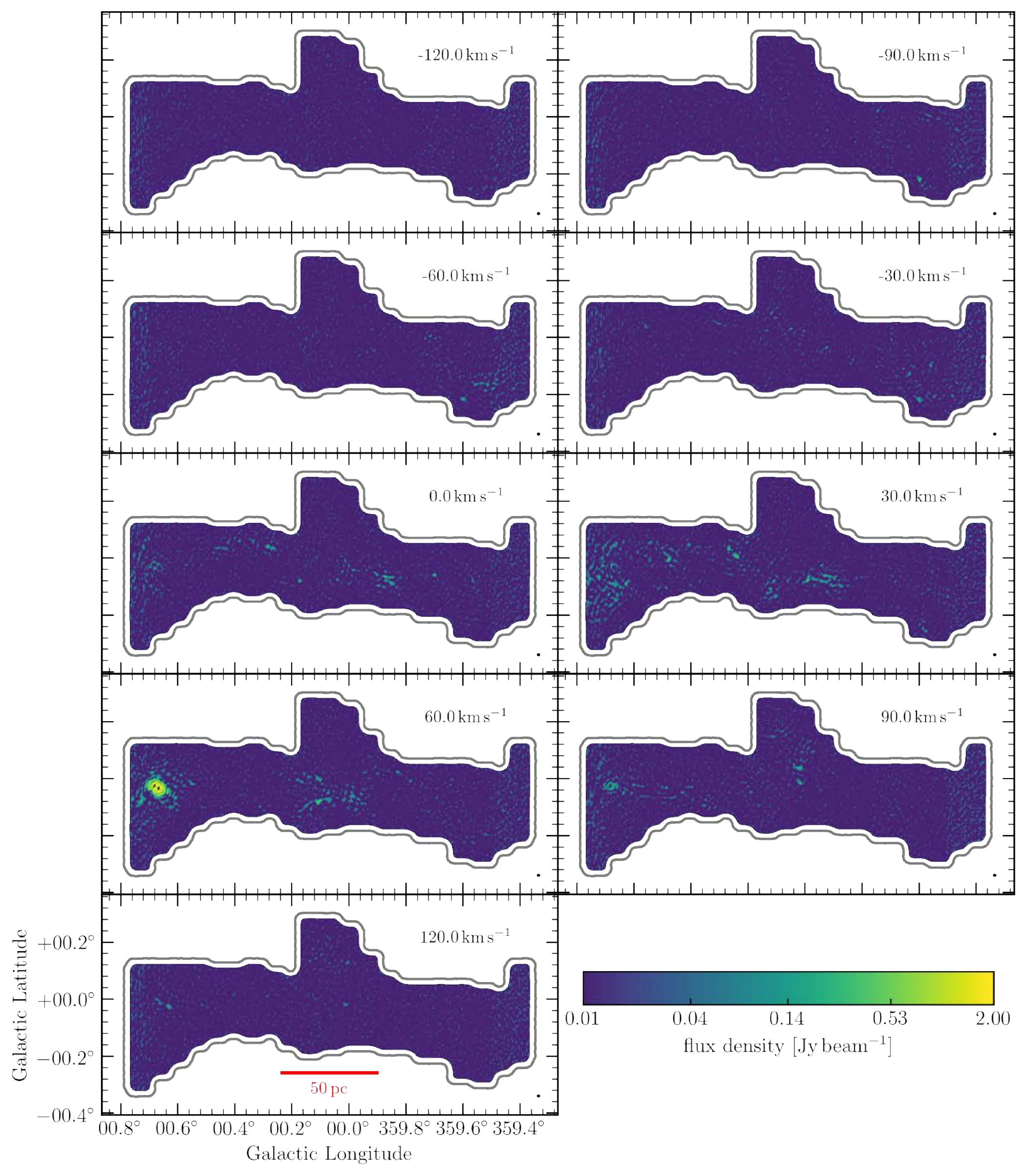}
	\caption{Channel maps of \nh366. Every 15$^{\mathrm{th}}$ channel of 2\,\kms (separated by 30\,\kms) in the range of -120\,\kms to +120\,\kms is shown. Velocity and beam ($24.80" \times 16.87"$, $89.3^\circ$) are indicated in the top and bottom left corner, respectively. The conversion factor from flux density in \jybeam to brightness temperature in K for this beam size at 25.06\,GHz is $\mathrm{T\ [K]}/\mathrm{S\ [Jy\,beam^{-1}]} = 4.65$.}
	\label{figure: nh366 channel map}
\end{figure}

\subsection{Sample Spectra}\label{appendix: SWAG example spectra}

\begin{figure}[H]
	\includegraphics[width=\textwidth]{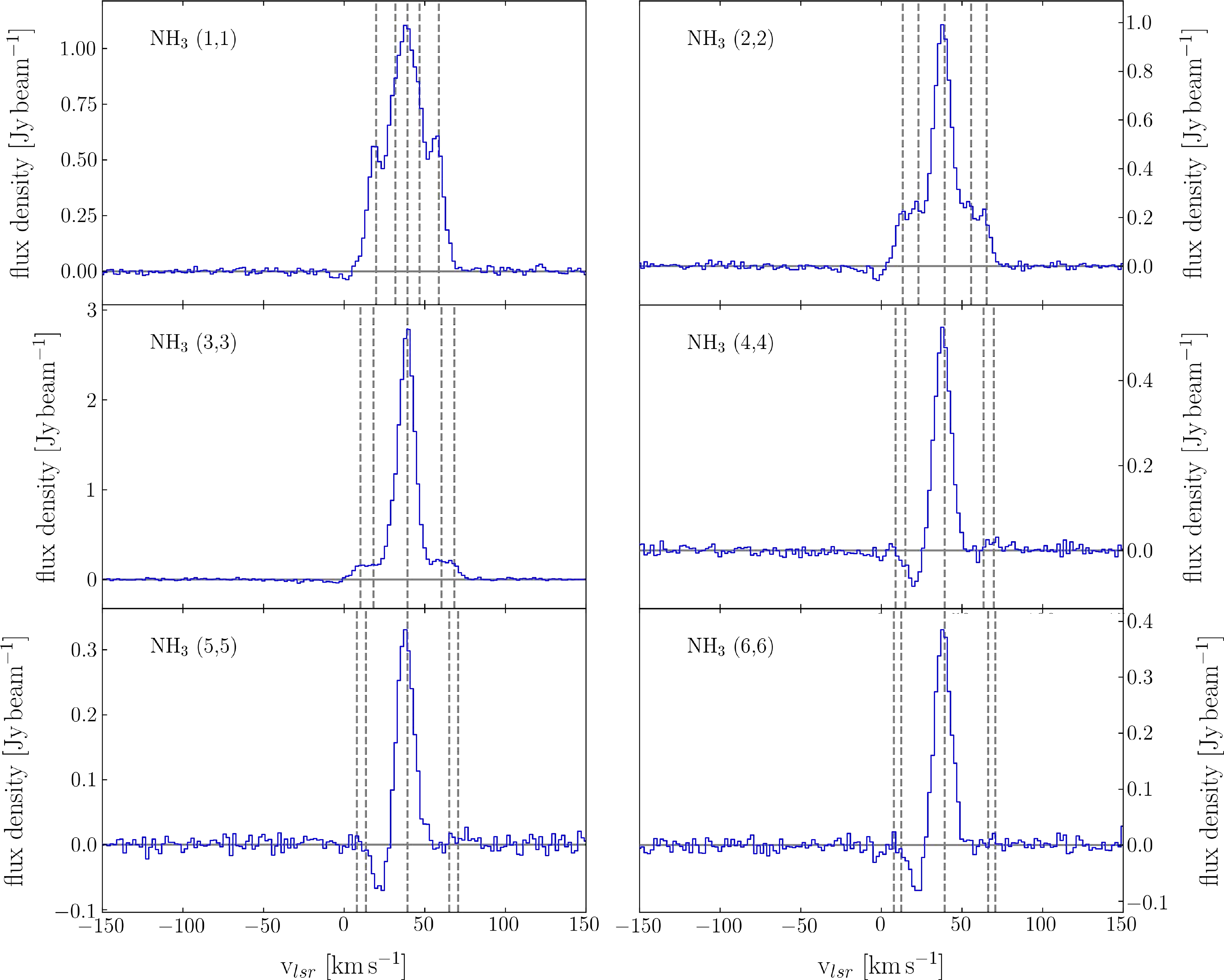}
	\caption{Example spectra with independent scaling of the y-axis towards the Brick extracted from the \nh311 to (6,6) data cubes. RMS noise is 0.062\,K, 0.061\,K, 0.061\,K, 0.060\,K, 0.056\,K and 0.061\,K, respectively. The positions of the hyperfine satellites according to Table~\ref{table: ammonia hyperfine components} are indicated by dashed vertical lines. Note the complex structure of broad line widths blending hyperfine components and an absorption component at $\sim 20$\,km\,$^{-1}$ in the higher lines.}
\end{figure}

\subsection{Moment Maps}\label{appendix: SWAG moment maps}

\begin{figure}[H]
	\centering
	\includegraphics[height=0.9\textheight]{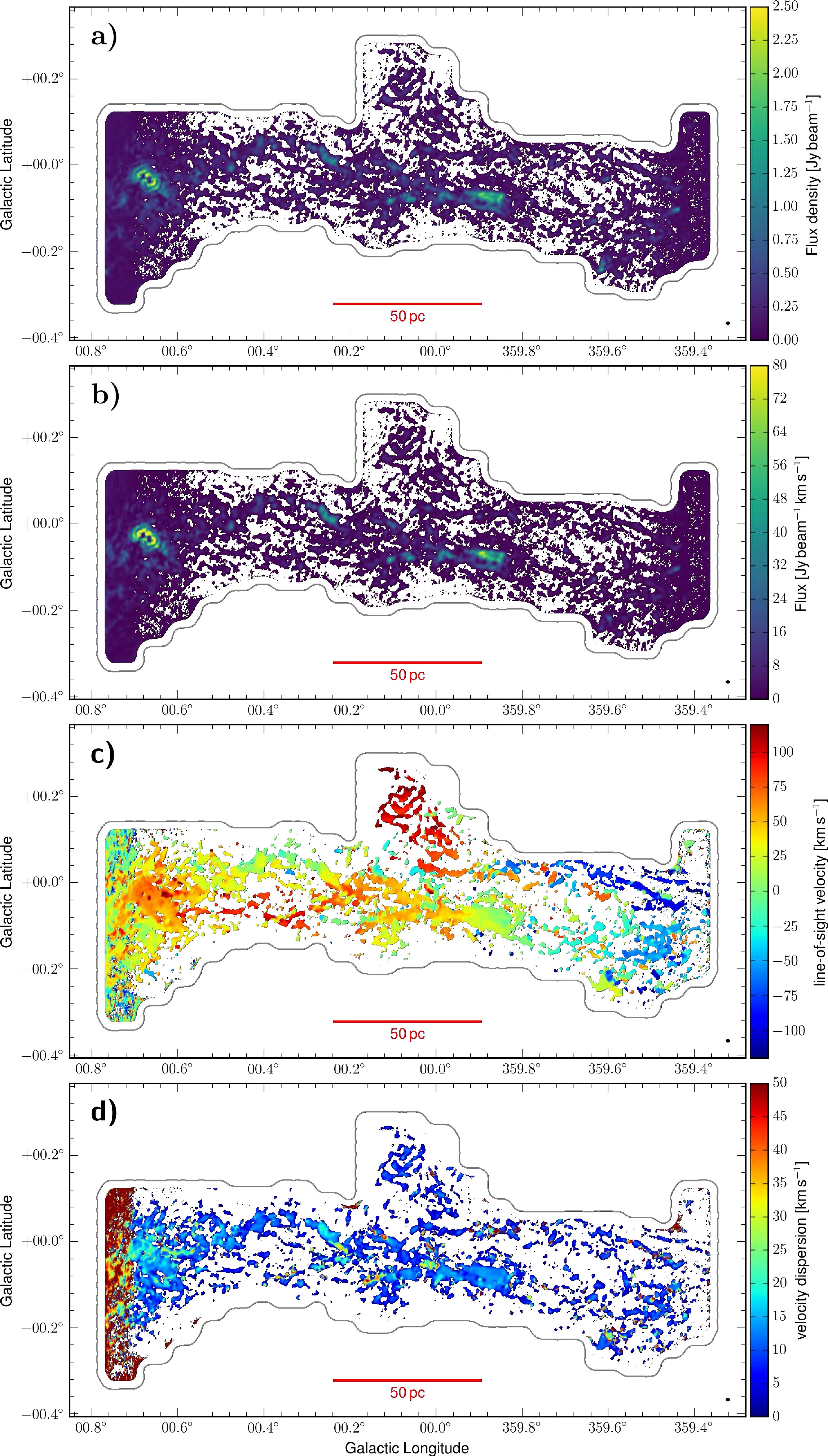}
	\caption{\nh311 moment maps: \emph{a}) peak intensity, \emph{b}) integrated intensity (moment 0), \emph{c}) intensity weighted mean velocity (moment 1), \emph{d}) intensity-weighted velocity dispersion (moment 2). The intensity maps a) and b) are masked at $3 \sigma$, velocity maps c) and d) are masked at $5 \sigma$ with an rms noise of $\sigma = 13.4$\,mJy. The beam of $26.2" \times 17.8"$ (1.06\,pc $\times$ 0.72\,pc) is indicated in the lower right corner of each panel.}
	\label{figure: nh311 moment map}
\end{figure}

\begin{figure}[H]
	\centering
	\includegraphics[height=0.9\textheight]{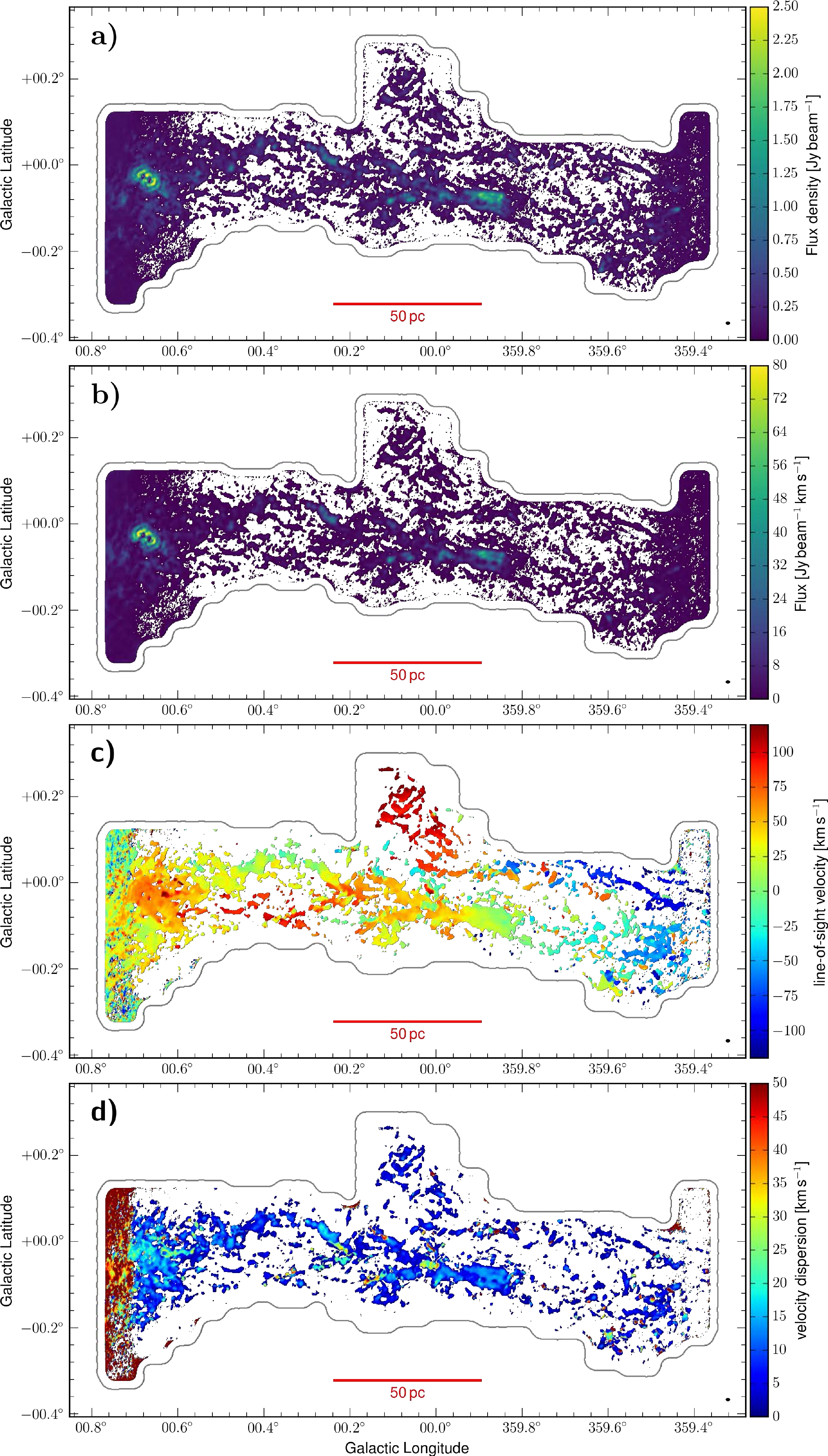}
	\caption{\nh322 moment maps: \emph{a}) peak intensity, \emph{b}) integrated intensity (moment 0), \emph{c}) intensity weighted mean velocity (moment 1), \emph{d}) intensity-weighted velocity dispersion (moment 2). The intensity maps a) and b) are masked at $3 \sigma$, velocity maps c) and d) are masked at $5 \sigma$ with an rms noise of $\sigma = 13.1$\,mJy. The beam of $26.2 \times 17.8"$ (1.05\,pc $\times$ 0.72\,pc) is indicated in the lower right corner of each panel.}
	\label{figure: nh322 moment map}
\end{figure}

\begin{figure}[H]
	\centering
	\includegraphics[height=0.9\textheight]{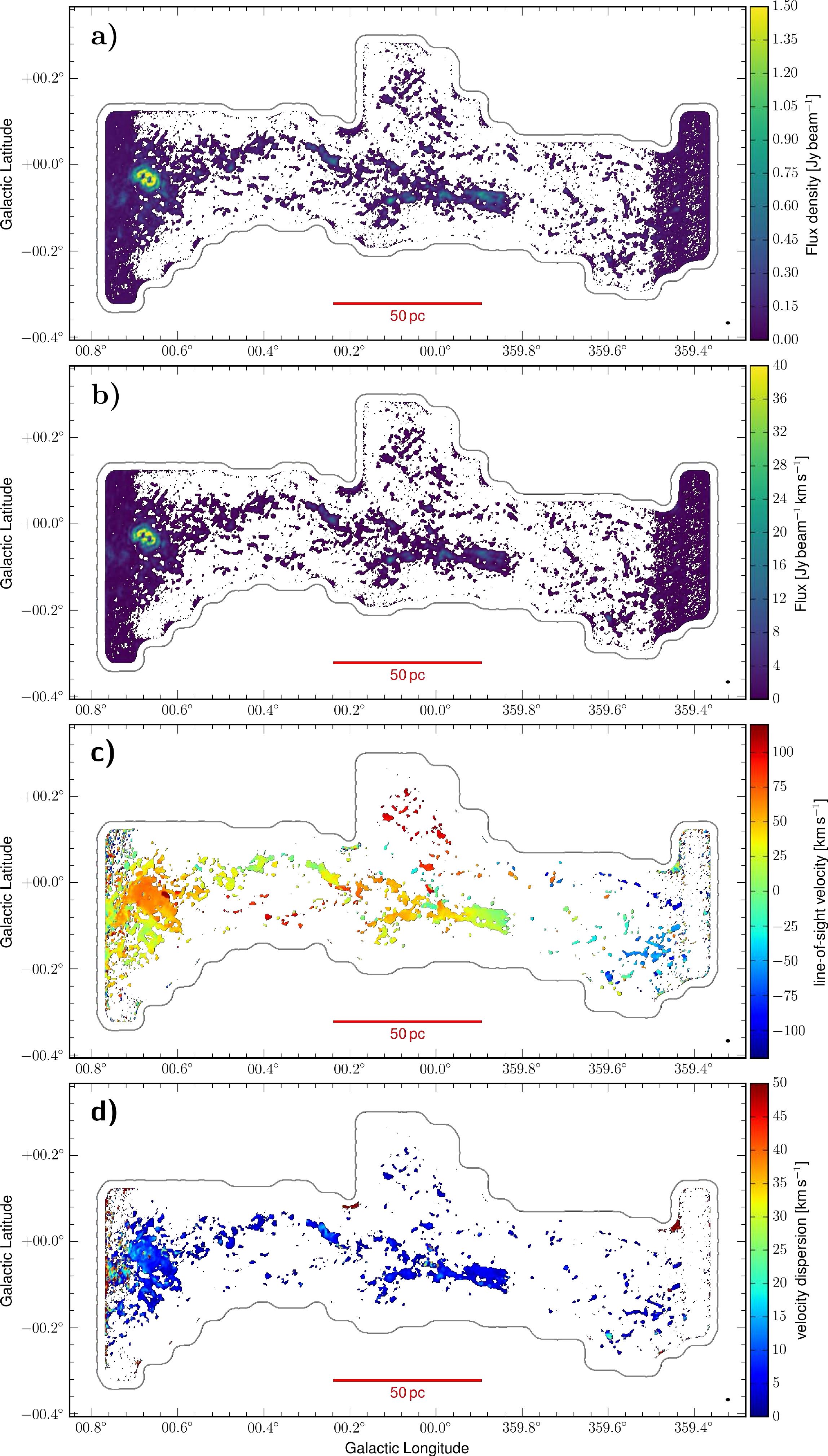}
	\caption{\nh344 moment maps: \emph{a}) peak intensity, \emph{b}) integrated intensity (moment 0), \emph{c}) intensity weighted mean velocity (moment 1), \emph{d}) intensity-weighted velocity dispersion (moment 2). The intensity maps a) and b) are masked at $3 \sigma$, velocity maps c) and d) are masked at $5 \sigma$ with an rms noise of $\sigma = 12.9$\,mJy. The beam of $25.7" \times 17.5"$ (1.04\,pc $\times$ 0.70\,pc) is indicated in the lower right corner of each panel.}
	\label{figure: nh344 moment map}
\end{figure}

\begin{figure}[H]
	\centering
	\includegraphics[height=0.9\textheight]{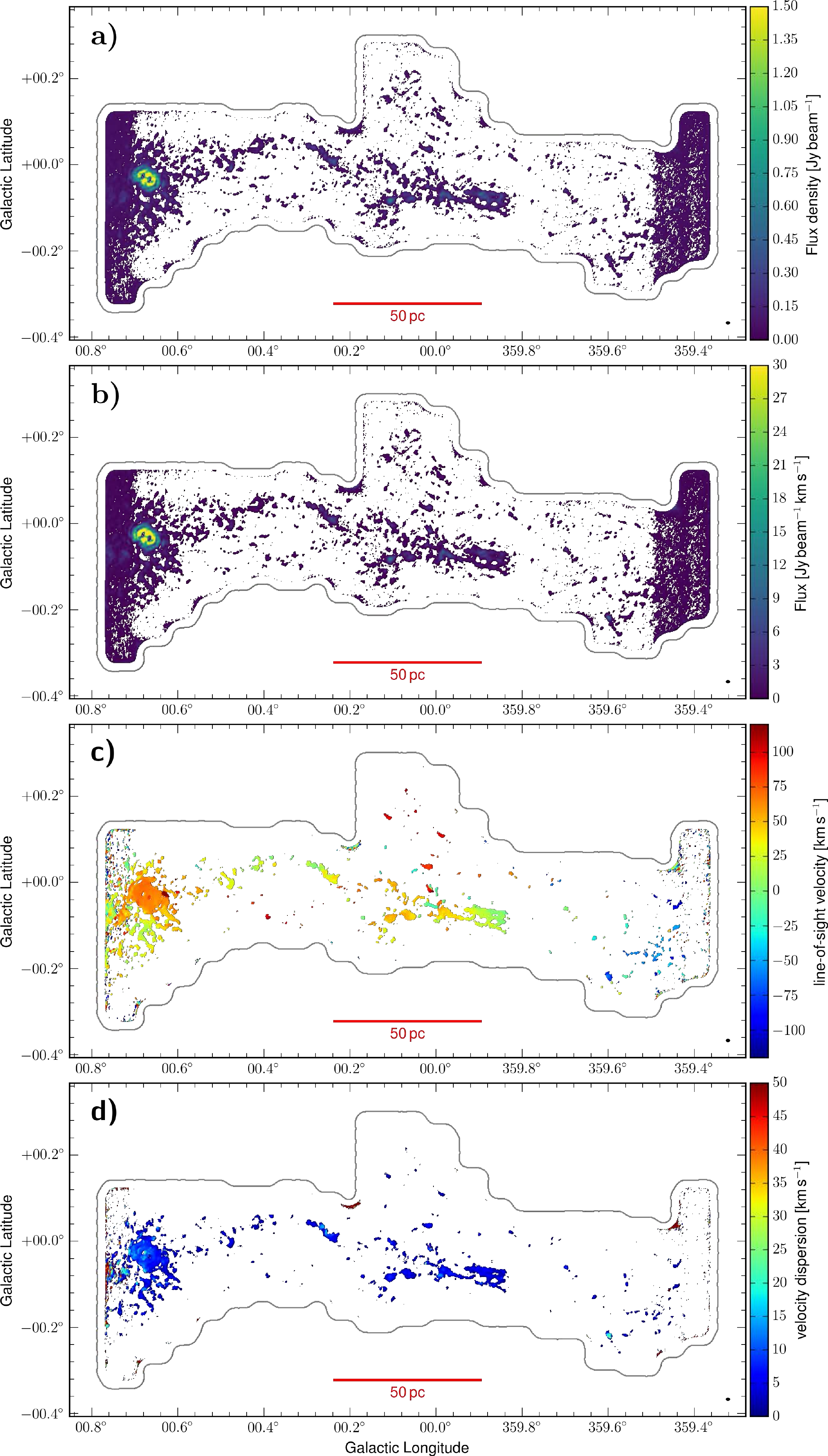}
	\caption{\nh355 moment maps: \emph{a}) peak intensity, \emph{b}) integrated intensity (moment 0), \emph{c}) intensity weighted mean velocity (moment 1), \emph{d}) intensity-weighted velocity dispersion (moment 2). The intensity maps a) and b) are masked at $3 \sigma$, velocity maps c) and d) are masked at $5 \sigma$ with an rms noise of $\sigma = 12.0$\,mJy. The beam of $25.3" \times 17.2"$ (1.02\,pc $\times$ 0.69\,pc) is indicated in the lower right corner of each panel.}
	\label{figure: nh355 moment map}
\end{figure}

\begin{figure}[H]
	\centering
	\includegraphics[height=0.9\textheight]{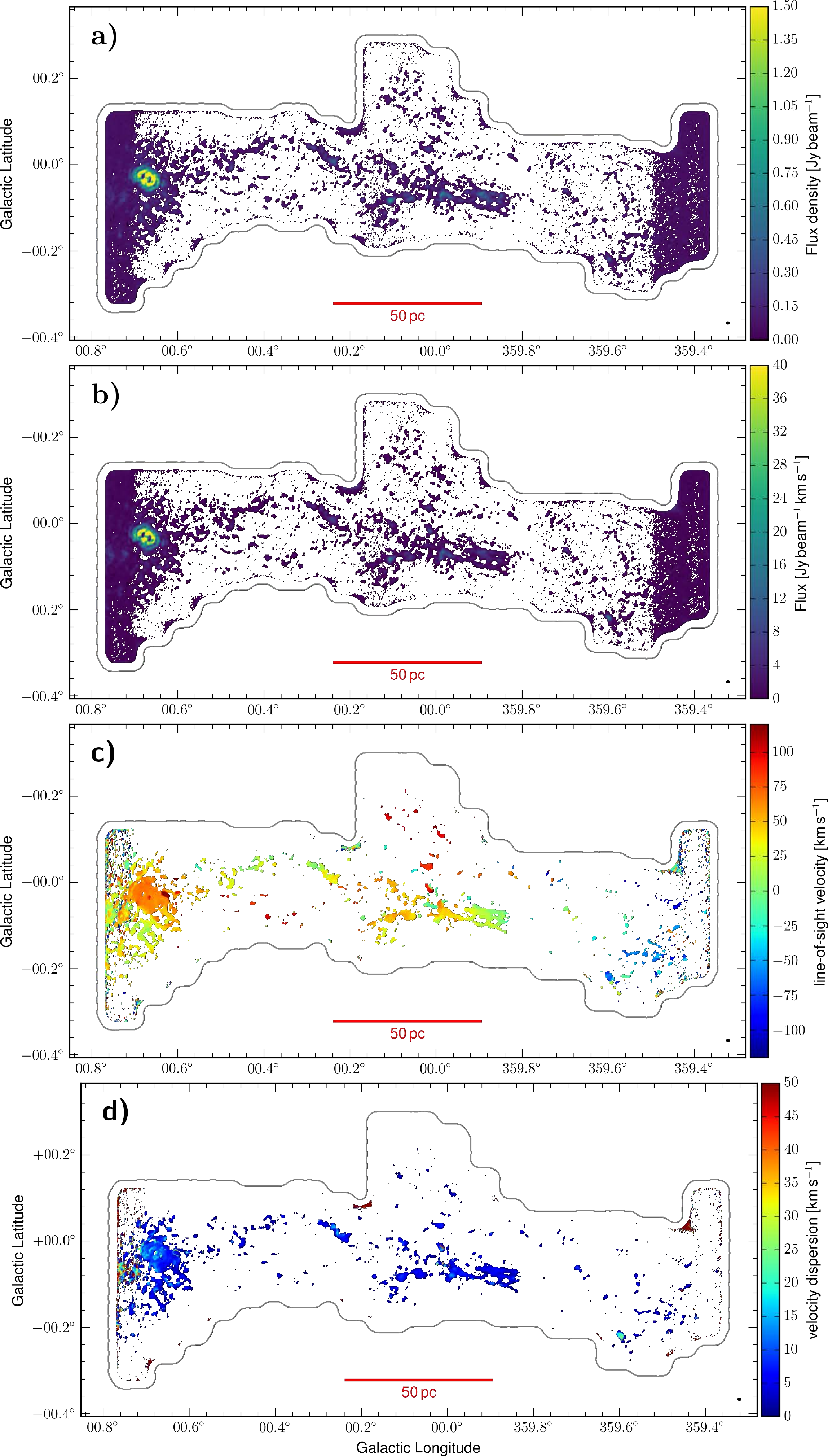}
	\caption{\nh366 moment maps: \emph{a}) peak intensity, \emph{b}) integrated intensity (moment 0), \emph{c}) intensity weighted mean velocity (moment 1), \emph{d}) intensity-weighted velocity dispersion (moment 2). The intensity maps a) and b) are masked at $3 \sigma$, velocity maps c) and d) are masked at $5 \sigma$ with an rms noise of $\sigma = 13.2$\,mJy. The beam of $24.8" \times 16.9"$ (1.00\,pc $\times$ 0.68\,pc) is indicated in the lower right corner of each panel.}
	\label{figure: nh366 moment map}
\end{figure}

\newpage
\section{Cleaning Depth in \texttt{MOSMEM}}\label{appendix: mosmem fails}

Version 1.6 of \texttt{mosmem} as of 2015 contained\footnote{As of August 2016 this bug is fixed and \texttt{mosmem} works as expected.} a bug that did not correctly scale the rms noise when emission is found which resulted in accumulation of the background flux density level and a non-constant baseline in spectra (M. Wieringa, private communication).
The strength of this effect depends on the amount of cleaned flux density and thus on the number of iterations.
Tests on line bearing and line-free channels of SWAG data showed that up to $\sim 50$ iterations no measurable effects occur while noticeable spectral slopes of tens of m\jybeam over the spectral range arise when $\gtrsim 100$ iterations are used and very high iteration counts of $250-500$ result in multiple sudden unpredictable jumps of $30-80$\,m\jybeam in the spectra.
These observations imply that the flux offsets are inherited from one channel to the next in the order of processing.
Channels containing large amounts of emission (e.g. ammonia lines) would need about $150-200$ iterations for \texttt{mosmem} to converge and thus still contain noticeable amounts of uncleaned sidelobe structure.
The applied restriction to 50 iterations in the generation of SWAG maps, therefore results in maps not affected by the software bug at the drawback of remaining not deconvolved flux.

\newpage
\section{The Ammonia Thermometer}\label{appendix: ammonia thermometer}

Ammonia is a good thermometer because of its inversion lines that cover rotational temperatures from close to absolute zero to thousands of Kelvin when the rotational quantum number J increases.
Its hyperfine structure allows to directly calculate opacities which are needed in order to derive correct temperatures without the need for a background source (on-off technique).
The simplest approach in deriving column densities and subsequently temperatures, is to assume optically thin emission, i.e. utilize moment 0 maps (Fig.~\ref{figure: nh333 moment map}).
As this technique cannot disentangle hyperfine satellite lines or multiple emission sources along the line-of-sight, it is problematic for ammonia and in the Galactic Center.
Therefore, we fit the ammonia spectra to resolve hyperfine lines and complex emission.
A comparison of both methods for temperature is given in §\ref{appendix: optical depth effects on temperature}.

\subsection{Rotational Temperature and Boltzmann Plot}\label{appendix: boltzmann plot}

The occupation of different rotational levels follows a Boltzmann distribution and as such the ratio of two column densities becomes a function of temperature \citep{Goldsmith1999}.
For a state of energy $E$ with excitation temperature $T_{ex}$ of the system, the probability of the system to be in that state is given by

\begin{equation}
p \propto \exp \frac{\mathrm{E}}{k_B \mathrm{T}_{ex}}
\end{equation}

where $k_B = 1.38 \times 10^{-23}$\,J/K is the Boltzmann constant.
For two metastable inversion lines with rotational numbers J = K, the relative probability is called Boltzmann factor and depends only the energy difference $\Delta \mathrm{E}$.
For rotational states, the temperature T$_{ex}$ is accordingly called rotational temperature T$_{\mathrm{JJ'}}$ for states J, J' after considering the correct conversion.
The probability of being in a certain state can be translated to the number of particles in that state which can be approximated by column densities because only the Boltzmann factor (relative ratio) is needed instead of absolute values.
Therefore, the ratio of column density N$_u$ in the upper rotational state $u$ with J or J' can be expressed as

\begin{equation}\label{equation: boltzmann factor}
\frac{\mathrm{N}'_u (\mathrm{J'},\mathrm{J'})}{\mathrm{N}_u (\mathrm{J},\mathrm{J})} = \frac{g(\mathrm{J'})}{g(\mathrm{J})} \frac{2\mathrm{J'}+1}{2\mathrm{J}+1} \exp \left( \frac{-\Delta \mathrm{E}}{\mathrm{T}_{\mathrm{JJ'}}} \right)
\end{equation}

with relative weight factors of $2\mathrm{J}+1$ and statistical weights $g$ that follow from quantum mechanical considerations as derived in \citet{Schilke1989}.
It must be taken into account that ammonia can exist in two different configurations of the hydrogen atoms nuclear spin, all spins parallel or one spin anti-parallel to the other two.
The configurations differ in energy and thus give rise to two ladders of ammonia states, ortho for rotational states $\mathrm{J} = 0, 3, 6, ...$ (all spins parallel, statistical weight $g = 2$) and para for $\mathrm{J} = 1, 2, 4, 5, ...$ (one spin anti-parallel, $g = 1$).
Eq. \ref{equation: boltzmann factor} can be rewritten as

\begin{equation}
\underbrace{\ln \left( \frac{\mathrm{N}'_u (\mathrm{J'},\mathrm{J'})}{g' (2\mathrm{J'}+1)} \right)}_{y'} - \underbrace{\ln \left( \frac{\mathrm{N}_u (\mathrm{J},\mathrm{J})}{g (2\mathrm{J}+1)} \right)}_{y} = - \underbrace{\frac{\Delta \mathrm{E}}{k_B \mathrm{T}_{\mathrm{rot, JJ'}}}}_{\Delta x / \mathrm{T}_{\mathrm{rot, JJ'}}}
\label{equation: Boltzmann plot}
\end{equation}

which is an equation of simple linear form for the inverse temperature with slope $m$.

\begin{equation}
m = \frac{\Delta y}{\Delta x} = \frac{y' - y}{\Delta x}
\label{equation: boltzmann slope}
\end{equation}

The corresponding plot is called Boltzmann plot, with energy above ground state on the x-axis and y-axis values derived from measured column densities.
For convenience, y is given in logarithms of base 10 instead of $e$ as

\begin{equation}
y = \log_{10} \left( \frac{\mathrm{N}(\mathrm{J},\mathrm{J})}{g(\mathrm{J}) (2\mathrm{J}+1)} \right)
\label{eqation: boltzmann y}
\end{equation}

which leads to a correction factor of $\log_{10}(e)$ when calculating temperatures.

\begin{equation}
\mathrm{T}_{\mathrm{rot, JJ'}} = \frac{-\log_{10}(e)}{m}
\label{equation: Trot}
\end{equation}

Overall, this means the rotational temperature of ammonia gas can be derived from the slope between two measurements of column density for the known state energies.

\subsection{Excitation temperature}

For radio frequencies and the expected ISM temperatures ($\mathrm{T}\ [K] \gtrsim 10\,\mathrm{K}$), the Rayleigh-Jeans approximation for black-body emission holds because $\exp \left( -\frac{h \nu}{k_B \mathrm{T}} \right) \approx 1-\frac{h \nu}{k_B \mathrm{T}}$.
With intensities expressed as line temperatures and considering the limited beam size of a telescope, radiative transfer can be expressed as

\begin{equation}
\mathrm{T}_L (\nu) = \eta_f \mathrm{T}_{ex} (1-e^{-\tau}) + \eta_f \mathrm{T}_{BG}\, e^{-\tau} + (1-\eta_f) \mathrm{T}_{BG}
\end{equation}

in the notation of \citet{Schilke1989}.
T$_L$ is the observed line temperature that results from source emission with excitation temperature T$_{ex}$ and background T$_{BG}$ that both passes through a medium with opacity $\tau$.
The beam filling factor $\eta_f$ denotes the fractional area of the beam filled by the source on the sky.
For completeness, it will be dragged along in further equations but set to unity in all calculations as this is expected to be approximately given for the beam size of SWAG ($\sim 0.9$\,pc) and the extended emission detected ($> 20"$).

Solving for T$_{ex}$ results in

\begin{equation}
\mathrm{T}_{\mathrm{rot, JJ'}} = \mathrm{T}_{ex} = \frac{\mathrm{T}_L}{f (1-e^{-\tau})}+\mathrm{T}_{BG}
\label{equation: excitation temperature}
\end{equation}

Background temperature T$_{BG}$ is set by the cosmic microwave background (CMB) at 2.7\,K.

\subsection{Column Density}

In order to calculate rotational temperatures via eq.~\ref{equation: Trot}, the column densities of two states needs to be known, which is given by

\begin{equation}
\mathrm{N} (\mathrm{J},\mathrm{K}) =  \frac{3h}{8 \pi^3} \frac{\mathrm{J} (\mathrm{J}+1)}{\mathrm{K}^2} \frac{1}{\mu^2} \frac{k_B}{h \nu_{\mathrm{JK}}} \int \mathrm{T}_{ex}\ \mathrm{d}v \label{equation: column density general}\\
\end{equation}

with Planck constant $h$, Boltzmann constant $k_B$, angular momentum J, projected angular momentum K, ammonia dipole moment $\mu$, frequency $\nu$ of the transition $\mathrm{J} \rightarrow \mathrm{J'}$ and excitation temperature $\mathrm{T}_{ex}$ integrated over velocity $\mathrm{d}v$.

The integral is solved differently when assuming optically thin emission or considering opacities.

\subsubsection{Column Density in the Optically Thin Limit}

Assuming optically thin emission, column densities can simply be derived from observed intensities with the appropriate scaling factors because every detected photon directly corresponds to one emitting molecule.
Following \citet{Ott2005,Ott2014a,Mills2013} whose work is based on \citet{Huettemeister1993,Huettemeister1995,Mauersberger2003}, the column density N in state (J,K) depends on electric dipole moment $\mu$ in Debye and line brightness temperature T$_{mb}$ as

\begin{eqnarray}
\mathrm{N} (\mathrm{J},\mathrm{K})\, \mathrm{[cm^{-2}]} &=& \ 1.6698 \cdot 10^{14} \cdot \frac{\mathrm{J}(\mathrm{J}+1)}{\mathrm{K}^2} \frac{1}{\mu^2\, \mathrm{[D^2]} \cdot \nu\, \mathrm{[GHz]}} \int_{line} \mathrm{T}_{mb}\ \mathrm{d}v\, \mathrm{[K\, km\,s^{-1}]}
\label{equation: column density thin general}\\
&=& \frac{7.77 \times 10^{13}}{\nu\, \mathrm{[GHz]}} \frac{\mathrm{J}(\mathrm{J}+1)}{\mathrm{K}^2} \int_{line} \mathrm{T}_{mb}\ \mathrm{d}v\, \mathrm{[K\, km\,s^{-1}]}
\label{equation: column density thin}
\end{eqnarray}

Eq. \ref{equation: column density thin general} lists the general formula for any symmetric top molecule with dipole moment $\mu$ whereas eq. \ref{equation: column density thin} already includes $\mu = 1.472$\,D for ammonia and shows the form given in the literature.

The data cube structure in discrete channels of width $\Delta v_{chan}$ simplifies the integration in velocity over the spectral line to a summation of spectral channels, which in turn is just the definition of moment 0 flux $\mathrm{F}_{mom0} = \int \mathrm{T}_{mb}\ \mathrm{d}v = \sum \mathrm{T}_{mb}\ \Delta v_{chan}$.

\begin{eqnarray}
\mathrm{N} (\mathrm{J},\mathrm{K})\, \mathrm{[cm^{-2}]} &=& \ 1.6698 \cdot 10^{14} \frac{\mathrm{J}(\mathrm{J}+1)}{\mathrm{K}^2} \cdot \frac{1}{\mu^2\, \mathrm{[D^2]} \cdot \nu\, \mathrm{[GHz]}} \cdot \sum_{line} \mathrm{T}_i\ \Delta v_{chan}\, \mathrm{[K\, km\,s^{-1}]}
\\[6pt]
&=& \frac{7.77 \times 10^{13}}{\nu\, \mathrm{[GHz]}} \frac{\mathrm{J}(\mathrm{J}+1)}{\mathrm{K}^2} \mathrm{F}_{mom0}\, \mathrm{[K\, km\,s^{-1}]}
\label{equation: column density thin moment}
\end{eqnarray}

\subsubsection{Column Densities Considering Opacity}

By solving the radiative transfer equation, it is possible to derive an expression for the column density without assuming high or low optical depth, because an exact solution for the relation of T$_{ex}$ ($\mathrm{T}_{\mathrm{JJ'}}$ for rotational states) and observed brightness temperature $\mathrm{T}_{mb}$ is known \citep{Schilke1989}.

The derivation starts again from the general approach \ref{equation: column density general} but then solves the integral over a spectral line of Gaussian shape (peak T$_L$ and FWHM $\Delta v_{int}$) with excitation temperature as in \ref{equation: excitation temperature}.
The integral over a single hyperfine component thus is given by

\begin{equation}
\mathrm{A} = \frac{\sqrt{\pi}}{2 \sqrt{ \ln 2 }}\cdot \mathrm{T}_L \cdot \Delta v_{int}
\end{equation}

and the  velocity integrated temperature becomes

\begin{eqnarray}
\int \mathrm{T}_{ex}\ dv &=& \frac{\sqrt{\pi}}{2 \sqrt{ \ln 2 }} \Delta v_{int} \frac{\mathrm{T}_{ex} \tau}{f_\mathrm{J}}\\
&=& \frac{\sqrt{\pi}}{2 \sqrt{ \ln 2 }} \Delta v_{int} \frac{\mathrm{T}_L \tau}{f f_\mathrm{J} \left( 1 - e^{- \tau} \right)}
\end{eqnarray}

T$_L$ denotes line peak intensity and must not be confused with integrated line intensity T$_{mb}$ used before.
$\Delta v_{int}$ stands for the fitted FWHM of a single hyperfine component.

Calculating all numerical factors and logically rearranging the parts to match eq. \ref{equation: column density thin general}-\ref{equation: column density thin moment} results in

\begin{eqnarray}
\mathrm{N} (\mathrm{J},\mathrm{K}) &=& \ 1.77745 \cdot 10^{14}\,\mathrm{cm}^{-2} \cdot \frac{\mathrm{J}(\mathrm{J}+1)}{\mathrm{K}^2} \frac{\Delta v_{int}\, \mathrm{[km\,s^{-1}]}}{(\mu\, \mathrm{[D])^2}\ \nu\, \mathrm{[GHz]}} \frac{\tau\ \mathrm{T}_L \,\mathrm{[K]}}{f_\mathrm{J} \left( 1-e^{-\tau} \right)}\\
\label{equation: column density opacity}
\end{eqnarray}

The pre-factor of $1.77745 \cdot 10^{14}$ applies when the beam filling factor is set to unity, line width $\Delta v_{int}$ is given in \kms, dipole moment $\mu$ in Debye, frequency of the inversion line in GHz and line peak intensity in Kelvin.
$f_\mathrm{J}$ is a factor to correct the fitted opacity $\tau$ for its line profile and is basically the ratio of flux in the central hyperfine component \citep[see][for details]{Schilke1989}.
Table~\ref{table: ammonia hyperfine components} lists $f_J$ for \nh311 to (6,6) for five hyperfine components per line whereof the central (``component 0'') enters eq.~\ref{equation: column density opacity}.

\begin{deluxetable}{ccccccc}
	\tablecaption{Ammonia hyperfine components as calculated from tabulated data in \citet{TownesSchawlow1975} and splatalogue.net. For each ammonia species, the top row lists spectral offset in \kms, the bottom row is fractional line strength.\label{table: ammonia hyperfine components}}
	\tablehead{line & \multicolumn{5}{c}{component} & \\
		& -2 & -1 & 0 & 1 & 2 & }
	\startdata
	\nh311 & -19.48 & -7.46 & 0 & 7.59 & 19.61 & [km\,s$^{-1}$]\\
	& 0.111 & 0.139 & 0.500 & 0.139 & 0.111 &\\
	\nh322 & -25.91 & -16.30 & 0 & 16.43 & 26.03 & [km\,s$^{-1}$]\\
	& 0.052 & 0.054 & 0.789 & 0.054 & 0.052 &\\
	\nh333 & -29.14 & -21.10 & 0 & 21.10 & 29.14 & [km\,s$^{-1}$]\\
	& 0.027 & 0.029 & 0.888 & 0.029 & 0.027 &\\
	\nh344 & -30.43 & -24.22 & 0 & 24.22 & 30.43 & [km\,s$^{-1}$]\\
	& 0.016 & 0.016 & 0.935 & 0.016 & 0.016 & \\
	\nh355 & -31.41 & -25.91 & 0 & 25.91 & 31.41 & [km\,s$^{-1}$]\\
	& 0.011 & 0.011 & 0.956 & 0.011 & 0.011 & \\
	\nh366 & -31.47 & -26.92 & 0 & 26.92 & 31.47 & [km\,s$^{-1}$]\\
	& 0.008 & 0.008 & 0.969 & 0.008 & 0.008 & \\
	\enddata
\end{deluxetable}

\begin{figure}
	\centering
	\includegraphics[width=\linewidth]{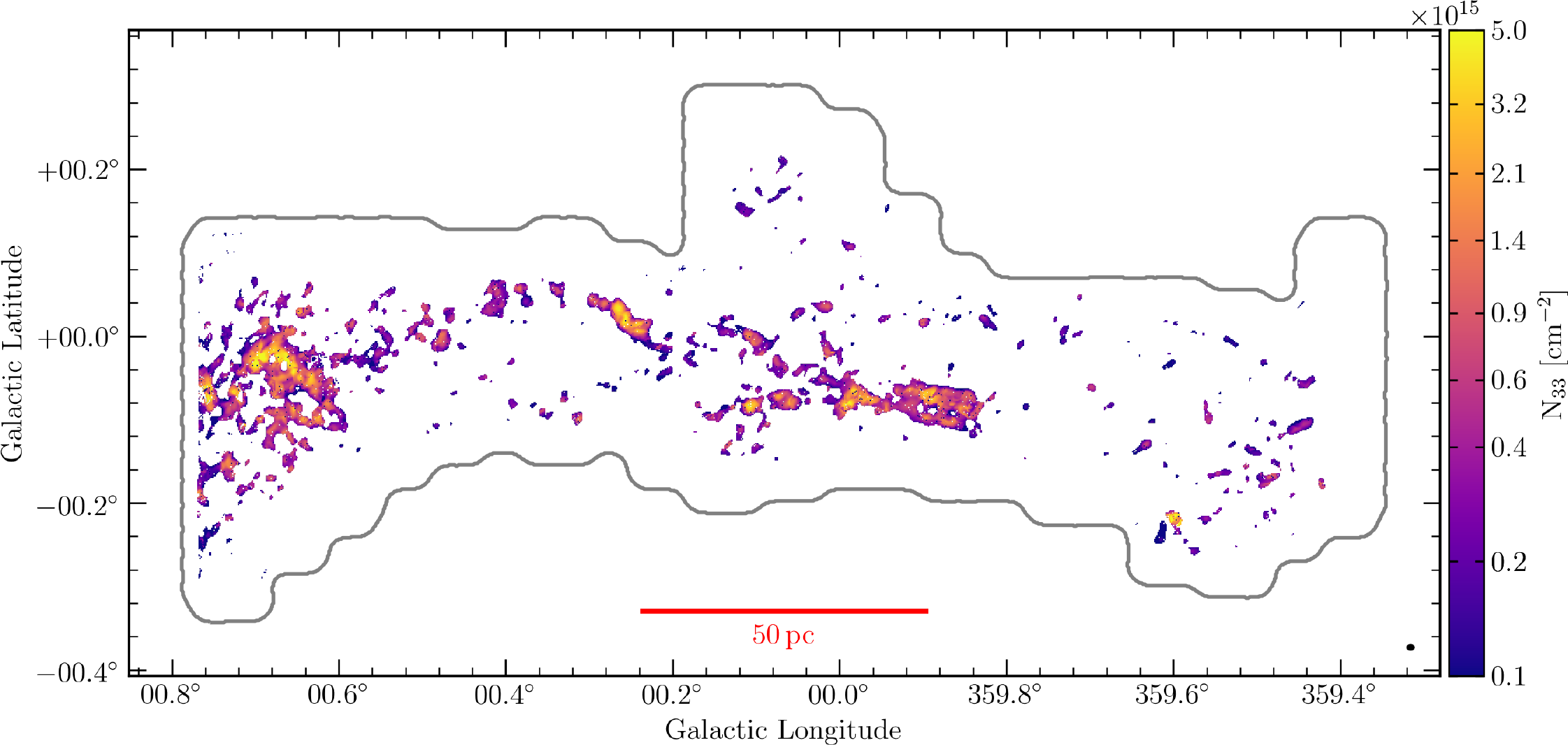}
	\caption{\nh333 column density N$_l$ under consideration of non-zero opacity as derived by eq. \ref{equation: column density opacity} from fitting the hyperfine structure.}
	\label{figure: nh333 column density}
\end{figure}

Fig.~\ref{figure: nh333 column density} shows the column density distribution of the upper state $\mathrm{N}_u$ of \nh333 as an example.

\subsubsection{Total Ammonia Column Density}

Note that Eqs.~\ref{equation: column density opacity} \& \ref{equation: column density thin} and Fig.~\ref{figure: nh333 column density} give the column density of ammonia in the lower transition state only.
The total column density of the state (J,K) must also take the molecules in the upper transition state into account.
In thermal equilibrium, the column densities of both states are related by a Boltzmann factor

\begin{eqnarray}
\mathrm{N (J,K)} &=&\mathrm{N}_l \mathrm{(J,K)} \left( 1 + \exp \left( -\frac{h \nu}{k_B\mathrm{T}} \right) \right)\\
&\simeq& 2 \times \mathrm{N}_l \mathrm{(J,K)} \qquad\qquad\qquad \mathrm{for\ T}\gg \frac{h\nu}{k_B}
\end{eqnarray}

In the ground state (J=0), the inversion states are degenerate and no transition can be observed which means the column density N$_{00}$ can not be derived directly but needs to be extrapolated from low metastable states.
As the slope in a typical Boltzmann plot (excitation state-scaled column density vs. excitation temperature) of ammonia emission of the Galactic Center starts to flatten around J = 4 - 6, the best estimate for N$_{00}$ is derived from the slope between measurements at J=1 and J=2, i.e. rotational temperature T$_{12}$.
If the ortho/para ammonia abundance ratio is not known, the most simple approach is to assume occupation by thermal equilibrium of states J=0, 1, 2 at temperature \temp12, hence extrapolate N$_{00}$ according to eq. \ref{equation: n00 extrapolation} \citep{Ungerechts1986}.
The factor 23.2\,K is given by the energy difference of \nh300 and \nh311.

\begin{eqnarray}
\mathrm{N}_{00} &=& \frac{1}{3} \exp\left(\frac{23.2\,\mathrm{K}}{\mathrm{T}_{kin,12}}\right) \mathrm{N}_{11}\label{equation: n00 extrapolation}
\end{eqnarray}

The ``total'' column density of ammonia is then derived by summing up the six directly measured column densities ($\mathrm{N} = \mathrm{N_l} + \mathrm{N}_u$) for states J = 1, ..., 6 and the extrapolated J = 0 column density.
In thermal equilibrium, higher states are increasingly less populated and can thus be approximated to  contribute negligibly to the total column density for low temperatures.
Eq.~\ref{equation: total column density} utilizes these assumptions, i.e. does not contain the contribution of ammonia in rotation states higher than J = 6 and consequently must be understood as a lower limit.
How close this limit is to the real ammonia column density can roughly be estimated by the ratio $\mathrm{N}_{00} / \mathrm{N}_{66}$ which, in SWAG, typically exceeds 10 with only few very hot clouds like Sgr~B2 containing relevant amounts of ammonia in higher excitation states as was also found by \citet{Mills2013}.
The assumption of thermal equilibrium which is required to extrapolate the total column density does not hold in all CMZ clouds whereas the lower limit is supported by direct measurements and thus more robust.

The total column density of ammonia (defined to include emission up to J=6) is thus given by
\begin{eqnarray}
\mathrm{N}_{tot} &=& \left[\frac{1}{3} \exp \left( \frac{23.2\,\mathrm{K}}{\mathrm{T}_{12,kin}}\right) +1 \right] \mathrm{N}_{11} + \mathrm{N}_{22} + \mathrm{N}_{33} + \mathrm{N}_{44} + \mathrm{N}_{55} + \mathrm{N}_{66}\label{equation: total column density}
\end{eqnarray}

\subsection{Rotational to Kinetic Temperature Conversion}\label{section: Trot Tkin conversion}

The rotational temperature is always lower than the kinetic temperature with negligible difference at low values. For increasing temperature, the deviation also increases and must be taken into account.
Reasons for this behavior are the increasing number of energetically available non-metastable levels for depopulation of higher metastable states and radiative decay \citep{WalmsleyUngerechts1983}.
LVG radiative transfer models allow to derive approximations to the conversion of rotational to kinetic temperature.
\citet{Ott2011} calculated LVG grids for temperatures up to 500\,K for metastable ammonia species up to (6,6) and derived exponential functions for the rotational-to-kinetic temperature conversion.
These fits agree with the modeled curves within less than 5\% for the temperature measures calculated in this work which are \temp12, \temp24, \temp45 and \temp36 denoting the kinetic temperature \temp{i}{j} derived from \nh3ii and \nh3jj.
Table~\ref{table: Trot Tkin} lists the conversion functions as partially published in \citet{Ott2011} and \citet{Gorski2017}.
Towards high kinetic temperatures, the conversion flattens out which decreases the ability to correctly discern temperatures.
Especially the temperature measurements of low ammonia states are affected, effectively setting an upper limit above which the tracer becomes unreliable as even small changes in T$_{rot}$ by random errors translate to substantial changes in T$_{kin}$.
A cut-off is thus applied to all rotational temperatures before conversion to exclude high values that cannot be reliably translated to kinetic temperatures.

With the $\mathrm{T}_{kin} - \mathrm{T}_{rot}$ conversion in hand, all further mentions of temperature are kinetic temperatures labeled T$_{ij}$ if not explicitly stated otherwise, e.g. by the subscript T$_{rot,ij}$.

\begin{deluxetable}{LCLRCL}
	\tablecaption{Conversion equations for the rotational to kinetic temperature conversion as obtained by \citet{Ott2011} and partially published in \citet{Ott2011} and \citet{Gorski2017}.\label{table: Trot Tkin}}
	\tablehead{\multicolumn{3}{c}{T$_{kin}$ - T$_{rot}$ conversion} & \multicolumn{3}{c}{temperature range}}
	\startdata
	\mathrm{T}_{kin, 12} &=& 6.05 \cdot \exp(0.06088\, \mathrm{T}_{rot, 12})\\[6pt]
	\mathrm{T}_{kin, 24} &=& 1.467\, \mathrm{T}_{rot, 24} - 6.984 & 0\,\mathrm{K} \lesssim &\mathrm{T}_{kin}& \lesssim 100\,\mathrm{K}\\
	&=& 27.085 \cdot \exp(0.019\, \mathrm{T}_{rot, 24}) & 100\,\mathrm{K} \lesssim &\mathrm{T}_{kin}& \lesssim 500\,\mathrm{K}\\[6pt]
	\mathrm{T}_{kin, 36} &=& \mathrm{T}_{rot, 36} & 0\,\mathrm{K} \lesssim &\mathrm{T}_{kin}& \lesssim 50\,\mathrm{K}\\
	&=& 28.87511 \cdot \exp(0.015\, \mathrm{T}_{rot, 36}) & &\mathrm{T}_{kin}& > 50\,\mathrm{K}\\[6pt]
	\mathrm{T}_{kin, 45} &=& 1.143\, \mathrm{T}_{rot, 45} - 1.611 & 0\,\mathrm{K} \lesssim &\mathrm{T}_{kin}& \lesssim 50\,\mathrm{K}\\
	&=& 21.024 \cdot \exp(0.0198\, \mathrm{T}_{rot, 45}) & 50\,\mathrm{K} \lesssim &\mathrm{T}_{kin}& \lesssim 500\,\mathrm{K}\\
	\enddata
\end{deluxetable}

\subsubsection{Optical depth effects on temperature}\label{appendix: optical depth effects on temperature}

\begin{figure}
	\centering
	\includegraphics[width=\linewidth]{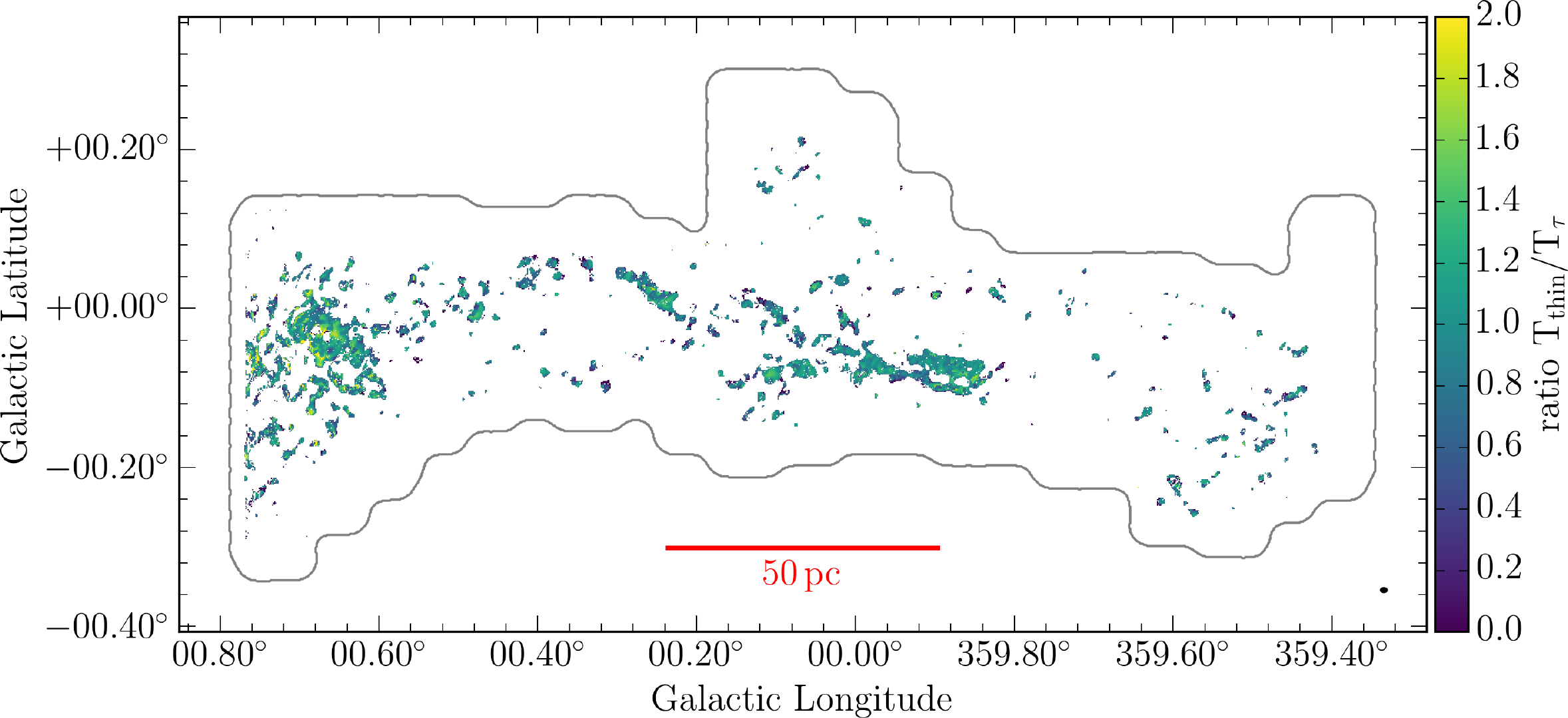}
	\caption{Ratio map $\mathrm{T}_{thin} / \mathrm{T}_\tau$ of ammonia \temp24 temperature assuming optically thin emission (T$_{thin}$) and considering opacity effects (T$_\tau$). The differences are minor (blue-green) in large areas of the map but the temperature can be overestimated (yellow) in dense regions by factors of $\sim 2.0$. Temperatures in less dense clouds and at the edges of fitted area are underestimated (dark blue).}
	\label{figure: T24 kin thin tau ratio}
\end{figure}

\begin{figure}
	\centering
	\includegraphics[width=0.5\linewidth]{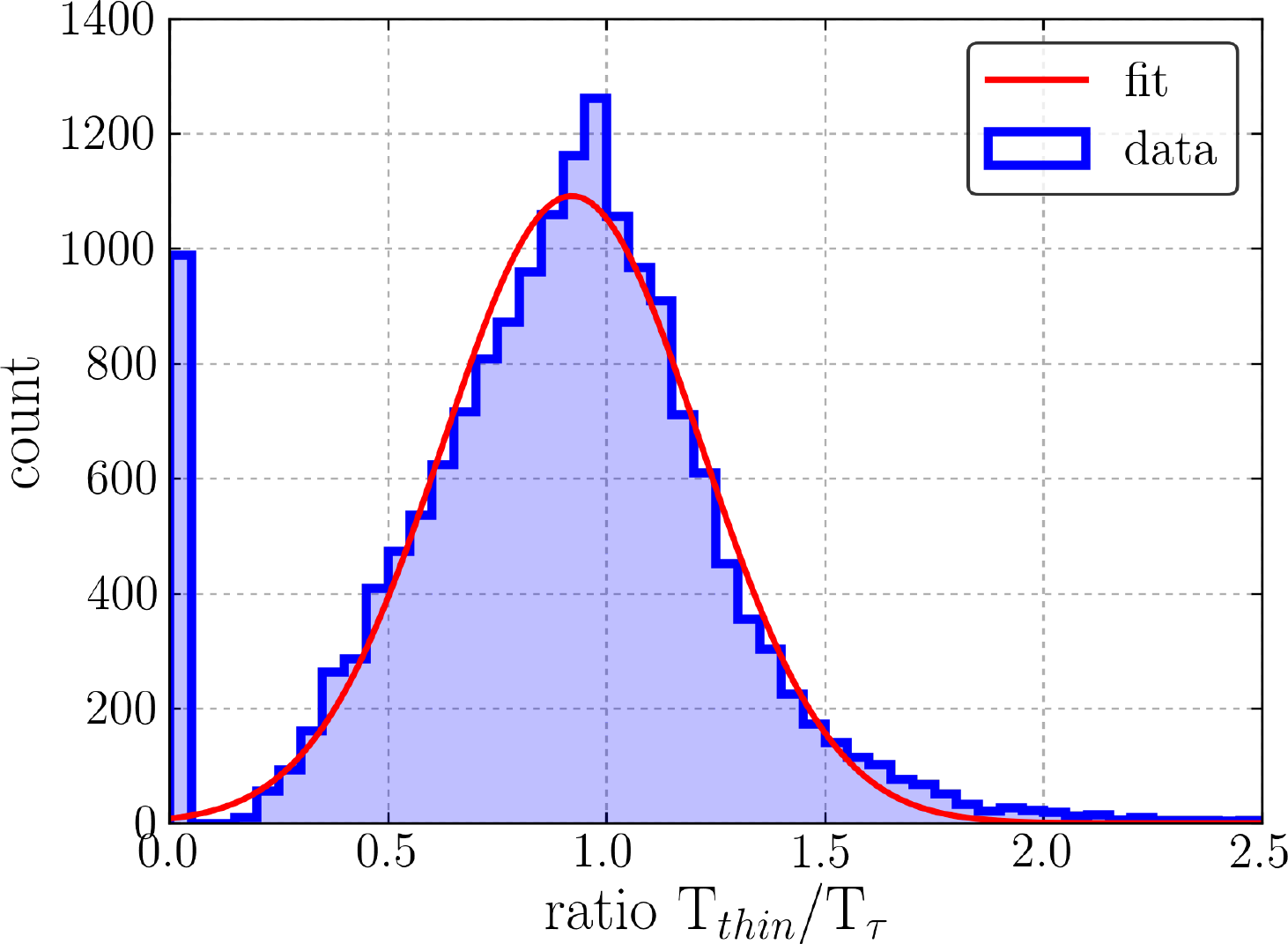}
	\caption{Histogram of the temperature ratio $\mathrm{T}_{thin} / \mathrm{T}_\tau$ between kinetic ammonia temperature \temp24 derived assuming optically thin emission and considering measured opacities (Fig.~\ref{figure: T24 kin thin tau ratio}). For $\sim 1000$ pixels at a ratio of 0.0, the hyperfine structure could be fitted and an opacity corrected temperature derived although the emission is faint enough to be masked while calculating image moment 0 on which the temperature under thin assumption is based. The inverse case (ratios of $+ \infty$) occurs more often but does not show up on this plot. A Gaussian fit generally describes the distribution well but does not fit an enhanced amount of temperature ratios of 1.0 and $1.5 - 2.0$. The latter is the effect of considering opacity for optically thick emission. As lower ammonia states are typically more occupied (falling slope in Boltzmann plot) and the opacity correction increases occupation over the thin approximation (N$_\tau > $N$_{thin}$), temperature decreases (T$_\tau < \mathrm{T}_{thin}$) and thus $\mathrm{T}_{thin} / \mathrm{T}_\tau > 1.0$.}
	\label{figure: T24 kin thin tau ratio histogram}
\end{figure}

As described in appendix~\ref{appendix: ammonia thermometer}, temperatures can also be calculated assuming optically thin emission as is often done (in extra-galactic context) with the advantage of being able to derive temperatures over a larger area.
From the fitted opacities (Fig.~\ref{figure: nh333 opacity}), it is already known that this assumption is not satisfied in a significant portion of the map.
The practical differences for derived temperatures are shown in Fig.~\ref{figure: T24 kin thin tau ratio} which plots the ratio $\mathrm{T}_{thin} / \mathrm{T}_{\tau}$ of kinetic temperature T$_{24}$ derived under the assumption of optically thin emission and considering fitted opacities, respectively.
The differences are minor in large areas of the map but temperature can be overestimated by $\mathrm{T}_{thin}$ in dense regions by factors of up to 1.6 in Sgr B2.
Temperatures in less dense clouds and at the edges of the fitted area are typically underestimated by factors of $\lesssim 2$.
This can also be seen in a histogram of the map (Fig.~\ref{figure: T24 kin thin tau ratio histogram}) as an increase of ratios > 1.0 over what is expected from statistical, normally distributed errors (Gaussian fit).
In an optically thick molecular cloud, not all emission can be detected due to self-absorption and scattering as would be in an optically thin cloud.
The derived column densities thus increase when opacity is considered.
Temperature is derived from the inverse slope of column density as a function of molecular energy level which results in lower temperatures for two conditions: If opacity is considered in calculating the column density and if opacity decreases for higher states which typically is the case.
Given that opacity typically decreases for higher states, the effect on column density is stronger for lower J transitions.
As temperature is derived from the inverse slope of column density as a function of molecular energy level, considering opacity results in lower temperatures.

\newpage
\section{Details on Hyperfine Structure Fitting in CLASS}\label{appendix: hyperfine fitting}

\begin{figure}[H]
	\centering
	\includegraphics[width=\linewidth]{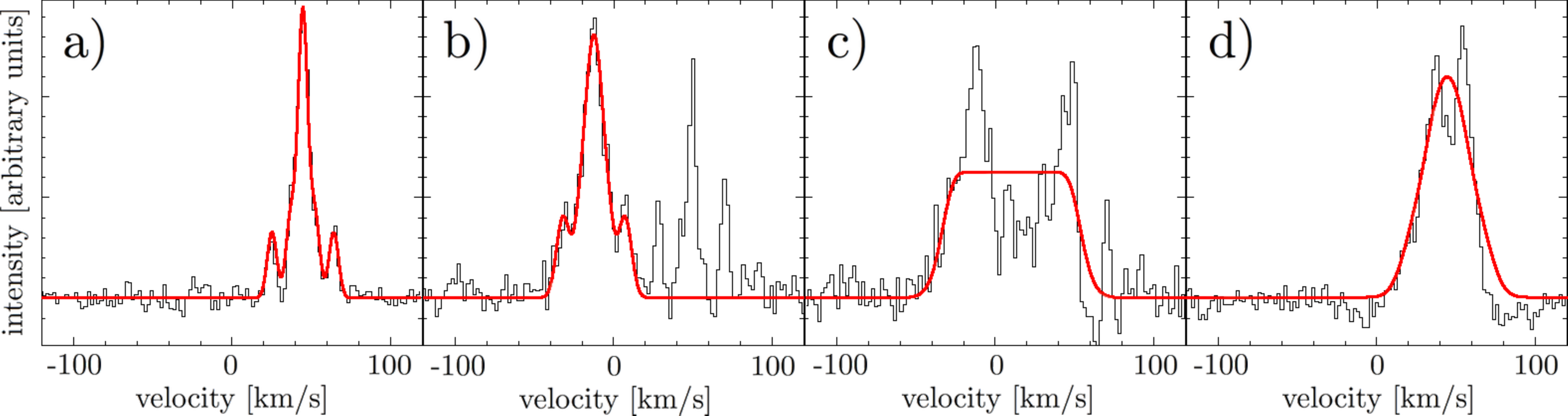}
	\caption{Examples of ammonia hyperfine structure per pixel fits (red) to observed SWAG spectra (black). Satellite components blend with the main line due to large line width (a, b, c) or blending of multiple line along the line-of-sight (c, d).  \textit{a)}: Typical good fit of a single emission component along the line-of-sight. \textit{b)}: For multiple components, the strongest is usually fitted well. \textit{c)}: Occasionally, two separated components of similar peak intensity may be fitted by an unphysically large line width. \textit{d)}: If two (or more) components of similar intensity blend, they might be fitted as a single component with large line width. The latter two cases are excluded from further analyses by the large line width (c, $\sim 80$\,\kms) and bad fit quality as given by reduced $\chi^2$ (d).\label{figure: good bad fit}}
\end{figure}

\begin{deluxetable}{lcccccc}
	\tablecaption{Fit success rates for the six ammonia inversion lines observed with SWAG. 24383 pixels were selected by a two-step SNR criterion to be fitable in at least one of the lines \nh311 to \nh366. The actual number of attempted fits per ammonia specie (``fitted pixels'', second row) is lower due to non-identical distributions and decreasing SNR for higher states. The difference is given in rows 3 and 4 (``failed to fit'' and success percentage) and lists the amount of pixels that could not be fitted due to exclusion by masking. Out of the attempted fits, the vast majority of $> 98.5\%$ did converge and yielded physical plausible results (``good fits'') ``Bad'' fits are defined by large uncertainty and physically implausible values (see §\ref{section: hyperfine fitting CLASS} for details).\label{table: fit success rates}}
	\tablehead{ & (1,1) & (2,2) & (3,3) & (4,4) & (5,5) & (6,6)}
	\startdata
	selected pixels & 24383 & 24383 & 24383 & 24383 & 24383 & 24383\\
	fitted pixels & 23790 & 23025 & 23646 & 19269 & 19261 & 19245\\
	failed to fit & 593 & 1358 & 917 & 5114 & 5122 & 5138\\
	\% fitted & 97.6 & 94.4 & 96.2 & 79.0 & 79.0 & 78.9\\[1mm]
	good fits & 23437 & 22822 & 23403 & 19137 & 18795 & 18971\\	
	bad fits & 353 & 203 & 63 & 132 & 466 & 274\\
	\% good & 98.5 & 99.1 & 99.7 & 99.3 & 97.6 & 98.6\\
	\enddata
\end{deluxetable}

\newpage
\section{Further Temperature Maps}\label{appendix: temperature maps}

\begin{figure}[H]
	\centering
	\includegraphics[width=\linewidth]{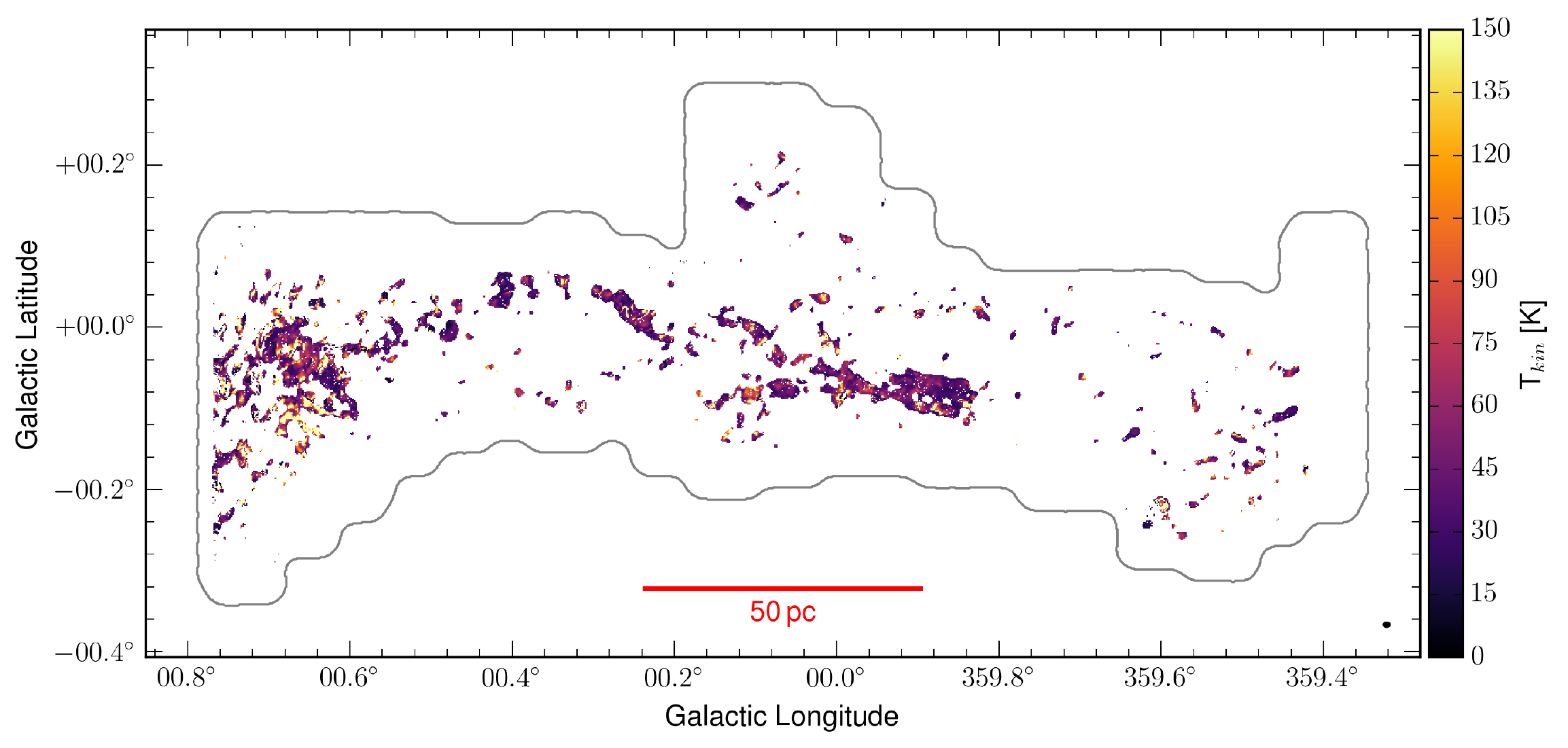}
	\caption{\temp12}
	\label{figure: T12 kin map}
\end{figure}

\begin{figure}[H]
	\centering
	\includegraphics[width=\linewidth]{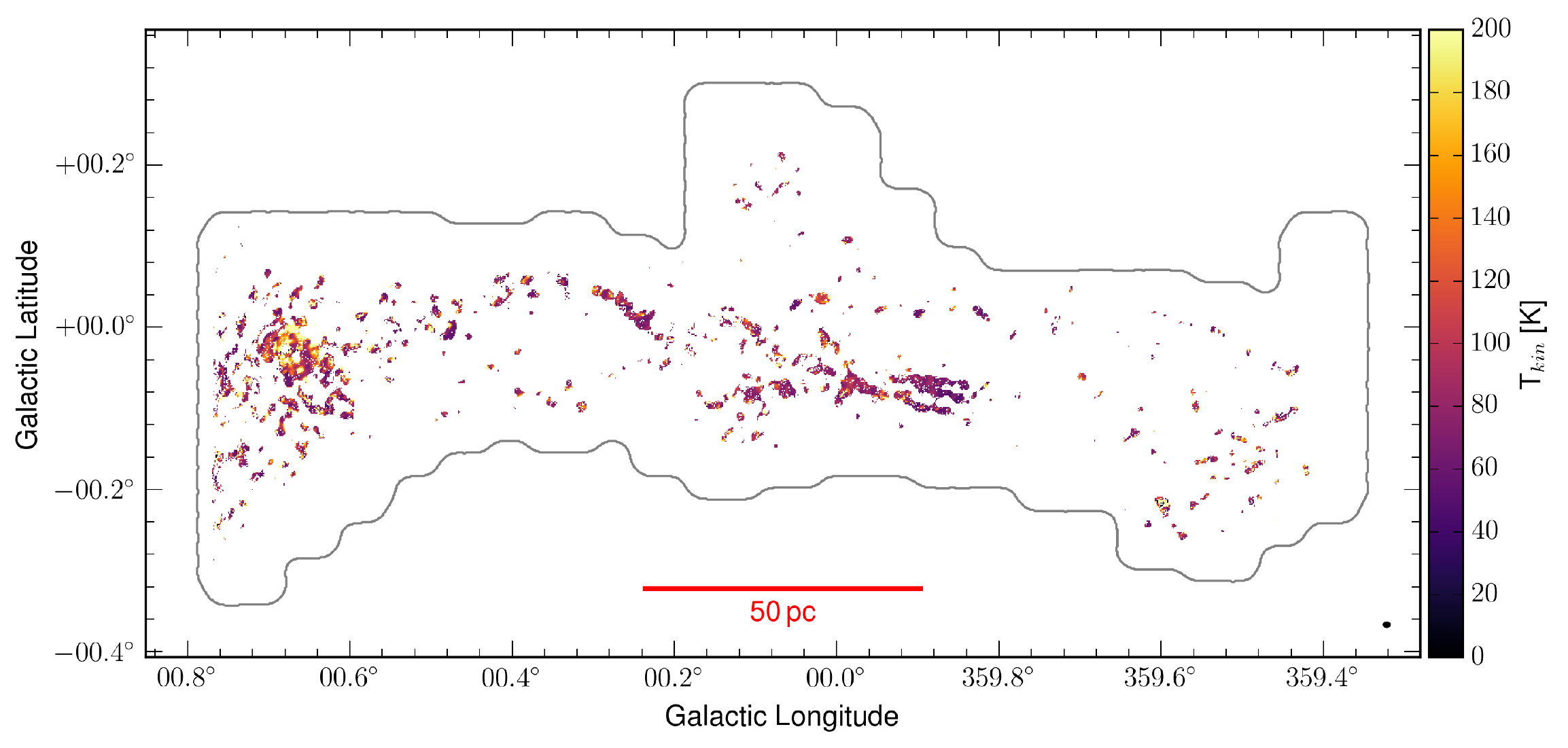}
	\caption{\temp45}
	\label{figure: T45 kin map}
\end{figure}

\begin{figure}[H]
	\centering
	\includegraphics[width=\linewidth]{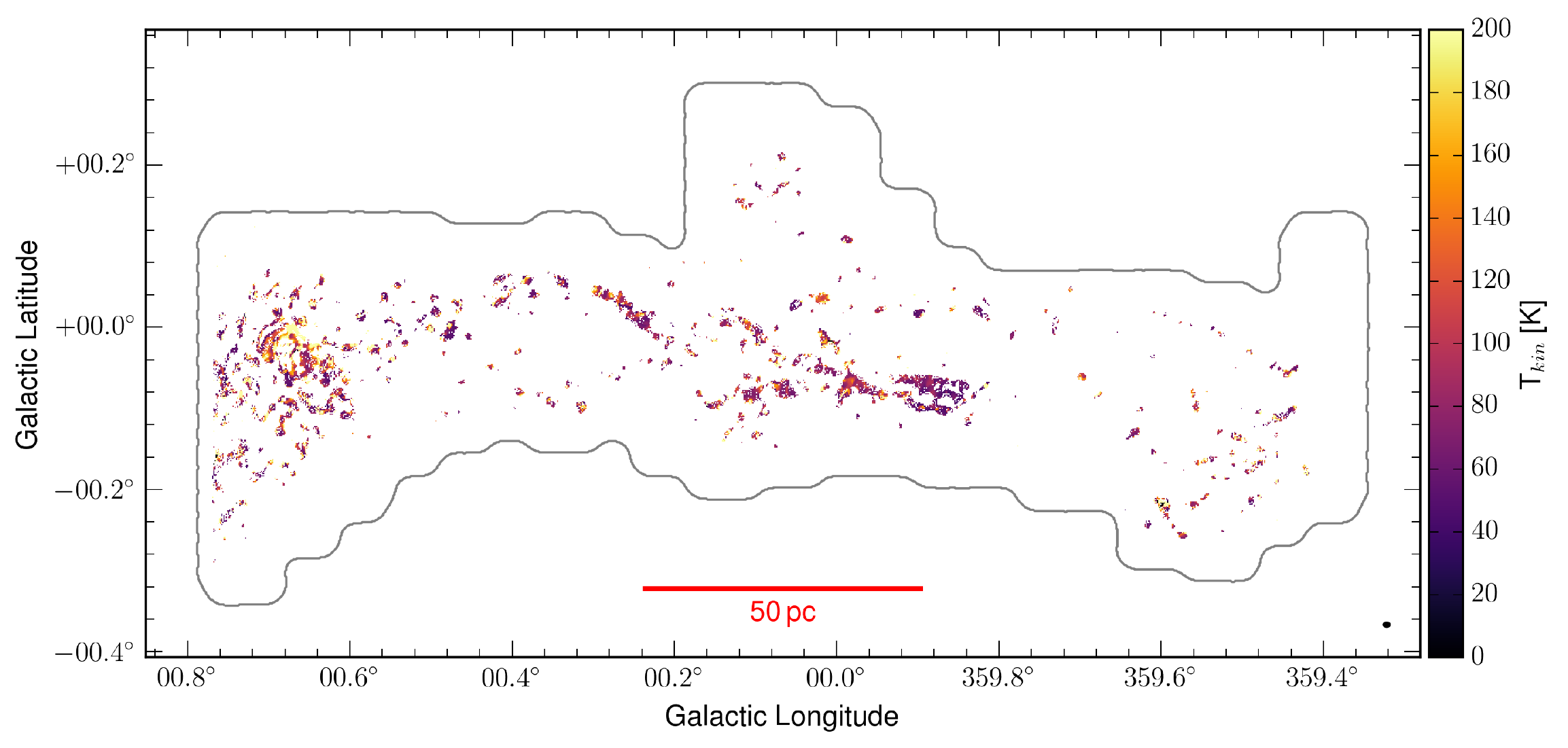}
	\caption{\temp36}
	\label{figure: T36 kin map}
\end{figure}

\newpage
\section{Further Temperature Comparisons}\label{appendix: further temperature comparisons}

\begin{figure}[H]
	\centering
	\includegraphics[height=0.65\textheight]{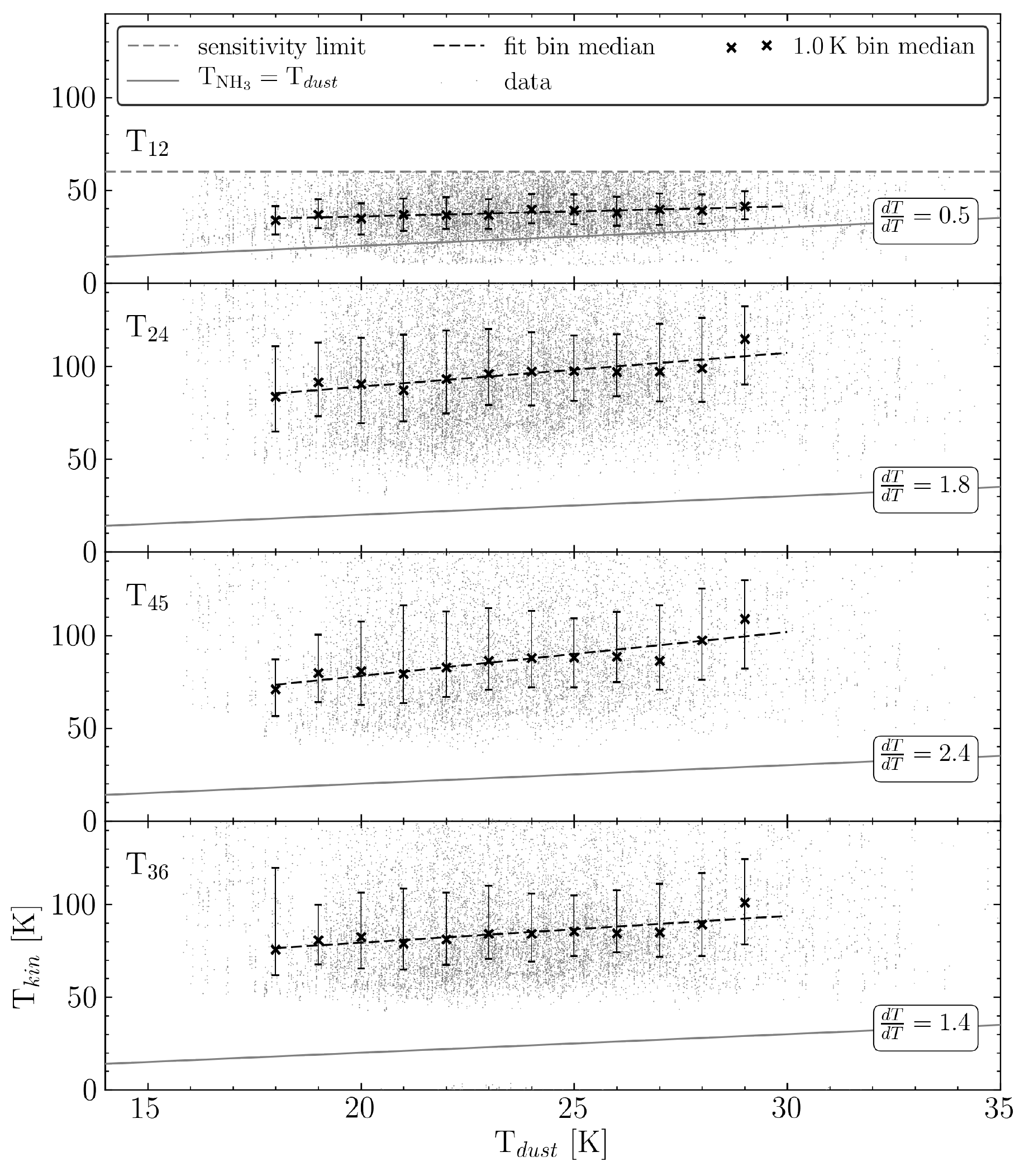}
	\caption{Correlation of Hi-Gal dust temperature \citep{Molinari2011} and kinetic ammonia gas temperature as in Fig.~\ref{figure: Tdust vs T24}. The temperature estimate is labeled in the top left corner of each panel. Panels are zoomed to display the same range to allow direct comparison.}
	\label{figure: Tdust vs TNH3}
\end{figure}

\newpage
\section{Further Sequence Plots}\label{appendix: further sequence plots}

\begin{figure}[H]
	\centering
	\includegraphics[height=0.85\textheight]{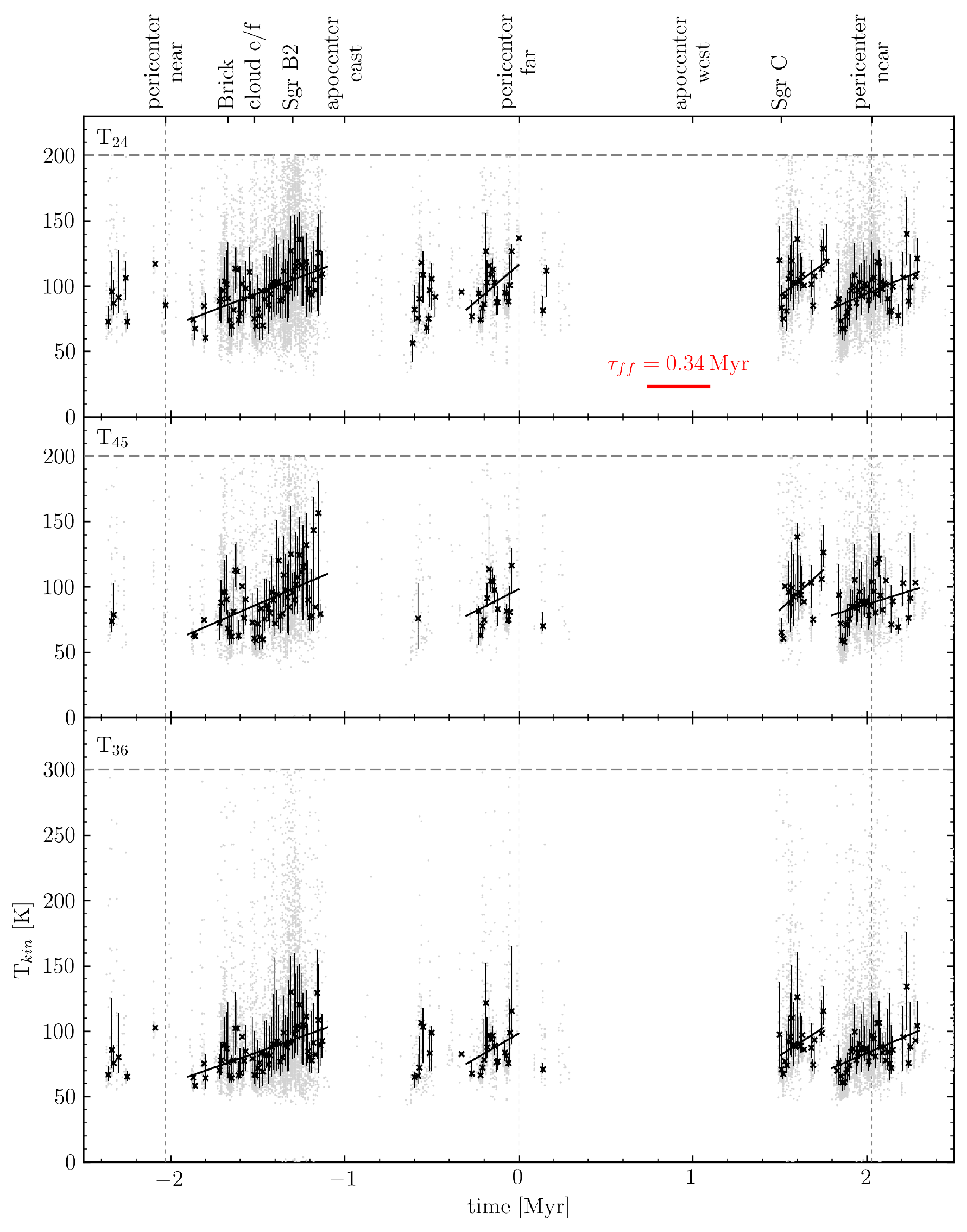}
	\caption{Kinetic ammonia temperature as a function of time since far side pericenter passage (bottom x-axis, top x-axis shows important dynamical points in the K15 model and massive molecular clouds). The temperature tracer is given in the top left corner of each panel. A horizontal dashed grey line denotes the sensitivity limit of the conversion from rotational to kinetic temperature. Vertical dashed lines highlight the occurrence of pericenter passages. Measurements for individual pixels (light grey points) are overlaid with black crosses denoting medians in bins of 0.01\,Myr and error intervals that include 50\% of the data in each bin.}
	\label{figure: time vs temperature}
\end{figure}

\begin{figure}[H]
	\centering
	\includegraphics[height=0.9\textheight]{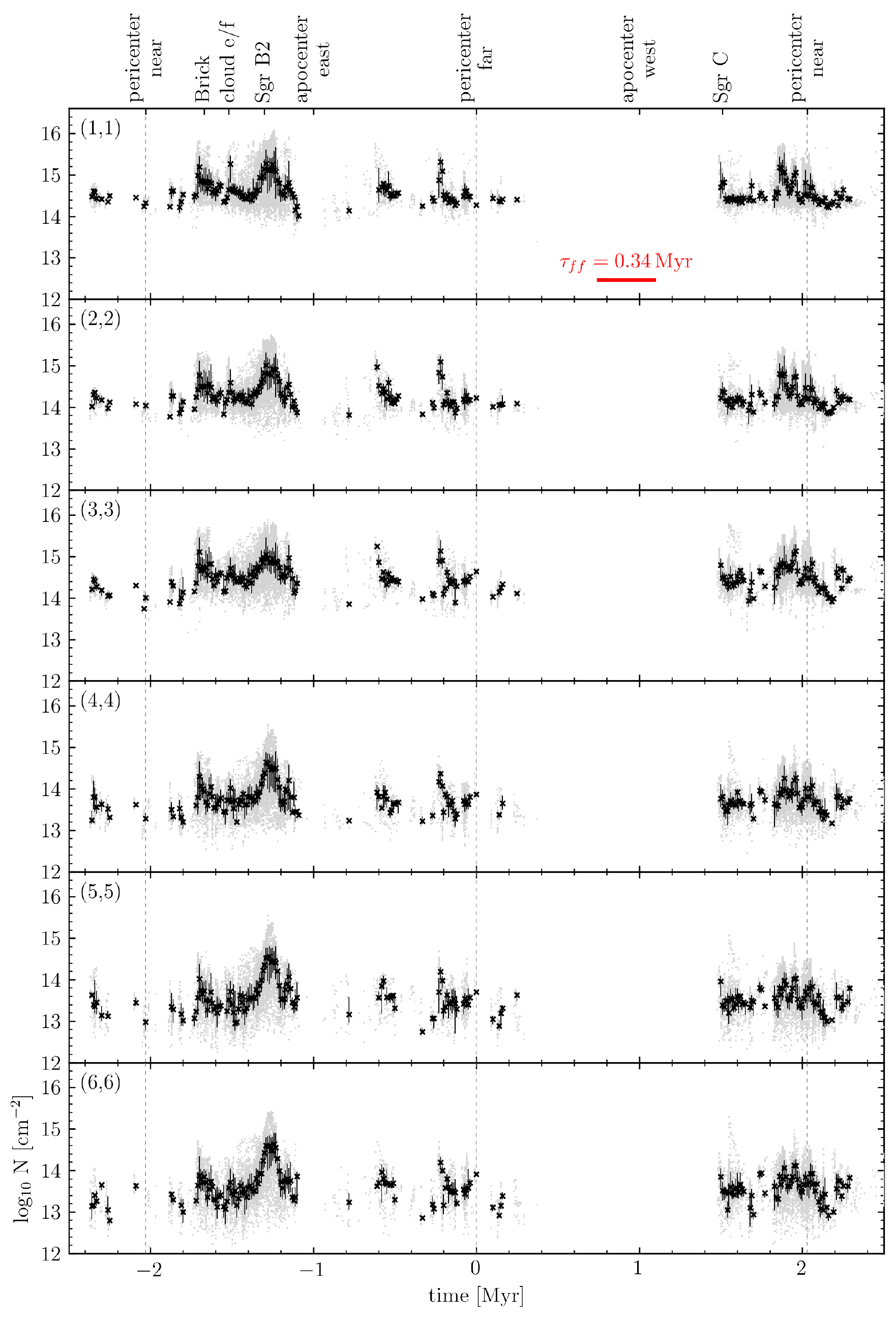}
	\caption{Ammonia column density as a function of time since far side pericenter passage (bottom x-axis, top x-axis shows important dynamical points in the K15 model and massive molecular clouds). The ammonia species is given in the top left corner of each panel. Vertical dashed lines highlight the occurrence of pericenter passages. Measurements for individual pixels (light grey points) are overlaid with black crosses denoting medians in bins of 0.01\,Myr and error intervals that include 50\% of the data in each bin.}
	\label{figure: time vs column density}
\end{figure}

\begin{figure}[H]
	\centering
	\includegraphics[height=0.9\textheight]{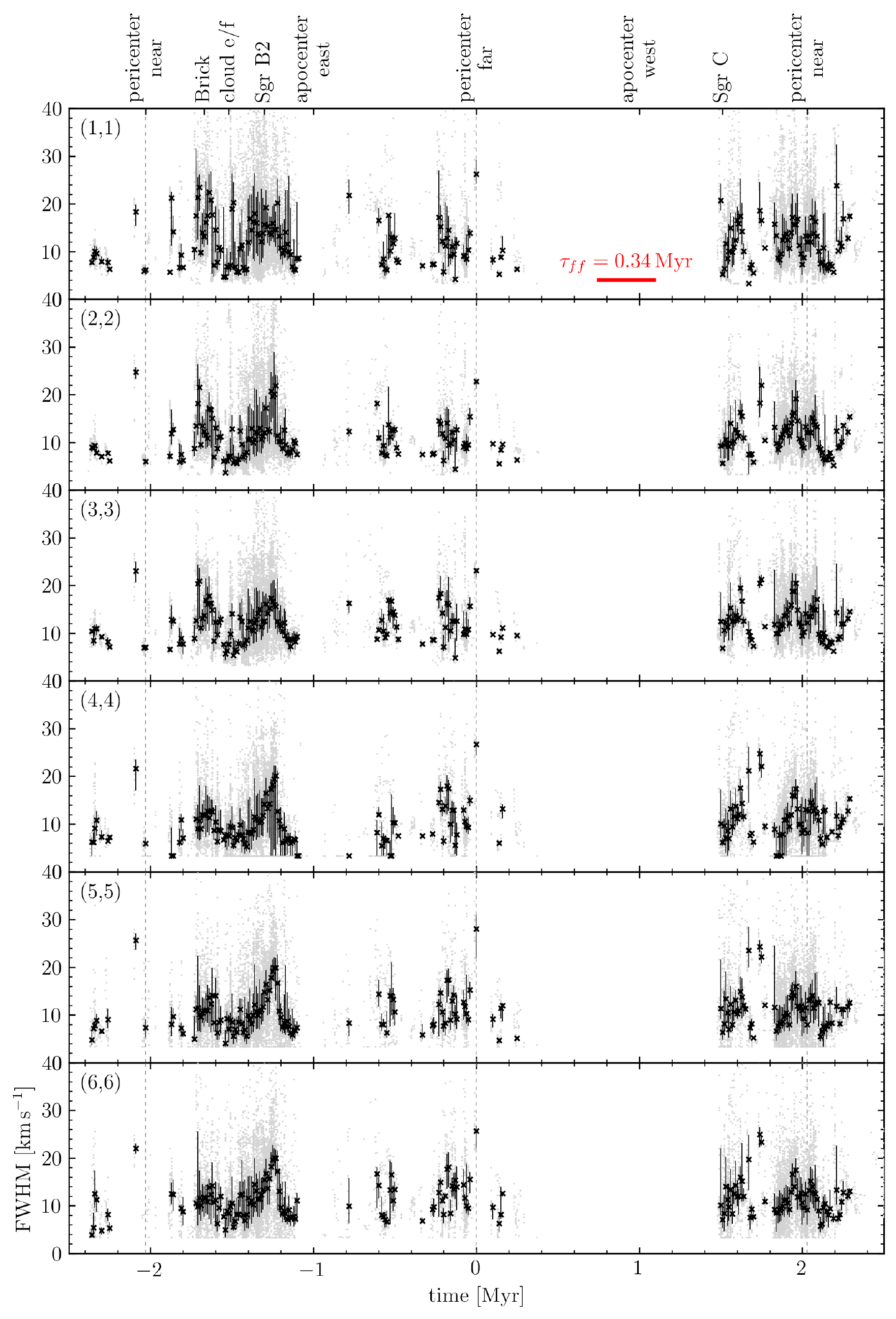}
	\caption{Ammonia line width as a function of time since far side pericenter passage (bottom x-axis, top x-axis shows important dynamical points in the K15 model and massive molecular clouds). The ammonia species is given in the top left corner of each panel. Vertical dashed lines highlight the occurrence of pericenter passages. Measurements for individual pixels (light grey points) are overlaid with black crosses denoting medians in bins of 0.01\,Myr and error intervals that include 50\% of the data in each bin.}
	\label{figure: time vs linewidth}
\end{figure}

\bibliography{bibliography.bib}

\end{document}